\documentclass[twocolumn, twocolappendix]{aastex631}
\usepackage{amsmath}
\usepackage{multirow}
\usepackage{enumitem}
\usepackage{tablefootnote}

\newcommand{\ContinuedFloat}{
  \addtocounter{figure}{-1}
  \renewcommand{\thefigure}{\arabic{figure}}
}

\newcommand{\sw}[1]{\texttt{#1}}

\shorttitle{VLASS Fast Luminous Extragalactic Transients}
\shortauthors{Sharma et al.}

\begin{document}

\title{Fast Luminous Extragalactic Transients in the VLA Sky Survey: Implications for the rates of Accretion-Induced Collapse Events, Fast Blue Optical Transients and Gamma Ray Burst Afterglows}

\correspondingauthor{Kritti Sharma}
\email{kritti@caltech.edu}

\author[0000-0002-4477-3625]{Kritti Sharma}
\affiliation{Cahill Center for Astronomy and Astrophysics, MC 249-17 California Institute of Technology, Pasadena CA 91125, USA.}

\author[0000-0002-7252-5485]{Vikram Ravi}
\affiliation{Cahill Center for Astronomy and Astrophysics, MC 249-17 California Institute of Technology, Pasadena CA 91125, USA.}
\affiliation{Owens Valley Radio Observatory, California Institute of Technology, Big Pine CA 93513, USA.}

\author[0000-0001-9584-2531]{Dillon Z. Dong}
\affiliation{National Radio Astronomy Observatory, 1003 Lopezville Road, Socorro, NM, 87801, USA.}

\author[0000-0002-7083-4049]{Gregg Hallinan}
\affiliation{Cahill Center for Astronomy and Astrophysics, MC 249-17 California Institute of Technology, Pasadena CA 91125, USA.}
\affiliation{Owens Valley Radio Observatory, California Institute of Technology, Big Pine CA 93513, USA.}

\author[0000-0002-4119-9963]{Casey Law}
\affiliation{Cahill Center for Astronomy and Astrophysics, MC 249-17 California Institute of Technology, Pasadena CA 91125, USA.}
\affiliation{Owens Valley Radio Observatory, California Institute of Technology, Big Pine CA 93513, USA.}

\author{Delina Levine}
\affiliation{Cahill Center for Astronomy and Astrophysics, MC 249-17 California Institute of Technology, Pasadena CA 91125, USA.}

\author[0000-0001-8426-5732]{Jean J. Somalwar}
\affiliation{Cahill Center for Astronomy and Astrophysics, MC 249-17 California Institute of Technology, Pasadena CA 91125, USA.}

\author{Jessie Miller}
\affiliation{Cahill Center for Astronomy and Astrophysics, MC 249-17 California Institute of Technology, Pasadena CA 91125, USA.}

\author{Nikita Kosogorov}
\affiliation{Cahill Center for Astronomy and Astrophysics, MC 249-17 California Institute of Technology, Pasadena CA 91125, USA.}

\author{Steven T. Myers}
\affiliation{National Radio Astronomy Observatory, 1003 Lopezville Road, Socorro, NM, 87801, USA.}

\begin{abstract}
Radio wavelengths offer a unique window into high-energy astrophysical phenomena that may be obscured or too rapidly evolving to be captured at other wavelengths. Leveraging data from the Very Large Array Sky Survey, we perform a systematic search for fast, luminous transients with characteristic timescales $\lesssim 3$ years in the nearby universe ($z \leq 0.3$). We report the discovery of five such transients, and classify them based on their synchrotron emission energetics and host galaxy properties. From this sample, we derive observational constraints on the volumetric rates of certain corresponding transient classes. We limit the rates of accretion-induced collapse of white dwarfs with dense circumstellar medium interaction (and those producing pulsar wind nebulae) at $\lesssim 1.10_{-0.90}^{+2.60}$\% ($\lesssim 0.20_{-0.10}^{+5.80}$\%) of the local Type Ia supernova rate, respectively, broadly consistent with theoretical predictions. For AT2018cow-like radio-bright luminous fast blue optical transients, we estimate a rare occurrence rate of $\lesssim 0.02_{-0.01}^{+0.32}$\% of the local core-collapse supernova rate. We constrain the local volumetric rates of long- and short-duration gamma-ray bursts (GRBs) to be $\lesssim 11.46_{-9.48}^{+26.28}$~Gpc$^{-3}$~yr$^{-1}$ and $\lesssim 80.88_{-66.90}^{+185.87}$~Gpc$^{-3}$~yr$^{-1}$, respectively. These estimates incorporate beaming corrections, with median detectable viewing angles derived from afterglow simulations of $\sim 0.4$ and $\sim 0.3$ radians for long- and short-duration GRBs. Our findings highlight the potential of radio surveys to uncover rare, energetic transients. We emphasize the critical role of coordinated multi-wavelength follow-up in fully characterizing these enigmatic events.
\end{abstract}

\section{Introduction}\label{sec:introduction}

Radio wavelengths offer a distinct perspective on high-energy astrophysical processes and their environments, often inaccessible in other bands. One advantage is their ability to penetrate dust and gas that can frustrate optical and X-ray observations. For instance, \citet{2007MNRAS.377..273S} found significant X-ray absorption and varying dust extinction in gamma-ray burst (GRB) afterglows, suggesting that GRBs with no optical afterglow detection might have higher visual extinction. Luminous Fast blue optical transients (LFBOTs), which exhibit blue colors with $g -r \lesssim -0.2$~mag, may also evade optical detection in dusty environments~\citep{2023ApJ...949..120H}.

The longer evolution timescales at radio wavelengths also allow for the discovery of transient classes that are faint and evolve rapidly at other wavelengths. For example, while the optical and X-ray afterglows of GRBs generally persist for days to weeks, their radio counterparts can evolve over months to years~\citep{2012ApJ...746..156C}. Such extended timescales are also valuable for discovering accretion-induced collapse (AIC) of white dwarfs, which are predicted to be $\gtrsim 5$ magnitudes fainter than typical supernovae (SNe) and fade rapidly, thus evading detection at optical wavelengths~\citep{2009MNRAS.396.1659M}. On the contrary, at radio frequencies, the spin-down of the newly formed magnetar can produce a luminous pulsar wind nebula (PWN), detectable for months to years, depending on the spin-down dynamics~\citep{2013ApJ...762L..17P}.

Radio wavelengths are especially useful for studying transients with beamed emission. For instance, GRBs observed at high energies are typically on-axis, limiting our ability to explore the angular structure of their emitting regions. Although low-luminosity GRBs and relativistic SNe are often thought to be connected to off-axis long-duration GRBs, decades of off-axis GRB searches have resulted in no confirmed discoveries~\citep{2022ApJ...938...85H}. Radio observations are more likely to detect GRBs with orphan afterglows, thus allowing for a broader investigation of the emission geometry and occurrence rate~\citep{2018ApJ...866L..22L}.

Finally, radio transients trace the presence of shock waves, sites of particle acceleration, and probe the transient environment. For stellar explosions, radio observations can unveil the interaction between ejected material and the surrounding medium, shedding light on the outflow properties and progenitor star's mass-loss history. Late-time radio emission can uncover signatures of binary companions, where the interaction with the companion results in enhanced mass-loss rates and asymmetric circumstellar medium (CSM), as often suggested for stripped-envelope SNe~\citep{2020ApJ...902...55C}.

In this work, we present a first-of-its-kind search for fast luminous extragalactic transients, evolving at timescales of a few months to a few years at gigahertz frequencies using the three epochs of publicly available Karl G. Jansky Very Large Array Sky Survey~\citep[VLASS;][]{2020PASP..132c5001L} observations. We specifically search for transients that are detected in the second VLASS epoch, but undetected in both the first and third epochs. VLASS is poised to advance studies of off-axis GRB afterglows and LFBOTs, mitigating the selection biases inherent in targeted radio follow-up observations of transients discovered at other wavelengths. The sensitivity of  VLASS also enables the discovery of PWNe formed by the AIC of white dwarfs~\citep{2013ApJ...762L..17P}. We classify the discovered sources in VLASS into various transient classes based on their host galaxy stellar population properties, host-normalized offsets, and synchrotron radio emission energetics, and leverage the contextual information from surveys at other wavelengths whenever possible. Based on our probabilistic transient classifications, we place upper limits on the volumetric rates of the underlying transient classes.

This article is structured as follows. In \S\,\ref{sec:fast_transients_search_in_the_vlass}, we present our search for fast luminous radio transients in VLASS, detailing the sample selection criteria and the methodology employed to estimate their volumetric rates. \S\,\ref{sec:optical_follow_up_archival_search} describes our optical follow-up campaign, the inferred properties of host galaxies, and the archival search for multi-wavelength counterparts. In \S\,\ref{sec:astrophysical_radio_transients_and_their_emission_mechanisms}, we place our findings within the broader context of known astrophysical radio transients and their emission mechanisms. We discuss the inferred volumetric rates for various transient classes and their implications in \S\,\ref{sec:discussion}. Finally, we summarize our key results and conclusions in \S\,\ref{sec:conclusion}. Throughout this work, we adopt the Planck18 cosmology~\citep{2020A&A...641A...6P}.

\section{Luminous Radio Transients Search in the VLA Sky Survey}\label{sec:fast_transients_search_in_the_vlass}

VLASS aims to cover the sky north of declination $-40$~degrees (33,885 square degrees) at a cadence of $\sim 3$~years, over three epochs from 2017 to 2024, and each epoch is divided into two halves. VLASS is conducted at S-band frequencies (2-4 GHz) and reaches a sensitivity of $\sim 130~\mu$Jy/beam, with an angular resolution of $\sim 2.5$\arcsec. In this section, we first discuss our forecasts for the lightcurve properties and expected luminosity distances of fast luminous transients that are detectable in VLASS (see \S\,\ref{subsec:expectations}). Based on the insights from these simulations, we define our sample selection (see \S\,\ref{subsec:sample_selection}) and elaborate on the host galaxy associations for the transients in our sample (see \S\,\ref{subsec:host_galaxy_association}). Finally, we also estimate their implied volumetric rate (see \S\,\ref{subsec:detection_rates}). 

\subsection{Expected Synchrotron Transient Signatures}\label{subsec:expectations}

The key classes of extragalactic transients of primary interest to this work that are expected to have evolution timescales of $\lesssim 3$~years include the AIC of white dwarfs in dense CSM environments or the formation of PWNe, LFBOTs, long- and short-duration GRBs. Given the aforementioned sensitivity and cadence of the VLASS, we first forecast the fast transients that may be detected in VLASS. For each transient class, we simulate a range of lightcurves given random draws from underlying physical-parameter distributions. We then identify a subset of lightcurves that will be detectable given the cadence and sensitivity of VLASS. The details of the lightcurve simulations will be discussed in Sharma et al. (in prep). Here we summarize the key insights from that analysis.
\begin{itemize}[leftmargin=*, itemsep=-0.5mm]
    \item \textit{AIC events in dense-CSM:} We use the \citet{2016ApJ...830L..38M} model for synchrotron emission from the interaction between the AIC ejecta and the dense CSM created by high mass loss rates from pre-AIC binary interactions. While the median luminosity of the simulated lightcurve suite at 3~GHz is $\log L_p = 27.5_{-1.0}^{+1.0}$ with median time above the half peak luminosity $t_{1/2} = 0.1_{-0.1}^{+0.9}$~years, the detected transients in VLASS are expected to be at a typical redshift $\bar{z} = 0.06$ with $\log L_p = 29.1_{-0.1}^{+0.1}$ and $t_{1/2} = 2.8_{-0.5}^{+0.9}$~years (where the super- and sub-script denotes the 68\% confidence interval).

    \item \textit{Formation of PWNe from AIC events:} We use the synchrotron emission model of a PWN powered by spin-down of the newly formed magnetar, as proposed by \citet{2013ApJ...762L..17P}. In contrast to the simulated lightcurves at 3~GHz with median $\log L_p = {28.8}_{-1.1}^{+0.9}$ and median $t_{1/2} = {0.1}_{-0.1}^{+0.1}$~years, the detections will be limited to luminous PWNe typically with $\log L_p = {30.5}_{-0.3}^{+0.2}$ and $t_{1/2} = {0.3}_{-0.1}^{+0.1}$~years at $\bar{z} = 0.21$.

    \item \textit{Luminous Fast Blue Optical Transients:} We use the synchrotron model of a jet launched into the dense CSM of LFBOTs from \citet{2019ApJ...871...73H}. The simulated lightcurves exhibit a median $\log L_p = {28.3}_{-0.9}^{+0.9}$ with median 
    $t_{1/2} = {0.4}_{-0.2}^{+0.6}$. On the contrary, the detected LFBOTs are expected to be more luminous ($\log L_p = {29.9}_{-0.2}^{+0.1}$) with longer evolution timescales ($t_{1/2} = {1.4}_{-0.6}^{+0.9}$~years) at $\bar{z} = 0.13$.

    \item \textit{Long-duration GRBs:} We use the relativistic numerical hydrodynamical jet simulations with synchrotron self-absorption to model the long-duration GRB radio afterglow lightcurves~\citep{2012ApJ...749...44V}. We simulate lightcurves with an isotropic distribution of observing angles, resulting in a median $\log L_p = {29.4}_{-1.0}^{+1.0}$ and a median $t_{1/2} = {2.8}_{-1.9}^{+4.7}$~years. We find that the median observing angle detectable is $\langle \theta_\mathrm{obs} \rangle = 0.41$~radians with 
    $\log L_p = {31.7}_{-0.5}^{+0.3}$ and $t_{1/2} = {1.1}_{-0.4}^{+0.6}$~years. The typical redshift of long-duration GRBs in VLASS is expected to be as high as $\bar{z} = 0.7$ and $\sim$85\% of the events are expected to be off-axis. 

    \item \textit{Short-duration GRBs:} We use the same simulation suite for short-duration GRBs as for long-duration GRBs, but switch the parameters to match those of observed short-duration GRBs. While we find a median 
    $\log L_p = {26.70}_{-1.89}^{+1.85}$ and 
    $t_{1/2} = {0.95}_{-0.77}^{+4.42}$~years of lightcurves in our simulation suite, the detectable transients will be the most luminous ones, typically with $\log L_p = {30.98}_{-0.53}^{+0.71}$ and $t_{1/2} = {0.69}_{-0.21}^{+0.48}$~years at $\bar{z} = 0.4$. For an isotropic distribution of observing angles, the detectable observing angles will be $\langle \theta_\mathrm{obs} \rangle = 0.34$~radians, with about 80\% of the events expected to be off-axis. 
    
\end{itemize}

\begin{figure*}
\includegraphics[width=\textwidth]{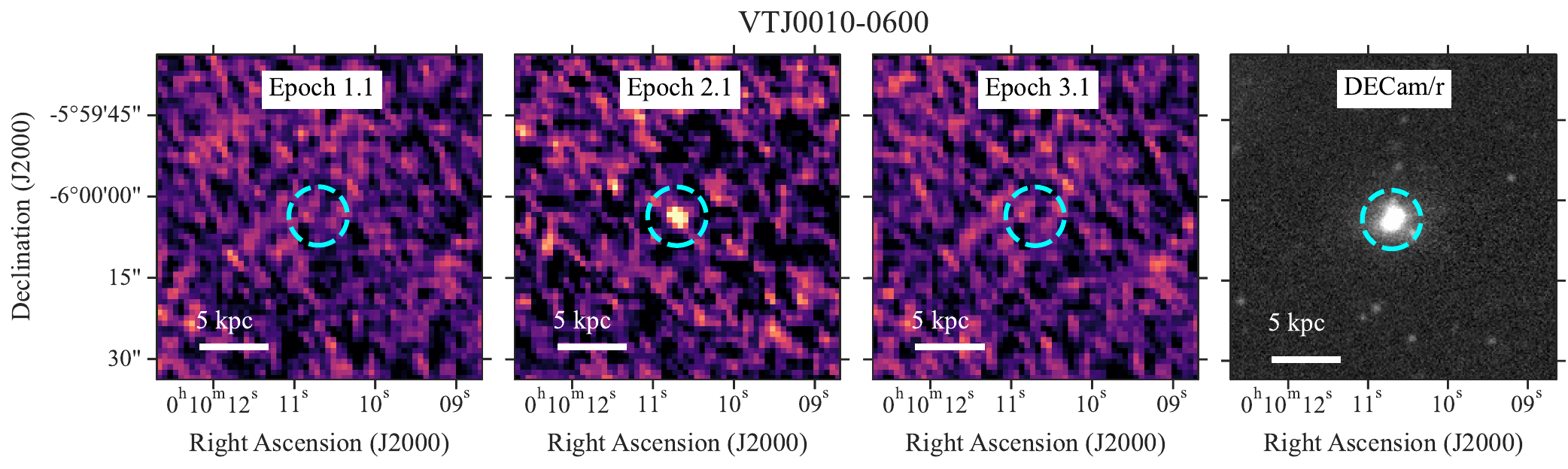}
\includegraphics[width=\textwidth]{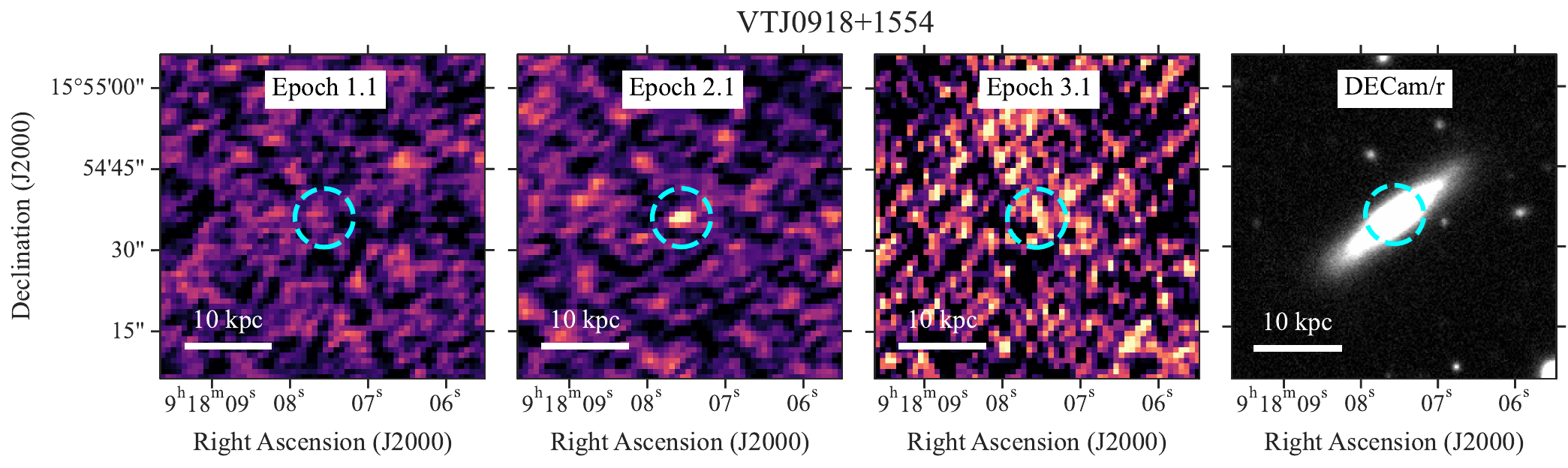}
\includegraphics[width=\textwidth]{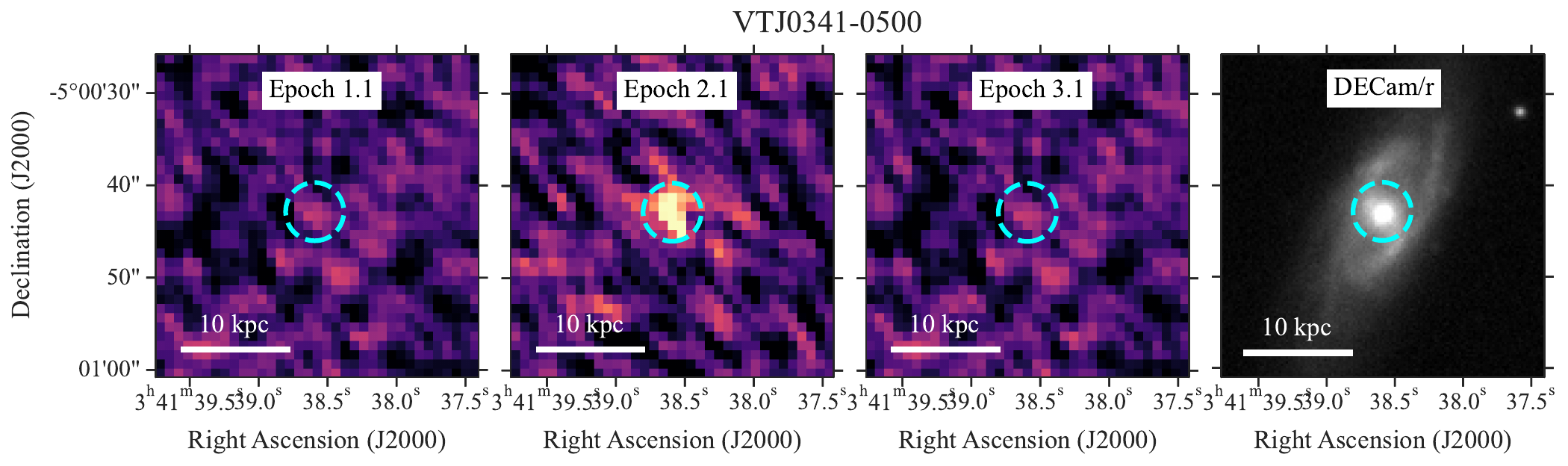}
\includegraphics[width=\textwidth]{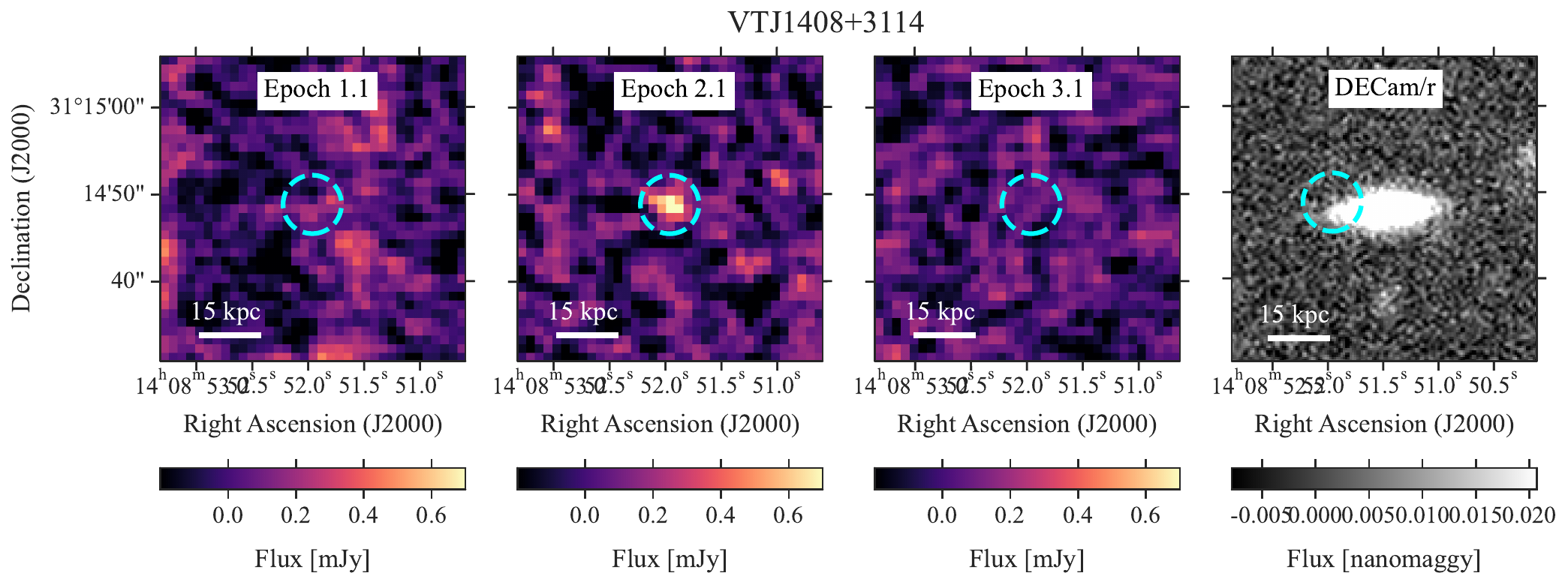}
\caption{The reference (E1.1), discovery (E2.1) and late-time (E3.1) imaging of VLASS fast luminous transients at 3~GHz in quicklook imaging. The VLASS images are centered at the transient location (blue circle) and the corresponding optical images are centered on host galaxies. The physical scales are marked on the panels for reference.}
\label{fig:cutouts}
\end{figure*}

\begin{figure*}
\ContinuedFloat
\includegraphics[width=\textwidth]{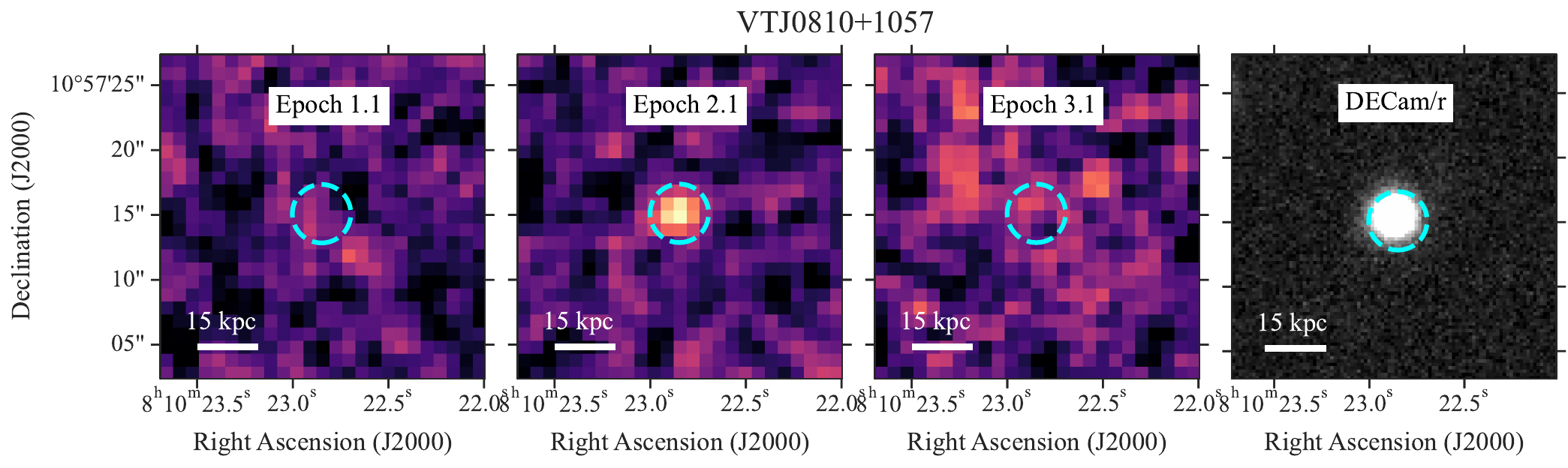}
\includegraphics[width=\textwidth]{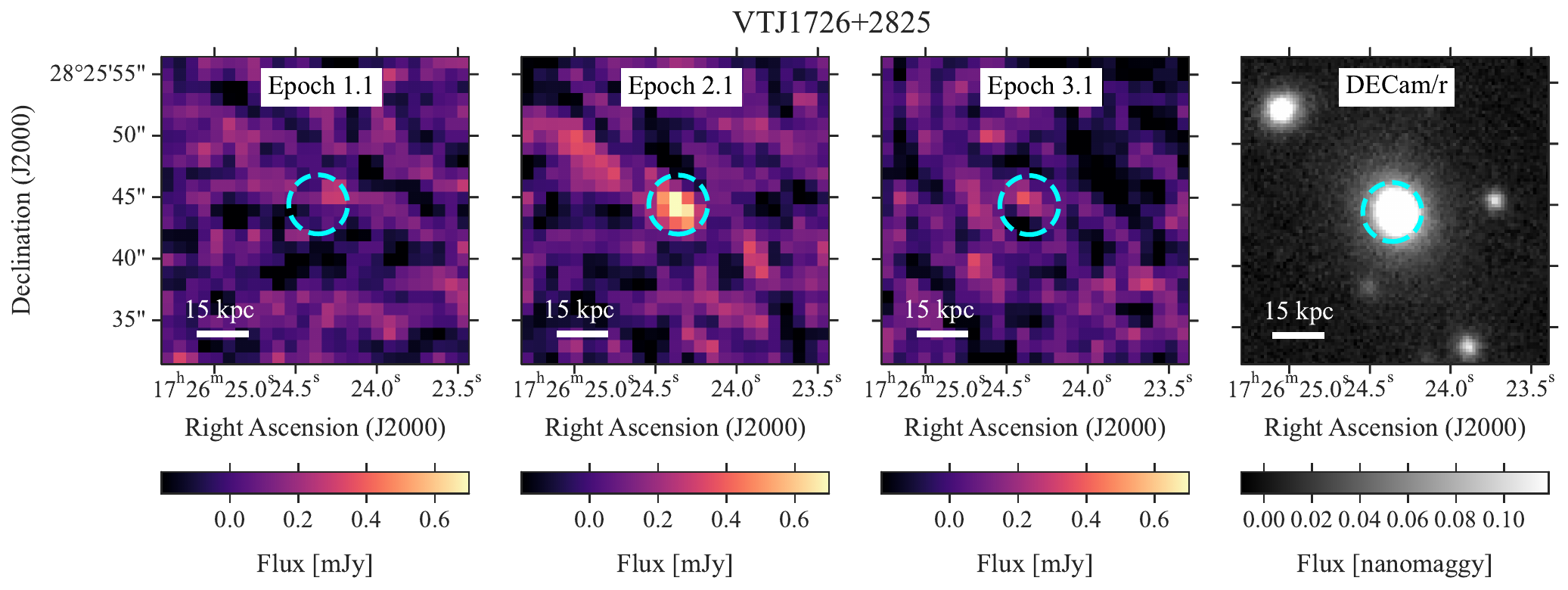}
\caption{(Continued.)}
\label{fig:cutouts_cont}
\end{figure*}

\begin{table*}
\setlength{\tabcolsep}{2.3pt}
    \centering
    \caption{S-band radio flux densities in the three epochs and host galaxies of VLASS luminous transients.}
    \begin{tabular}{lrrrrrr}
        \toprule
        Parameter & VTJ0010-0600 & VTJ0918+1554 &  VTJ0341-0500 & VTJ1408+3114 & VTJ0810+1057 & VTJ1726+2825 \\
        
        \hline

        RA (Transient) & 00:10:10.70 & 09:18:07.56 & 03:41:38.59 & 14:08:51.96 & 08:10:22.85 & 17:26:24.36 \\

        Decl. (Transient) & -06:00:03.60 & 15:54:36.00 & -05:00:42.84 & 31:14:48.84 & 10:57:15.12 & 28:25:44.40 \\
        
        $t_\mathrm{obs}$ (E1.1) & 2017-11-28 & 2017-12-28 & 2017-12-01 & 2017-10-02 & 2017-10-18 & 2017-10-02 \\

        $S_{\mathrm{3~GHz}}$ (E1.1) & $0.33^{+0.16}_{-0.15}$~mJy & $0.35^{+0.14}_{-0.15}$~mJy & $0.32^{+0.12}_{-0.13}$~mJy & $0.18^{+0.13}_{-0.14}$~mJy & $0.28^{+0.12}_{-0.12}$~mJy & $0.18^{+0.12}_{-0.13}$~mJy\\

        $t_\mathrm{obs}$ (E2.1) & 2020-07-10 & 2020-10-11 & 2020-08-07 & 2020-09-18 & 2020-09-21 & 2020-09-11 \\

        $S_{\mathrm{3~GHz}}$ (E2.1) & $1.48^{+0.17}_{-0.17}$~mJy & $1.09^{+0.13}_{-0.14}$~mJy & $1.19^{+0.16}_{-0.13}$~mJy & $0.85^{+0.12}_{-0.14}$~mJy & $0.79^{+0.11}_{-0.12}$~mJy & $0.85^{+0.11}_{-0.14}$~mJy\\

        $t_\mathrm{obs}$ (E3.1) & 2023-03-19 & 2023-01-16 & 2023-05-09 & 2023-01-22 & 2023-01-25 & 2023-01-18 \\

        $S_{\mathrm{3~GHz}}$ (E3.1) & $0.12^{+0.14}_{-0.18}$~mJy & $0.67^{+0.35}_{-0.37}$~mJy & $0.00^{+0.12}_{-0.13}$~mJy & $0.13^{+0.11}_{-0.12}$~mJy & $0.37^{+0.15}_{-0.14}$~mJy & $0.30^{+0.11}_{-0.13}$~mJy\\

        \hline

        RA (Host) & 00:10:10.68 & 09:18:07.56 & 03:41:38.59 & 14:08:51.48 & 08:10:22.87 & 17:26:24.34 \\

        Decl. (Host) & -06:00:03.24 & 15:54:35.64 & -05:00:43.20 & 31:14:48.12 & 10:57:15.48 & 28:25:44.40 \\

        $z$ (Host) & $0.0154	\pm 0.0001$ & $0.0311 \pm	0.0001$ & $0.0429	\pm 0.0001$ & $0.1180 \pm	0.0001$ & $0.1964 \pm	0.0000$ & $0.2224 \pm	0.0002$ \\

        M$_\mathrm{AB}$ & $18.61 \pm 0.20$ & $15.52 \pm 0.20$ & $15.45 \pm 0.21$ & $18.47 \pm 0.2$ & $18.45 \pm 0.01$ & $17.71 \pm 0.20$ \\
        
        E(B-V) & 0.031 & 0.034 & 0.049 & 0.012 & 0.042 & 0.047 \\

        Offset & 0.508\arcsec & 0.360\arcsec & 0.360\arcsec & 6.197\arcsec & 0.504\arcsec & 0.317\arcsec \\

        & 0.164~kpc & 0.235~kpc & 0.324~kpc & 15.333~kpc & 2.077~kpc & 1.476~kpc \\

        Offset [$R_e$] & 0.143 & 0.037 & 0.035 & 1.998 & 1.035 & 0.650 \\

        \hline
    \end{tabular}
    \label{table:vlass_summary}
\end{table*}

\subsection{Sample Selection}\label{subsec:sample_selection}

We compile our radio-selected fast luminous transients sample using half of the sky from three VLASS epochs and identify transients as sources that were detected with $>7\sigma$ significance in the second epoch (E2.1) but not detected at $3\sigma$-level in first (E1.1) and third (E3.1) epochs. The $7\sigma$ detection threshold was chosen to minimize the false alarm rate to 1~yr$^{-1}$~\citep{2015ApJ...806..224M}. The transient detection algorithm employed in the preparation of this catalog will be presented in Dong et al. (in prep). In the past, this pipeline has been successfully used to uncover longer-duration transients, such as a candidate extragalactic PWN~\citep{2023ApJ...948..119D}, and a merger-driven core-collapse SN~\citep{2021Sci...373.1125D}.

We identified 245 transient candidates that pass our detection criterion in the first half epochs of VLASS. We remove the transients with no plausible optical source (galaxies/stars) detectable down to an imaging depth of $23.5$~mag within $10$\arcsec in archival $r$-band data from PanSTARRS1~\citep[PS1;][]{2016arXiv161205560C}, Beijing-Arizona Sky Survey~\citep[BASS;][]{2017PASP..129f4101Z} and Mayall z-band Legacy Survey~\citep[MzLS;][]{2019AJ....157..168D} data from the Dark Energy Survey~\citep{2018ApJS..239...18A}. This filter removes 107 transient candidates (we note that the surveys used for host galaxy searches are fairly complete to low-mass galaxies in the redshift range of interest, as discussed below). Next, we remove transients where the associated source is classified as a star in the Legacy Survey catalog, the PS1-Point Source Catalog \citep{2018PASP..130l8001T, 2005cs........2072O} or GAIA public catalogs~\citep{2016A&A...595A...1G, 2023A&A...674A...1G}. We also remove potential Galactic sources by rejecting candidates with Galactic latitude $|b| \leq 15$ degrees. This filter removes 111 transients, which may be stellar flares.

Leveraging the insights from the overview in last section, we limit our sample to redshift $z \leq 0.3$. We use photometric redshifts from Legacy survey catalogs and PS1~\citep{2021MNRAS.500.1633B} to remove sources at higher redshifts. Out of 27 luminous fast radio transients with a potential host galaxy, 25 have Legacy photometric redshifts, which are known to be fairly accurate over the redshift range of interest~\citep{2022MNRAS.512.3662D}. This redshift cut leaves us with 6 transient sources, which forms our luminous radio transients sample. We summarize the observed flux densities in E2.1 and flux density limits in E1.1 and E3.1 for these transients in Table~\ref{table:vlass_summary} and Figure~\ref{fig:cutouts} shows the reference (E1.1), discovery (E2.1), and late-time (E3.1) radio imaging of the sources, together with their host galaxies.

\subsection{Host Galaxy Association} \label{subsec:host_galaxy_association}

We quantify the host-galaxy association probability for our transients to determine whether they are likely associated with the presumed host galaxy or merely background sources. To achieve this, we use the framework developed by \citet{2002AJ....123.1111B} and \citet{2017ApJ...849..162E}, which calculates the probability of chance coincidence ($P_{cc}$) between a transient and a nearby galaxy. 

We begin by fitting the $r$-band galaxy number counts as presented in \citet{2016ApJ...827..108D}. This allows us to calculate the projected areal number density of galaxies, denoted as $\sigma(\leq m)$, for galaxies brighter than a given r-band magnitude ($m$). The projected areal density is essential for estimating the likelihood of random alignment of galaxies with transients. The $P_{cc}$ for a transient occurring within a radius $R$, is computed under the assumption of a Poisson distribution of galaxies across the sky and is given by $P_{cc} = 1 - e^{-\pi R^2 \sigma(\leq m)}$, where $R = \max \left(2R_{\mathrm{loc}},~\sqrt{R_0^2 + 4R_h^2}\right)$, $R_{\mathrm{loc}}$ represents the localization limit, set by the VLASS angular resolution, $R_0$ is the radial angular separation between the transient and its presumed host galaxy, and $R_h$ is the half-light radius of the galaxy in question. The uncertainty in the offsets is determined by the subsampling of the VLASS quick look images used in this work, which we assume to contribute an additional 0.5\arcsec uncertainty.

Our analysis reveals that the probability of chance coincidence for all VLASS fast luminous transients in our sample is $P_{cc} < 0.001$. This low probability indicates that these transients are indeed associated with their respective host galaxies with more than $6\sigma$ confidence. Consequently, the observed transients are not likely to be background sources but are physically associated to their presumed host galaxies.

\begin{figure}
\includegraphics[width=\columnwidth]{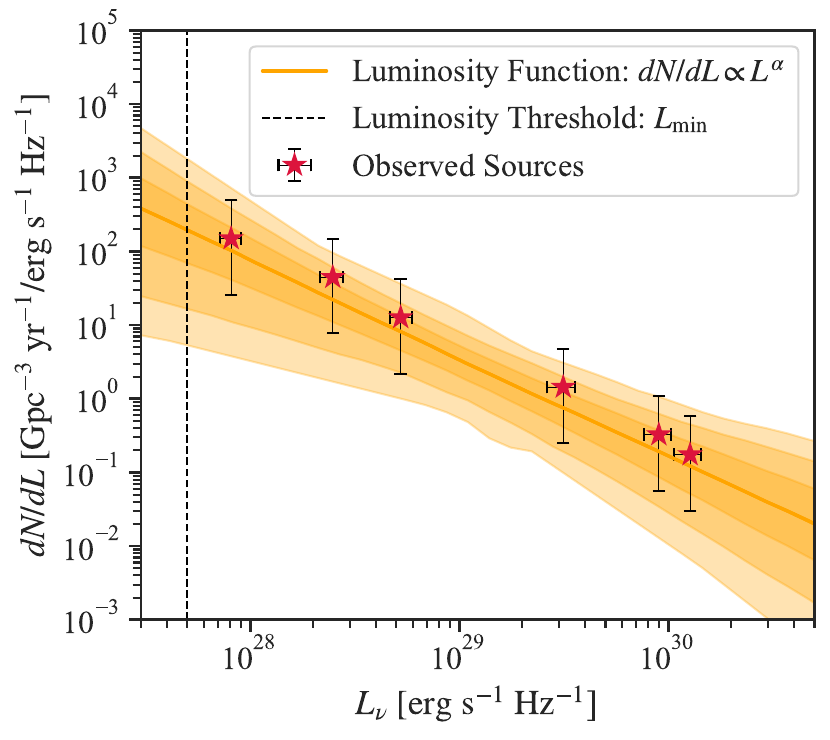}
\caption{Luminosity function of VLASS fast luminous radio transients. The red data points represent the observed number densities of transients as a function of luminosity, with error bars indicating the total Poisson and measurement uncertainties. The plot illustrates the luminosity function fitted with a power-law model $dN/dL \propto L^\alpha$. The solid orange line denotes the median fit with power-law index $\alpha = -1.32^{+0.25}_{-0.28}$ derived from our Monte Carlo fitting procedure. The shaded region represents the $1, 2, 3\sigma$ bands of the fit. The vertical dashed line indicates the minimum luminosity threshold ($L_\mathrm{min}$) used for volumetric rate estimation, corresponding to the luminosity of a $7\sigma$ detection at the distance of the closest transient candidate. The estimated volumetric rate of such  transients is $\mathcal{R} = 39.7_{-18.2}^{+46.4}$~Gpc$^{-3}$~yr$^{-1}$, assuming a transient evolution timescale of 1~yr in VLASS.}
\label{fig:rates_all}
\end{figure} 

\subsection{Detection Rates}\label{subsec:detection_rates}

For each transient with observed luminosity $L$, the expected rate in a differential luminosity bin $dL$ at $L$ can be approximated as $\left( f_\mathrm{eff} \cdot \frac{4\pi}{3} d_\mathrm{L,max}^3 \cdot t_\mathrm{span} \right)^{-1} \cdot \frac{t_\mathrm{span}}{\tau}$, where $t_\mathrm{span}$ represents the VLASS observing cadence of approximately 3 years, and $\tau$ denotes the evolution timescale of the transient phenomena. The maximum luminosity distance $d_\mathrm{L,max}$, out to which each transient would be detectable by the VLASS, given their observed luminosities are estimated to be approximately 112 Mpc, 168 Mpc, 256 Mpc, 528 Mpc, 866 Mpc, and 1 Gpc, respectively. Even at these distances, their host galaxies would have been detectable in archival optical surveys, thus justifying our sample selection. Since our search includes data from each first half epoch of VLASS, the area of the sky covered in our search is $\approx 16,940~\mathrm{deg}^2$. The effective search volume fraction, therefore, is $f_\mathrm{eff} \approx {16,940~\mathrm{deg}^2}/{41,253~\mathrm{deg}^2} \approx 0.41$.

Following the \citet{Dong2023} formulation, we assume a power-law form for the luminosity function $dN/dL \propto L^\alpha$, where $\alpha$ is the power-law index, applicable for luminosities $L > L_\mathrm{min}$, where the minimum luminosity $L_\mathrm{min}$ is defined as the luminosity of a transient with flux at the $7\sigma$ detection threshold, placed at the distance of the closest transient candidate. In our fitting procedure, we also account for Poisson uncertainties. The luminosity function fit to our sample is shown in Figure~\ref{fig:rates_all}, which yields a power-law index of $\alpha = -1.32^{+0.25}_{-0.28}$. The transient event rate, obtained by integrating the luminosity function down to the minimum luminosity, $L_\mathrm{min} = S_{\nu,7\sigma} \cdot 4\pi d_{\mathrm{min, obs}}^2 \approx 5 \times 10^{27}$ erg s$^{-1}$ Hz$^{-1}$ is $\mathcal{R} = 39.68_{-18.21}^{+46.43} \left(\frac{\tau}{1~\mathrm{yr}} \right)^{-1}~\mathrm{Gpc}^{-3}~\mathrm{yr}^{-1}$, where the evolution timescale of 1~yr was chosen based on our expectations for fast luminous transients in VLASS (see \S\,\ref{subsec:expectations}) and the significant margin of error is primarily attributed to the Poisson noise.

\begin{figure*}[ht]
\includegraphics[width=0.5\textwidth]{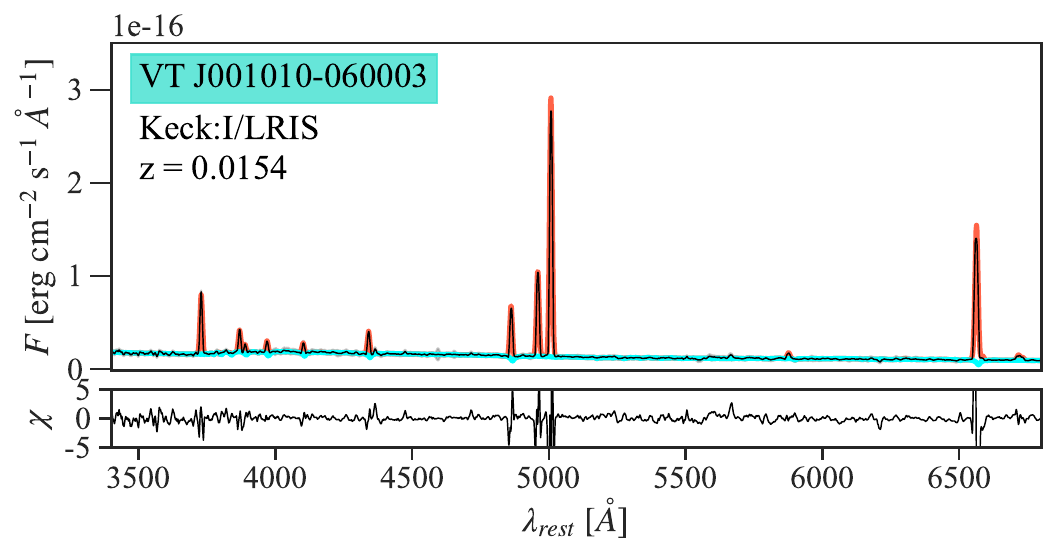}
\includegraphics[width=0.5\textwidth]{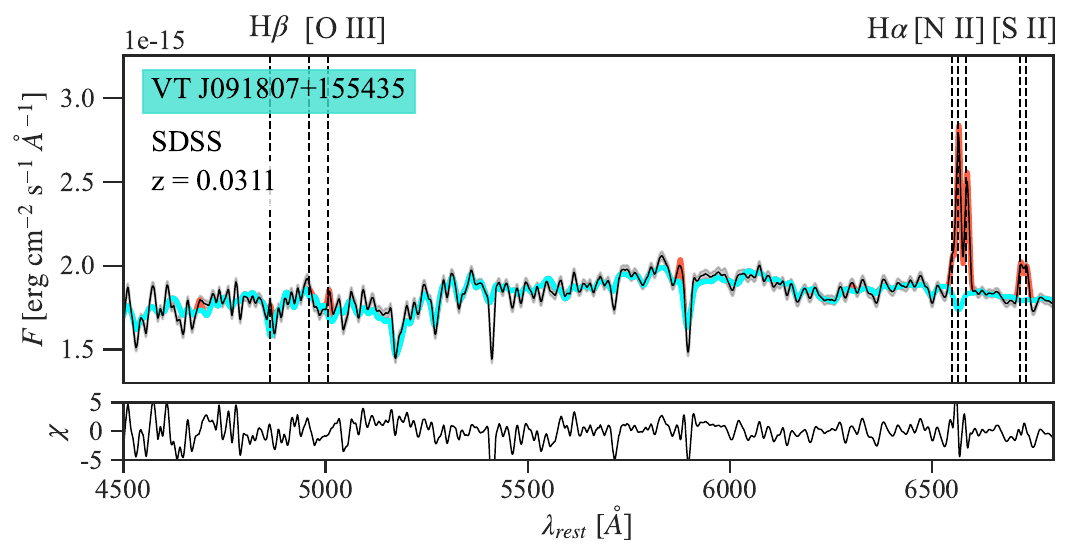}
\includegraphics[width=0.5\textwidth]{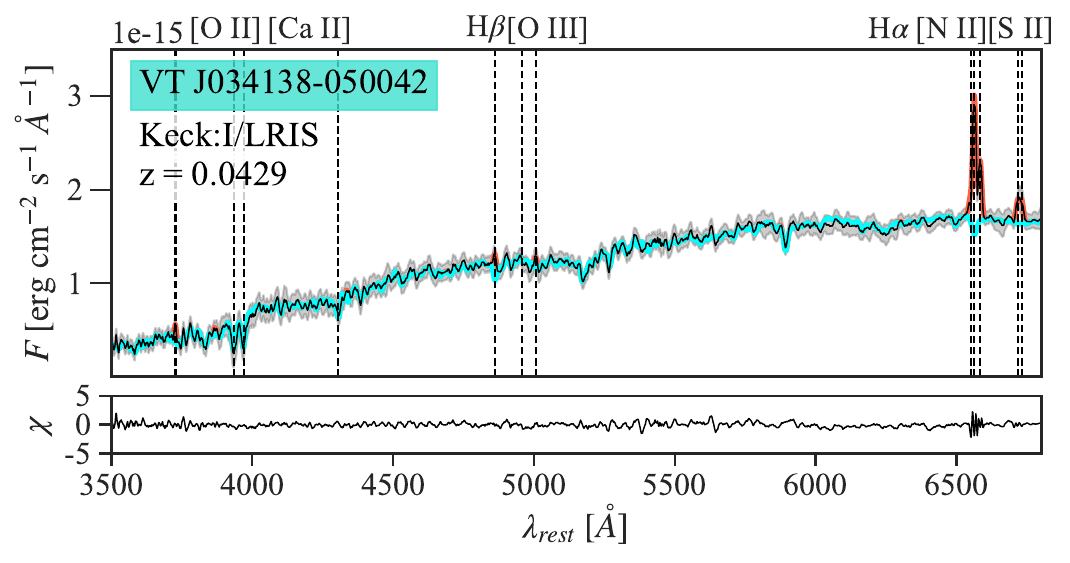}
\includegraphics[width=0.5\textwidth]{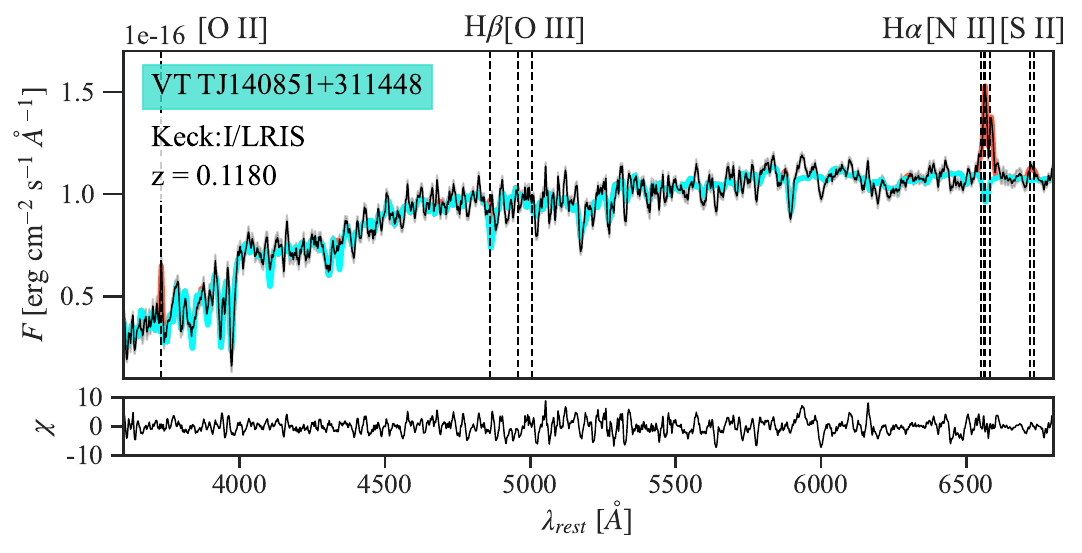}
\includegraphics[width=0.5\textwidth]{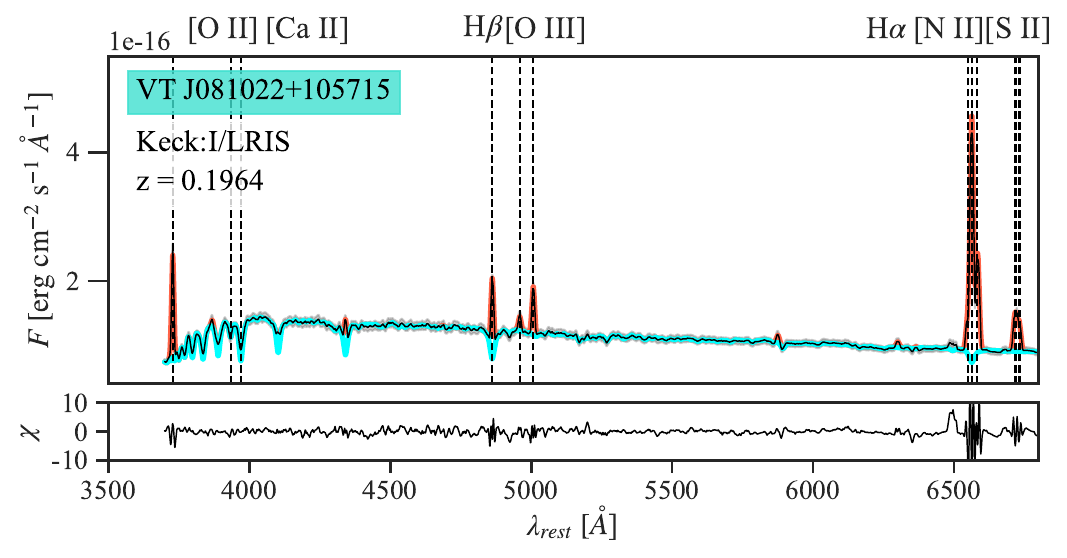}
\includegraphics[width=0.5\textwidth]{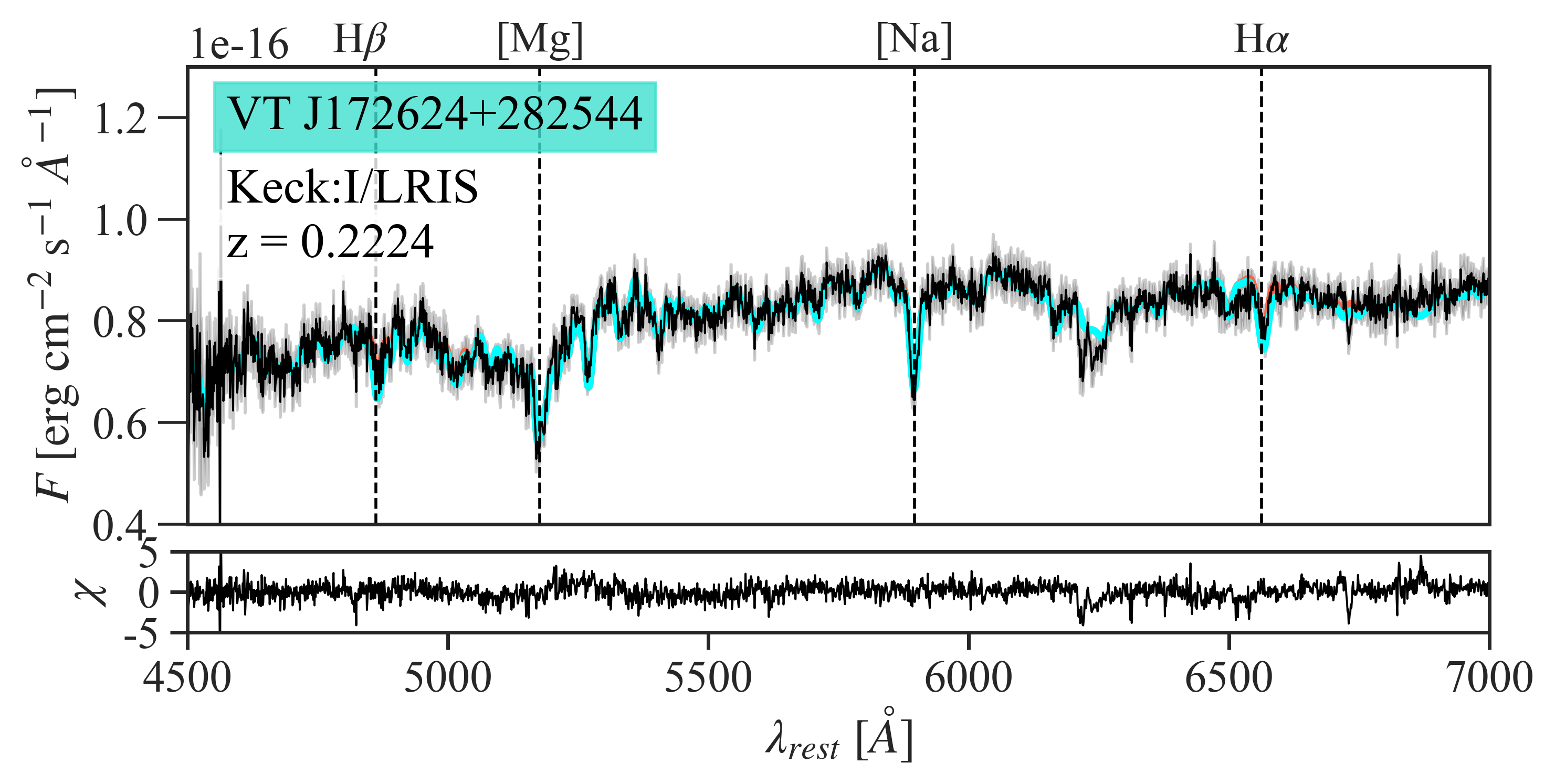}
\caption{Optical spectra of VLASS transients host galaxies. For VTJ0918+1554, the archival SDSS spectrum provides a $\Delta \lambda/\lambda \sim 2000$ resolution. The spectra for the other five host galaxies were obtained using the Keck-I:LRIS (see Table~\ref{table:spectroscopy_summary}) with a $\Delta \lambda/\lambda \sim 1000$ resolution. Each spectrum is shown in the rest-frame of the host galaxy, with key emission lines and stellar absorption features labeled. The fitted models using \sw{pPXF} are overlaid, demonstrating the separation of stellar continuum (cyan) and nebular emission (red). The spectrum of VTJ0010-0600 is dominated by SN emission.}
\label{fig:spectra_ppxf_fits}
\end{figure*}

\section{Optical Follow-up and Archival Multi-wavelength Signatures}\label{sec:optical_follow_up_archival_search}

In this section, we analyze properties of the host galaxies associated with our sample of fast luminous VLASS transients. We examine the optical spectroscopy (see \S\,\ref{subsec:optical_spectroscopy}) to compute the star formation rates (SFRs), dust content, and identify any potential active galactic nuclei (AGN) activity within these galaxies (see \S\,\ref{subsec:optical_emission_line_features}). We delve into the stellar population properties of these host galaxies, employing advanced modeling techniques (see \S\,\ref{subsec:host_galaxy_stellar_population_modeling}). Finally, we conduct a thorough search for archival transient counterparts, incorporating multi-wavelength data to understand the nature of these transients (see \S\,\ref{subsec:search_for_archival_transient_counterparts}).

\begin{table}
    \centering
    \caption{Log of Spectroscopic Observations obtained with Keck:I/LRIS.}
    \begin{tabular}{lll}
        \toprule
        Transient & UT Date & Exp Time [s] \\
        \hline

        VTJ0010-0600 & 2023-10-15 & 600 \\
        
        VTJ0341-0500 & 2023-10-15 & 600 \\
        
        VTJ1408+3114 & 2024-05-29 & 900 \\ 
        
        VTJ0810+1057 & 2023-12-17 & 600 \\
        
        VTJ1726+2825 & 2024-06-10 & 900 \\

        \hline
    \end{tabular}
    \label{table:spectroscopy_summary}
\end{table}

\begin{figure*}[ht]
\includegraphics[width=\textwidth]{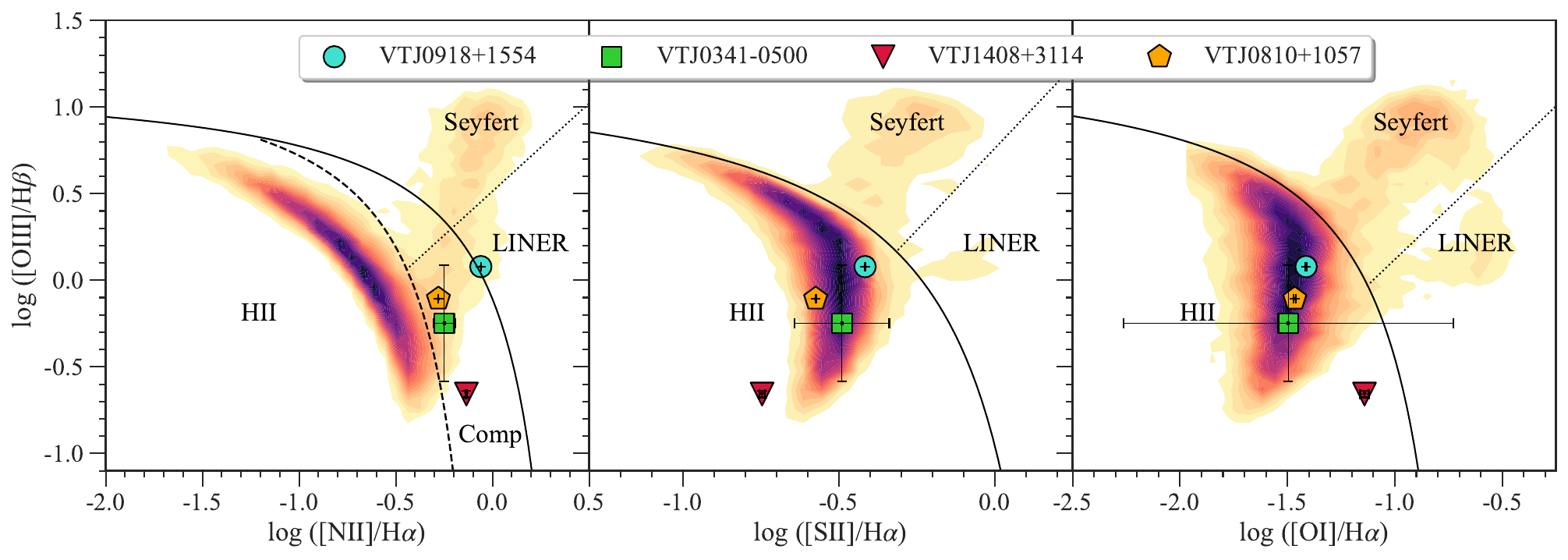}
\caption{Determining the primary ionization mechanism within the host galaxies of our VLASS transients sample using the BPT empirical optical emission-line diagnostic diagrams~\citep{1981PASP...93....5B}. We include the maximum starburst line~\citep[solid;][]{2001ApJ...556..121K}, the demarcation between pure star-forming galaxies and Seyfert-H II composite objects~\citep[dashed;][]{2003MNRAS.346.1055K}, and the line separating Seyfert galaxies from LINERs~\citep[dotted;][]{2006MNRAS.372..961K}. For context, the emission line diagnostics for SDSS galaxies are displayed in the background~\citep{2015ApJS..219...12A}. We omit VTJ0010-0600 from this plot because its emission lines are dominated by an SN. We also omit VTJ1726+2825 due to a lack of emission lines in the spectrum. We observe that the 4 galaxies are consistent with the locus of star-forming galaxies.}
\label{fig:BPT_analysis}
\end{figure*}

\newpage
\subsection{Optical Spectroscopy}\label{subsec:optical_spectroscopy}

The optical spectrum of a galaxy provides vital information about its recent SFR, the age of its stellar population, dust content, and potential AGN activity. For the host galaxies of all VLASS transients, we searched the Sloan Digital Sky Survey~\citep[SDSS;][]{2015ApJS..219...12A} for archival spectra. The host galaxy of VTJ0918+1554 was found to have an archival spectrum with a resolution of $\Delta \lambda/\lambda \sim 2000$, obtained on 24 March 2006, approximately 14 years before the VLASS detection. For the other five transients, we conducted follow-up observations to obtain spectra, using the Low Resolution Imaging Spectrometer~\citep[LRIS;][]{1995PASP..107..375O} at the W.M. Keck Observatory (see Table~\ref{table:spectroscopy_summary} for a log of observations). In our observations, we employed the 400/3400 grism on the blue arm and the 400/8500 grating centered at 7830~$\AA$ on the red arm, achieving a spectral resolution of $\Delta \lambda/\lambda \sim 1000$. The 1\arcsec wide long slit was oriented along the direction of each transient. Data reduction was performed using the \sw{LPipe} software~\citep{2019PASP..131h4503P}. All spectra were flux-calibrated using observations of standard stars. We correct for slit losses by scaling the spectra to match the photometry of the galaxies.

To analyze the spectra, we measured the spectroscopic redshifts and emission line fluxes using the Penalized PiXel-Fitting (\sw{pPXF}) software~\citep{2022arXiv220814974C, 2017MNRAS.466..798C}. This method allows for the simultaneous fitting of the stellar continuum and nebular emission lines, using templates from the MILES stellar library~\citep{2006MNRAS.371..703S}. The \sw{pPXF} fits to the reduced spectra in the rest-frame of the host galaxies are shown in Figure~\ref{fig:spectra_ppxf_fits}. For the emission line analysis, we subtract the stellar continuum to isolate the nebular lines, which are then fitted with Gaussian profiles to measure their fluxes. 

\setlength{\tabcolsep}{2.4pt}

\begin{table*}[ht!]
    \centering
    \caption{Host galaxy properties of VLASS fast luminous radio transients.}
    \begin{tabular}{lllllll}
        \toprule
        Parameter & VTJ0010-0600 & VTJ0918+1554 &  VTJ0341-0500 & VTJ1408+3114 & VTJ0810+1057 & VTJ1726+2825 \\
        
        \hline

        $z$ & $0.0154 \pm 0.0001$ & $0.0311 \pm 0.0001$ & $0.0429 \pm 0.0001$ & $0.1180 \pm 0.0001$ & $0.1964 \pm 0.0001$ & $0.2224 \pm 0.0002$ \\
        
        $\log M_\ast/M_\odot$ & $7.81_{-0.01}^{+0.01}$ & $10.53_{-0.01}^{+0.01}$ & $10.32_{-0.01}^{+0.01}$ & $10.64_{-0.01}^{+0.01}$ & $10.20_{-0.01}^{+0.01}$ & $11.36_{-0.01}^{+0.01}$ \\

        $\log Z/Z_\odot$ & $0.19_{-0.01}^{+0.01}$ & $-0.21_{-0.01}^{+0.01}$ & $-0.34_{-0.01}^{+0.01}$ & $-0.24_{-0.01}^{+0.01}$ & $-0.65_{-0.01}^{+0.01}$ & $-0.09_{-0.01}^{+0.01}$ \\

        SFR$_\mathrm{100~Myr}$ & $0.04_{-0.01}^{+0.01}$ & $0.19_{-0.01}^{+0.01}$ & $0.48_{-0.01}^{+0.01}$ & $3.47_{-0.08}^{+0.09}$ & $5.42_{-0.13}^{+0.13}$ & $0.25_{-0.03}^{+0.03}$ \\

        SFR$_{\mathrm{H}\alpha}$ & - & 10.27 & 10.98 & 0.03 & 2.82 & $\approx 0$ \\
        $\mathrm{H}\alpha/\mathrm{H}\beta$ & - & 5.03 & 4.62 & 2.23 & 2.79 & -- \\

        $t_m$~[Gyr] & $6.56_{-0.09}^{+0.09}$ & $8.65_{-0.02}^{+0.02}$ & $7.66_{-0.08}^{+0.06}$ & $9.03_{-0.11}^{+0.10}$ & $2.98_{-0.06}^{+0.09}$ & $7.13_{-0.08}^{+0.06}$ \\

        $M_r^0$~[mag] & $-15.70_{-0.34}^{+0.02}$ & $-20.13_{-0.08}^{+0.06}$ & $-20.69_{-0.08}^{+0.09}$ & $-19.84_{-0.16}^{+0.19}$ & $-21.33_{-0.23}^{+0.08}$ & $-22.07_{-0.04}^{+0.46}$ \\

        $(g-r)^0$~[mag] & $-0.02_{-0.01}^{+0.01}$ & $0.78_{-0.01}^{+0.01}$ & $0.52_{-0.01}^{+0.01}$ & $0.94_{-0.01}^{+0.01}$ & $0.61_{-0.01}^{+0.01}$ & $1.32_{-0.01}^{+0.01}$ \\

        $(u-r)^0$~[mag] & $0.64_{-0.01}^{+0.01}$ & $2.38_{-0.01}^{+0.01}$ & $1.78_{-0.01}^{+0.01}$ & $2.31_{-0.01}^{+0.01}$ & $1.63_{-0.02}^{+0.02}$ & $3.21_{-0.01}^{+0.03}$ \\

        \hline
    \end{tabular}

    {\raggedright N{\scriptsize{OTE}}: The tabulated values are the median and 68\% confidence intervals. $\log M_{\ast}$ denotes the stellar mass and $\log (Z/Z_\odot$) represents stellar metallicity. SFR$_{\mathrm{H}\alpha}$ and SFR$_{\mathrm{100~Myr}}$ represents the SFR of the galaxy measured using dust extinction corrected H$\alpha$ flux and SFR averaged over the 100~Myr, as measured from stellar population modeling, respectively. t$_\mathrm{m}$ is the mass-weighted age of the galaxy. M$_r^0$, $(g-r)^0$ and $(u-r)^0$ denotes the rest-frame $r$-band magnitude and colors.
    \par}
    \label{table:host_properties}
\end{table*}

\begin{figure}
\includegraphics[width=\columnwidth]{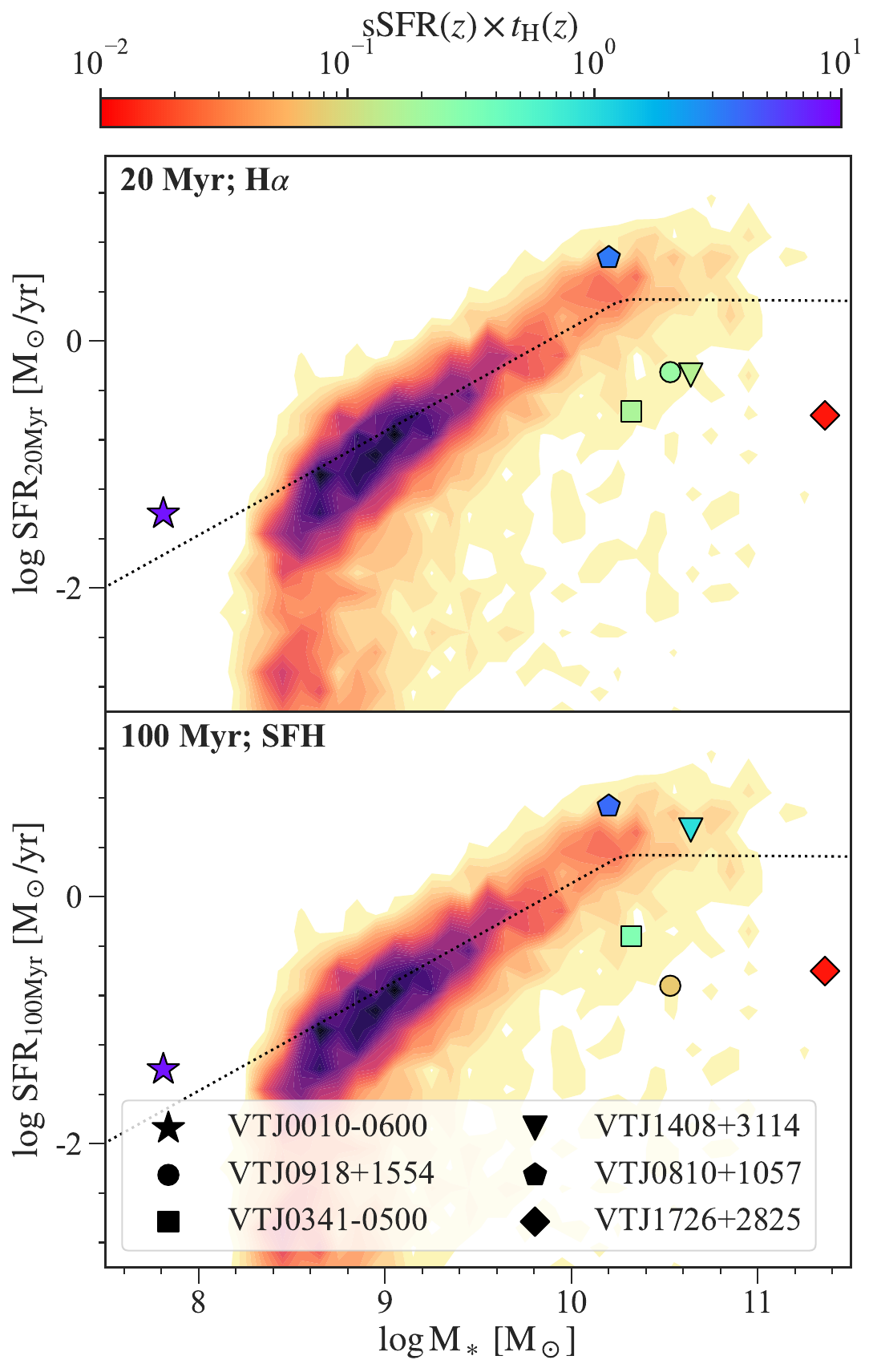}
\caption{Comparison of VLASS host galaxies with background galaxy population. We use the COSMOS-2015~\citep{2016ApJS..224...24L, 2020ApJ...893..111L} and 3D-HST~\citep{2014ApJS..214...24S} galaxy catalogs at $z \leq 0.3$ to represent the field galaxy population. For reference, the center of the star-forming main sequence at $z = 0.15$ is marked~\citep{2022ApJ...936..165L}. The markers denote the SFR averaged over the past 20~Myr (100 Myr) in the top (bottom) panel and are colored by the degree of star-formation~\citep{2022ApJ...926..134T}. 
Notably, all but the host galaxy of VTJ1726+2825 are star-forming galaxies.
}
\label{fig:gal_sfms}
\end{figure}

\subsection{Optical Emission Line Features}\label{subsec:optical_emission_line_features}

We employ the optical emission line ratio diagnostics to investigate the presence of strong AGN activity in the host galaxies of VLASS transients. This analysis is crucial, given that many VLASS transients are situated near the centers of their respective galaxies, where AGN activity is most likely to be observed. To assess potential AGN signatures, we use the Baldwin, Phillips, and Terlevich (BPT) empirical optical emission-line diagnostic diagrams \citep{1981PASP...93....5B}, as shown in Figure~\ref{fig:BPT_analysis}. We exclude two host galaxies from this analysis. The host galaxy of VTJ0010-0600 was omitted due to the dominance of SN emission (see Section~\ref{subsec:search_for_archival_transient_counterparts}). The broad emission line features in its spectrum are indicative of a Type Ib/c SN~\citep{2020ATel14020....1H}. Additionally, the host galaxy of VTJ1726+2825, being a quiescent galaxy devoid of emission line features, was also excluded.

For the remaining galaxies, our analysis indicates that all four host galaxies exhibit line ratios consistent with the locus of background star-forming galaxies. We observed an offset in the [NII]/H$\alpha$ measurements, deviating from the star-forming galaxy locus. This discrepancy is likely attributed to systematic errors, possibly due to H$\alpha$ flux contamination in the [NII] measurements. Notably, our findings align with the galaxies' classification using WISE color-color diagnostics~\citep{2010AJ....140.1868W}, reinforcing the conclusion that none of these galaxies exhibit observable AGN activity.

The host galaxies of VTJ0918+1554, VTJ0341-0500, VTJ1408+3114, and VTJ0810+1057 demonstrate active star formation. To quantify this, we use their H$\alpha$ fluxes to estimate SFRs based on the \citet{1994ApJ...435...22K} calibration. The host galaxies of VTJ0918+1544 and VTJ0341-0500 display significant Balmer decrements of 5.03 and 4.62, respectively. These decrements suggest substantial dust extinction, necessitating corrections to their SFRs. After applying dust extinction corrections, the SFRs over the past 20~Myr, derived from these H$\alpha$ line fluxes, are summarized in Table~\ref{table:host_properties}. Our results indicate that the host galaxies of VTJ0918+1554, VTJ0341-0500, and VTJ0810+1057 are actively star-forming, with SFRs $\gtrsim 3$~M$_\odot$ yr$^{-1}$. In contrast, the host galaxy of VTJ1408+3114 is predominantly quiescent, exhibiting an SFR of only 0.03~M$_\odot$ yr$^{-1}$, and the host galaxy of VTJ1726+2825 exhibits no measurable ongoing star formation.

\subsection{Host Galaxy Stellar Population}\label{subsec:host_galaxy_stellar_population_modeling}

We measure the host galaxy stellar population properties using the \sw{Prospector} software~\citep{2021ApJS..254...22J} to perform Bayesian forward modeling of photometric and spectroscopic data for host galaxies. We obtain photometry from archival optical to near-infrared (NIR) imaging surveys such as PS1, BASS, MzLS, SDSS, Two Micron All Sky Survey~\citep[2MASS;][]{2006AJ....131.1163S}, and Wide-field Infrared Survey Explorer~\citep[WISE;][]{2014yCat.2328....0C}. We use models similar to \citet{2024Natur.635...61S} and briefly summarize the key components here. We use \sw{dynesty}, a dynamic nested sampling routine known for its robustness in exploring complex parameter spaces, for posterior sampling~\citep{Speagle_2020}.

To accurately capture the SFH of these galaxies, we opt for a non-parametric modeling approach, which uses a piecewise constant SFH to define the stellar mass formed during each time bin. We adopted the Kroupa initial mass function~\citep{2001MNRAS.322..231K} and incorporated the mass-metallicity relation~\citep{2005MNRAS.362...41G} to mitigate the age-metallicity degeneracy. Our model also accounts for dust attenuation effects, with free parameters for the wavelength-dependent optical depth normalization, additional attenuation toward young stars, and slope deviation from a canonical attenuation curve~\citep{2000ApJ...533..682C}. We include dust emission~\citep{2007ApJ...657..810D} in our model when data at wavelengths above 1~$\mu$m are available. Furthermore, due to significant mid-infrared emission, we incorporated an AGN dust torus model~\citep{2008ApJ...685..160N}. Additionally, our model includes a nebular emission model, linking gas-phase and stellar metallicity and allowing for variations in nebular ionization parameters.

Given the inherent systematics in photometric measurements, spectrum calibration, and uncertainties in stellar and photoionization models, we assume 10\% additional photometric errors~\citep{2021ApJS..254...22J} and a 12th-order Chebyshev polynomial as multiplicative calibration function. Our model includes spectral smoothing to account for line-of-sight stellar velocity distributions and instrument resolution. The key properties derived from our analysis are presented in Table~\ref{table:host_properties}, providing the median values and 68\% credible intervals. The host galaxies of all except one transient (VTJ0010-0600) are massive. Since our sample is restricted to a redshift of $< 0.3$ and archival optical surveys are fairly complete down to $0.01-0.1~L_\ast$ galaxies in this redshift range, it is unlikely that our sample is biased due to optical selection effects~\citep{2024Natur.635...61S}.

Consistent with the optical emission line ratio diagnostics, the fraction of galaxy luminosity contributed by dust-heated torii around AGN is constrained by our models to be negligible. The location of these galaxies with respect to the background galaxies star-forming main sequence and the evolution of these galaxies over the past 20~Myr to 100~Myr is shown in Figure~\ref{fig:gal_sfms}. While the host galaxy of VTJ0810+1057 has been stably star-forming over the past 100~Myr, the host galaxies of VTJ0341-0500 and VTJ0918+1554 recently transitioned from quiescent phase to star-forming, probably due to a recent star burst, and on the other hand the host galaxy of VTJ1408+3114 and VTJ1726+2825 transitioned from star-forming to quiescent phase and quiescent to quenched phase, respectively.

\subsection{Search for Archival Transient Counterparts}\label{subsec:search_for_archival_transient_counterparts}

We search archival multi-wavelength catalogs to identify potential transient counterparts to our observed radio transients. We began with searching the Transient Name Server\footnote{\url{https://www.wis-tns.org/}} for any recorded transient that spatially coincides with the locations of our radio transients. In this search, we identified a match for the transient VTJ0010-0600, which has been reported to be associated with SN~2019xhb~\citep{2020ATel14020....1H}. SN~2019xhb was initially discovered on 21 December 2019, approximately seven months prior to its detection in the VLASS Epoch 2.1. Initially classified as a Type II SN based on its optical characteristics~\citep{2019TNSTR2663....1F}, further spectroscopic analysis revealed broad emission line features indicative of a Type Ib/c SN~\citep{2020ATel14020....1H}. Further investigation into the radio and optical characteristics of SN~2019xhb will be presented in a forthcoming study. 

We conducted forced photometry at the location of the transient in archival data from the Zwicky Transient Factory~\citep[ZTF;][]{2019PASP..131a8002B, 2023arXiv230516279M}, the Asteroid Terrestrial-impact Last Alert System~\citep[ATLAS;][]{2018PASP..130f4505T, 2021TNSAN...7....1S}, ALLWISE~\citep{2014yCat.2328....0C} and NEOWISE~\citep{2011ApJ...731...53M}. Our analysis revealed no $>3\sigma$ detections at the transient location across all epochs of ZTF and ATLAS, except for the notable detection of SN~2019xhb. The WISE lightcurves also do not indicate any statistically significant variability.

In addition to optical searches, we extended our investigation to high-energy emission catalogs. We examined data from the Monitor of All-sky X-ray Image~\citep[MAXI;][]{2014PASJ...66...87S}, the International Gamma-Ray Astrophysics Laboratory~\citep[INTEGRAL;][]{2005A&A...438.1175R}, the Fermi Gamma-ray Space Telescope~\citep[FERMI;][]{2021ApJS..256...12A}, and the Neil Gehrels Swift Observatory~\citep{2016ApJ...829....7L}. Despite a thorough search, no significant matches were found. The absence of detectable high-energy counterparts suggests that the transient events, including SN~2019xhb, are either not associated with prominent X-ray or gamma-ray emissions typically expected from highly energetic processes, such as those involving relativistic jets or compact object interactions, or they may involve beamed off-axis emission. Both of these scenarios remain plausible, given the poorly understood completeness of these surveys.

\setlength{\tabcolsep}{2.4pt}

\begin{table*}[ht!]
    \centering
    \caption{Constrained transient parameters for dense CSM interaction-powered synchrotron radiation model.}
    \begin{tabular}{lllllll}
        \toprule
        Parameter & VTJ0010-0600 & VTJ0918+1554 &  VTJ0341-0500 & VTJ1408+3114 & VTJ0810+1057 & VTJ1726+2825 \\
        
        \hline

        $L_p$ [erg~s$^{-1}$~Hz$^{-1}$] & $\approx 8.09 \times 10^{27}$ & $\approx 2.48 \times 10^{28}$ & $\approx 5.24 \times 10^{28}$ & $\approx 3.14 \times 10^{29}$ & $\approx 9.00 \times 10^{29}$ & $\approx 1.27 \times 10^{30}$ \\

        $t_p$ [yr] & $\lesssim 2.62$ & $\lesssim 2.79$ & $\lesssim 2.69$ & $\lesssim 2.96$ & $\lesssim 2.93$ & $\lesssim 2.94$ \\

        $R_p$ [cm] & $\gtrsim 3.62 \times 10^{16}$ & $\gtrsim 6.15 \times 10^{16}$ & $\gtrsim 8.77 \times 10^{16}$ & $\gtrsim 2.05 \times 10^{17}$ & $\gtrsim 3.37 \times 10^{17}$ & $\gtrsim 3.38 \times 10^{17}$ \\

        $B_p$ [G] & $\lesssim 0.28$ & $\lesssim 0.25$ & $\lesssim 0.23$ & $\lesssim 0.19$ & $\lesssim 0.17$ & $\lesssim 0.17$ \\

        $U$ [erg] & $\gtrsim 9.57 \times 10^{47}$ & $\gtrsim 3.71 \times 10^{48}$ & $\gtrsim 9.18 \times 10^{48}$ & $\gtrsim 8.01 \times 10^{49}$ & $\gtrsim 2.87 \times 10^{50}$ & $\gtrsim 4.37 \times 10^{50}$ \\

        $\beta$ & $\gtrsim 0.014$ & $\gtrsim 0.022$ & $\gtrsim 0.033$ & $\gtrsim 0.070$ & $\gtrsim 0.116$ & $\gtrsim 0.136$ \\

        \hline
    \end{tabular}

    {\raggedright N{\scriptsize{OTE}}: The constraints on the shock radius ($R_p$), magnetic field strength ($B_p$), shock energy ($U$) and shock velocity ($\beta$) estimated by assuming that the transient was observed in VLASS at its peak luminosity ($L_p$) at 3~GHz and transient rise timescale $t_p$ is less than the time between epoch E1.1 and E2.1.
    \par}
    \label{table:dense_csm_interaction_constraints}
\end{table*}

\section{Radio Synchrotron Transients}\label{sec:astrophysical_radio_transients_and_their_emission_mechanisms}

In this section, we explore radio transients that could explain the sources in our sample of fast extragalactic transients in VLASS. We explore synchrotron emission mechanisms, focusing on interactions with dense CSM in \S\,\ref{subsec:dense_csm_interaction} and the formation of PWNe in \S\,\ref{subsec:neutron_star_nebulae}. Lastly, we examine the accretion dynamics of black hole jets which can lead to relativistic outflows in \S\,\ref{subsec:black_hole_jets}.\footnote{Given the minimum expected brightness temperature of our VLASS transients sample ($T_B \gtrsim 6\times 10^6$K) and minimum thermal energy deposited ($U \gtrsim 5.8 \times 10^{53}$~erg), which greatly exceeds the typical energies released by SNe ($U \lesssim 10^{50}$~erg) and GRBs ($U \lesssim 10^{52}$~erg), a free-free emission mechanism is insufficient to account for the observed characteristics.} 

\begin{figure}
\centering
\includegraphics[width=\columnwidth]{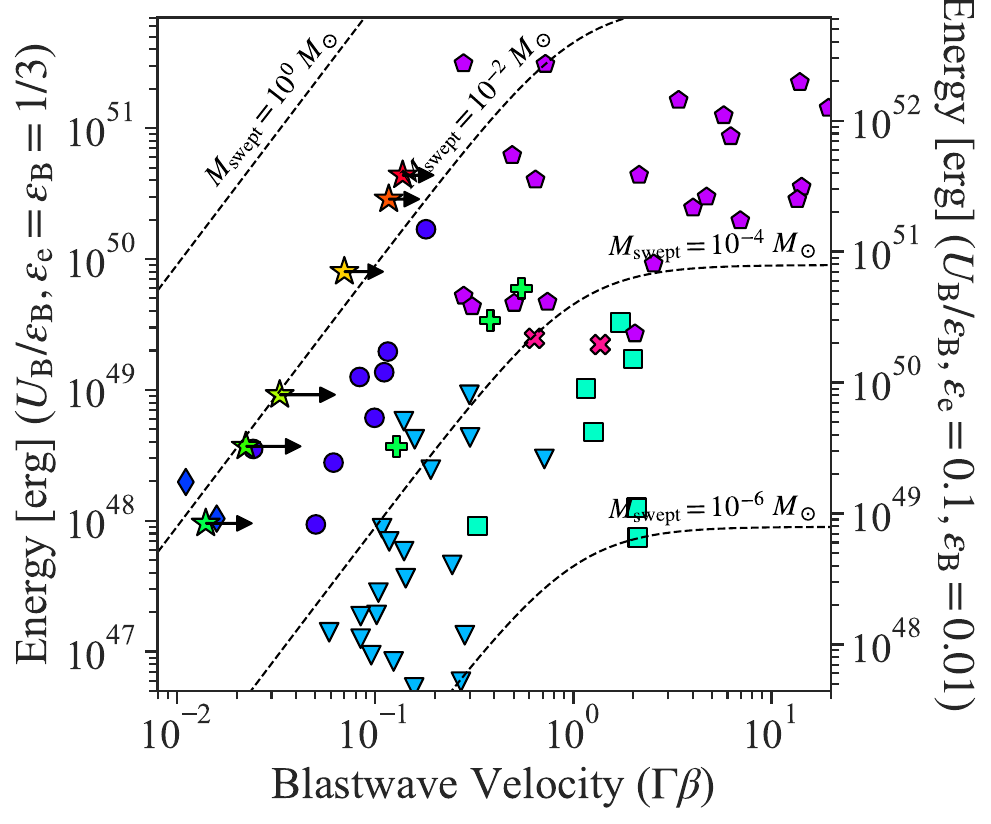}
\includegraphics[width=\columnwidth]{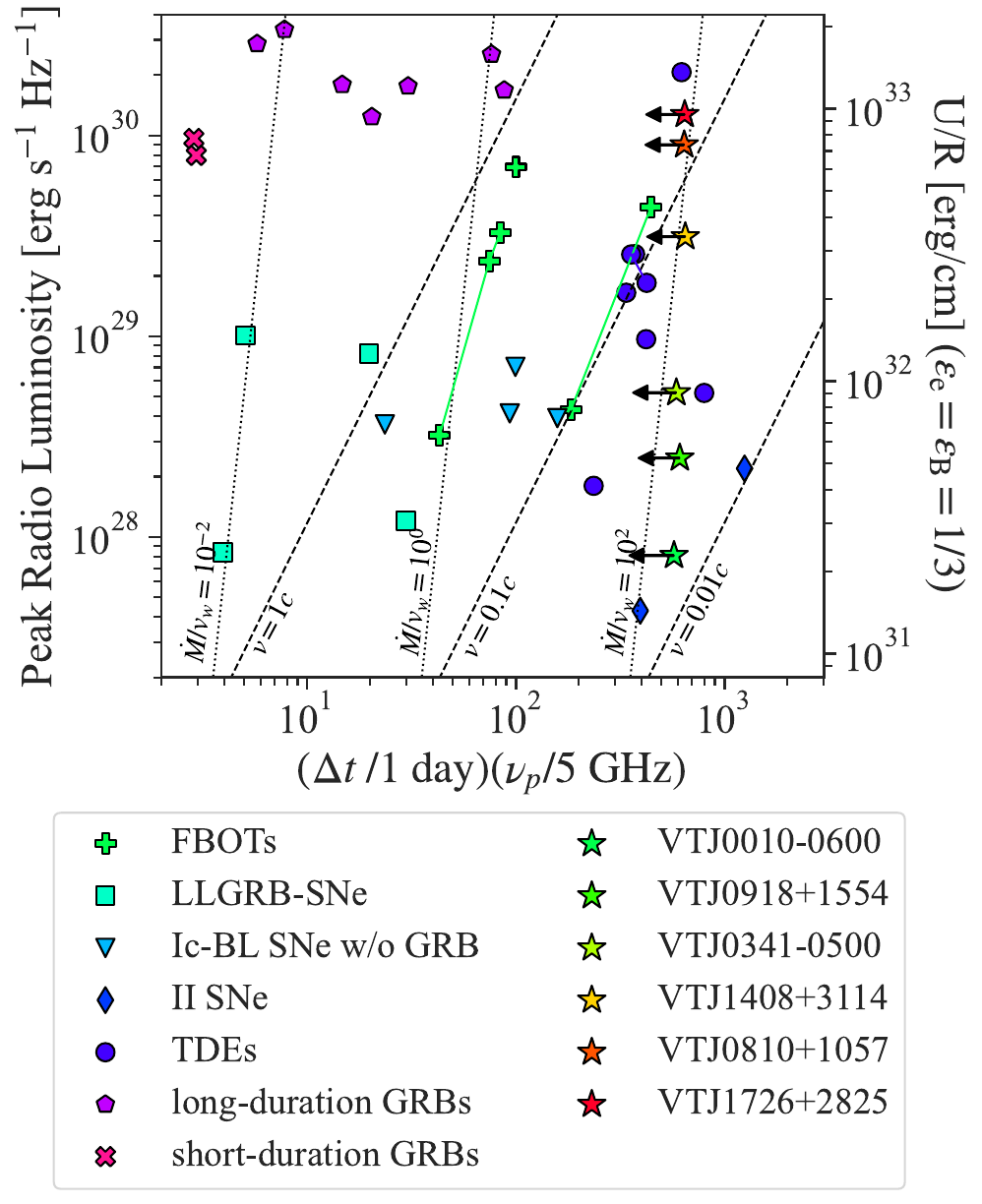}
\caption{Constraints from the dense CSM interaction-powered synchrotron radiation model~\citep{2019ApJ...871...73H}. The constraints for our 6 VLASS transients in velocity-energy space (top panel) and peak luminosity-timescale space (bottom panel) are shown in contrast to other classes of radio transients~\citep{2019ApJ...871...73H}. Since the energy estimates strongly depend on the partition fractions, we show energy estimates for two sets of their values in the top panel, where we also show lines of constant swept-up mass ($M_\mathrm{swept}$). In the bottom panel, we show the lines of constant wind velocity normalized mass loss rates ($\dot{M}/v_w$) and average shock velocity ($v$).}
\label{fig:energy_velocity_peak_luminosity_timescale}
\end{figure}

\subsection{Interaction with a dense shell in the circumstellar medium: SNe, AICs, LFBOTs}\label{subsec:dense_csm_interaction}

Assuming that the synchrotron self-absorbed spectrum has a peak frequency $\nu_p \approx 3$~GHz (the observing frequency of VLASS) and a peak flux density $F_p \approx F_\mathrm{obs}$ (the observed flux density of the transient in epoch 2), following the methodology of \citet{1998ApJ...499..810C} and \citet{2019ApJ...871...73H}, we estimate the shock properties, such as the outer shock radius $R_p$, the magnetic field strength $B_p$, the energy swept up by the shock $U$ and the mean velocity $v = R_p/t_p$ (see Appendix~\ref{appendix:dense_csm_interaction} for more details). Since the total amount of energy in the shock critically depends on the energy partition fractions $\epsilon_e$ and $\epsilon_B$, which are difficult to constrain, we follow the methodology of \citet{2019ApJ...871...73H} and choose to compare the energy scales of all transients for two conventions: $\epsilon_e = \epsilon_B = 1/3$ and $\epsilon_e = 0.1, \epsilon_B = 0.01$. When quoting constraints on the model parameters, we remain mindful that the peak flux could actually be larger, and the peak frequency could be different. Our measurements are summarized in Table~\ref{table:dense_csm_interaction_constraints}. 

The lower limits on the blastwave velocity ($\Gamma \beta$) and energy swept up by the shock, as inferred from the constraints on the peak luminosity and the time since the transient ($t_p \lesssim 3$~yr), allows us to put the VLASS fast extragalactic transients sample within the broader context of radio-luminous synchrotron transients. We compare the swept up energy and the blastwave velocity of our VLASS transients with several other classes of transients, including LFBOTs~\citep{2020ApJ...895L..23C, 2020ApJ...895...49H, 2019ApJ...871...73H}, tidal disruption events~\citep[TDEs;][]{2023arXiv231003791S}, short-duration GRBs~\citep{2021ApJ...914L..20B, 2015ApJ...815..102F}, long-duration GRBs~\citep{2012ApJ...746..156C} and SNe~\citep{2005ApJ...621..908S, 2006Natur.442.1014S, 2006ApJ...651.1005S, 2010Natur.463..513S, 2014ApJ...782...42C, 2018ApJ...854...86E, 2019ApJ...871...73H} in the top panel of Figure~\ref{fig:energy_velocity_peak_luminosity_timescale}. The blastwave velocities of $\Gamma \beta \gtrsim 0.07$ and energy constraints $8 \times 10^{49} \lesssim U \lesssim 5 \times 10^{50} $~erg neatly puts VTJ1726+2825, VTJ0810+1057 and VTJ1408+3114 in the parameter space of relativistic transients such as GRBs and mildly relativistic transients such as LFBOTs, while ruling out similarities with SNe and sub-relativistic TDEs. The lower constraints on the blastwave velocity $\Gamma \beta \gtrsim 0.01$ and energy $9 \times 10^{47} \lesssim U \lesssim 9 \times 10^{48} $~erg implies that the transients VTJ0341-0500, VTJ0918+1554 and VTJ0010-0600 are consistent with SNe, LFBOTs and TDEs.

Despite the broad constraint on the transient timescale ($t_p \lesssim 3$yr), which limits our ability to tightly constrain the wind velocity-normalized mass-loss rate ($\dot{M}/v \lesssim 10^2M_\odot$yr$^{-1}$/kms$^{-1}$), the limits on peak luminosity shown in the lower panel of Figure~\ref{fig:energy_velocity_peak_luminosity_timescale} support conclusions consistent with those drawn from the swept-up energy and blastwave velocity.

\subsection{Neutron Star Nebulae: PWNe, AICs}\label{subsec:neutron_star_nebulae}

\begin{figure}
\includegraphics[width=\columnwidth]{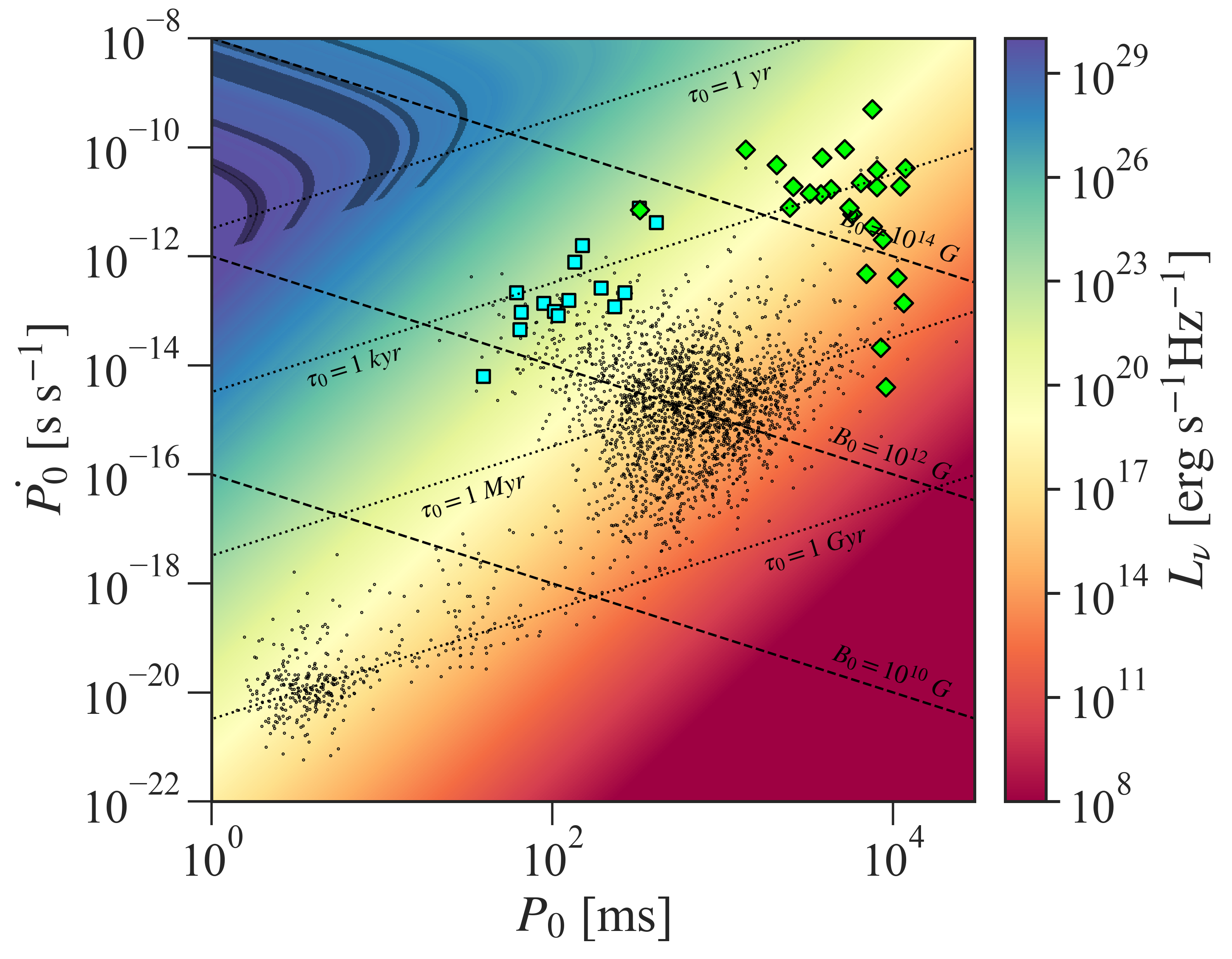}
\caption{Constraints on the magnetar spin down-powered PWN model~\citep{2013ApJ...762L..17P}. The dashed and dotted lines show the lines of constant initial surface dipole magnetic field $B_0$ and spin-down timescale $\tau_0$, respectively. The color-map in the background shows the expected peak PWN luminosity $L_\nu$, as given by eqn.~\ref{eqn:pwn_lum}. The Galactic pulsars~\citep[black points;][]{2005AJ....129.1993M}, pulsars with well-characterized nebulae~\citep[cyan squares;][]{2019JApA...40...36G} and magnetars~\citep[green diamonds;][]{2014ApJS..212....6O} are shown for reference. The black shaded tracks represent the allowed parameter space, given the luminosity of our VLASS transients. The high luminosity and fast evolution of these transients necessitate them to have $1-10$~ms periods and $10^{12}-10^{14}$~G magnetic fields. The lack of Galactic sources in this parameter space is perhaps a consequence of their rarity and short spin-down timescales $\tau_0 \lesssim 1$~yr, thus evading their discovery.}
\label{fig:ppdot_pulsar_wind_nebulae}
\end{figure}

Using the \citet{2013ApJ...762L..17P} formulation\footnote{In the observed parameter space of short spin-down timescales $\tau_0 \lesssim 1$~yr and high dipole magnetic field strengths $10^{12}-10^{14}$~G, the synchrotron emission peaks after $\sim 3$~months of the event at the observing frequencies of the VLASS. At such post-AIC timescales, the synchrotron cooling timescale is too short and the effects of synchrotron cooling can be safely ignored.} (see Appendix~\ref{appendix:neutron_star_nebulae} for details), which predicts that the spin down dynamics of a newly formed magnetar after an AIC event can power a radio luminous PWNe, we show the expected peak luminosity $L_\nu$ of the PWNe thus formed in $P - \dot{P}$ parameter space in Figure~\ref{fig:ppdot_pulsar_wind_nebulae}. We highlight the parameter space permitted by the five VLASS transients (excluding the known SN). These transients are consistent with some of the most luminous possible PWNe with short spin-down timescales ($\tau \lesssim 1$~year), powered by millisecond magnetars. The brevity of these timescales, and the rarity of (millisecond) magnetar formation, could explain why Galactic PWNe in this parameter space have thus far escaped detection. However, future sensitive surveys may permit a better exploration of this parameter space.

Such millisecond magnetars are theoretically well-motivated and consistent with certain observations in a variety of contexts~\citep{2022ASSL..465..245D}. Although classical magnetars typically exhibit surface dipole fields exceeding $10^{14}$~G, several neutron stars displaying magnetar-like behavior (bursting activity) have been found with sub-critical dipole fields closer to $10^{13}$~G, placing them within the upper range of ordinary radio pulsars \citep{2010Sci...330..944R, 2017ARA&A..55..261K}. These lower-field magnetars may form through the preservation of progenitor magnetic flux (the fossil-field scenario; \citealt{2015SSRv..191...77F}), or via more moderate dynamo amplification in the proto-neutron star, driven by differential rotation and magneto-rotational instabilities \citep{1992ApJ...392L...9D, 1993ApJ...408..194T}.

Astrophysical modeling of GRB afterglows and superluminous supernovae (SLSNe) has further substantiated the viability of such objects. In fits to GRB X-ray plateau phases, inferred magnetic field strengths span $10^{14} - 10^{16}$~G, but some events require only $B \sim 10^{13}$~G and initial spin periods of 1--10~ms~\citep{2013MNRAS.430.1061R, 2015MNRAS.448..629G}. Similarly, SLSNe light curve models frequently invoke central engines with $B \sim 10^{13} - 10^{14}$~G and spin periods of a few milliseconds to explain peak luminosities and durations~\citep{2017ApJ...850...55N}. These findings suggest that millisecond magnetars with moderate magnetic fields may constitute a natural extension of the magnetar population.

\subsection{Black Hole Jets: X-ray binaries, ULX sources, TDEs and AGNs}\label{subsec:black_hole_jets}

\begin{figure}
\includegraphics[width=\columnwidth]{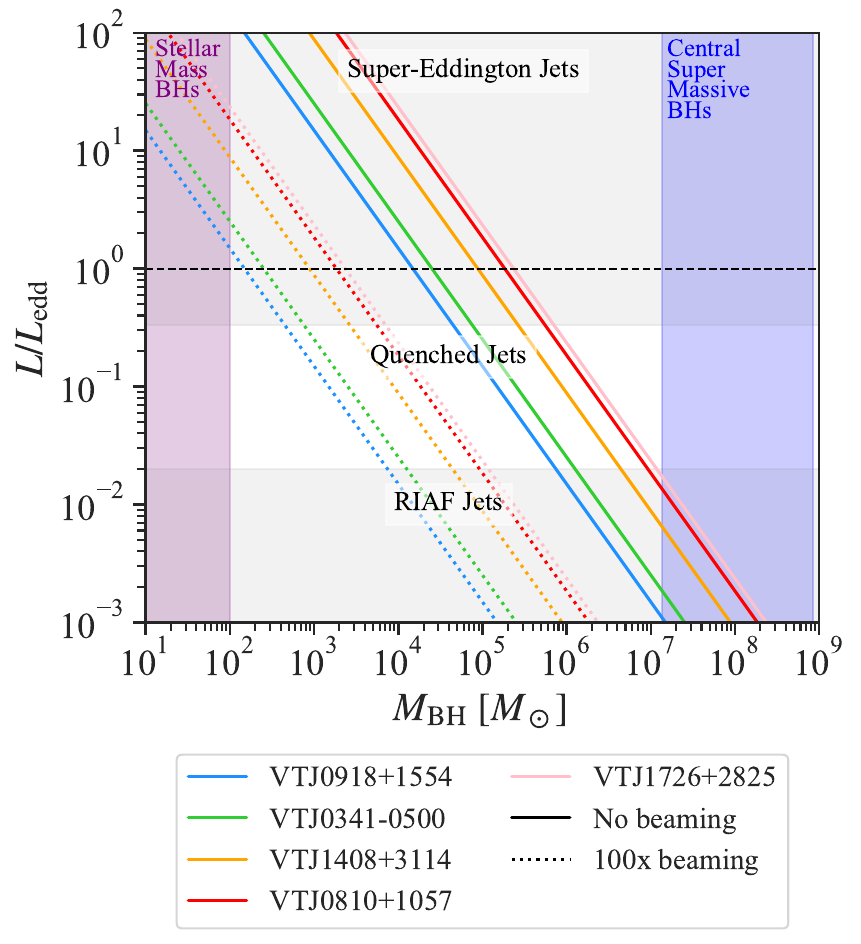}
\caption{Constraints on the black hole accretion rates required to power black hole jet models for VLASS transients. The solid lines show the predicted jet power from a radio power-jet power relation~\citep{2010ApJ...720.1066C} and the dotted lines show the predicted jet power with a factor of 100 magnification due to relativistic beaming. The blue shaded region shows the range of central supermassive black hole mass in the host galaxies of these transients~\citep{2020ARA&A..58..257G}. The quenched jet zone rules out sub-Eddington accretion onto intermediate mass black holes for all VLASS transients. They can however be explained by sub-Eddington accretion in the radiatively inefficient accretion flows (RIAFs) regime onto the central supermassive black hole or super-Eddington accretion onto stellar mass black holes.}
\label{fig:blackhole_jets}
\end{figure}

We explore the possibility that the observed transients may be black hole jets in three mass ranges: stellar-mass ($M_{\mathrm{BH}} \lesssim 100~M_\odot$), intermediate-mass ($M_{\mathrm{BH}} \sim 10^2 - 10^5~M_\odot$), and supermassive ($M_{\mathrm{BH}} \gtrsim 10^5~M_\odot$) black holes. The luminosity of our transients are 4-6 orders of magnitude higher than those from even the most extreme Galactic X-ray binaries \citep{2012MNRAS.421.2947C}. Calculating the jet power using the radio power-jet power relation proposed by \citet{2010ApJ...720.1066C}, 
\begin{equation}
    P_\mathrm{jet} \approx 5.8 \times 10^{43}(P_\mathrm{radio}/10^{40})^{0.7}~\mathrm{erg~s}^{-1},
\end{equation}
we find the jet powers of our transients to be in the range of $10^{42} - 10^{43}$~erg~s$^{-1}$. Given that the Eddington luminosity for a 100~$M_\odot$ black hole is approximately $10^{40}$~erg~s$^{-1}$, the required jet power would need to exceed this by a factor of $10^2-10^3$, which is highly improbable.

Relativistic beaming offers another potential explanation for the observed luminosities, where the Doppler factor $\delta = (\Gamma - \sqrt{\Gamma^2 -1} \cos \theta)^{-1}$~\citep{1979rpa..book.....R}, can significantly amplify the observed luminosity at an angle $\theta$ from our line of sight. Figure~\ref{fig:blackhole_jets} illustrates the beaming fraction and Eddington ratios necessary to account for the observed transient luminosities. While relativistic beaming combined with super-Eddington accretion could theoretically account for the stellar-mass black hole jets, such extreme conditions are unlikely.

For intermediate-mass black holes, we consider the role of TDEs as potential drivers of super-Eddington accretion and jet formation. TDEs can launch powerful jets and sustain high accretion rates over extended periods~\citep{2025ApJ...982..196Y, 2023ApJ...957L...9T, 2022MNRAS.511.3795N}. However, sub-Eddington accretion coupled with beaming in this mass range is improbable due to the jet quenching effect, where the accretion disk becomes geometrically thin, inhibiting the vertical magnetic fields necessary for jet collimation at Eddington ratios between a few percent and $\sim$30\%~\citep{2014MNRAS.437.2744T}. While near-Eddington accretion is less likely due to the absence of AGN signatures and lack of variability in WISE, extremely strong beaming or accretion in the radiatively inefficient accretion flow (RIAF) regime could potentially bypass this quenched jet zone. However, these scenarios are also unlikely.

\begin{figure*}[ht]
\centering
\includegraphics[width=0.95\textwidth]{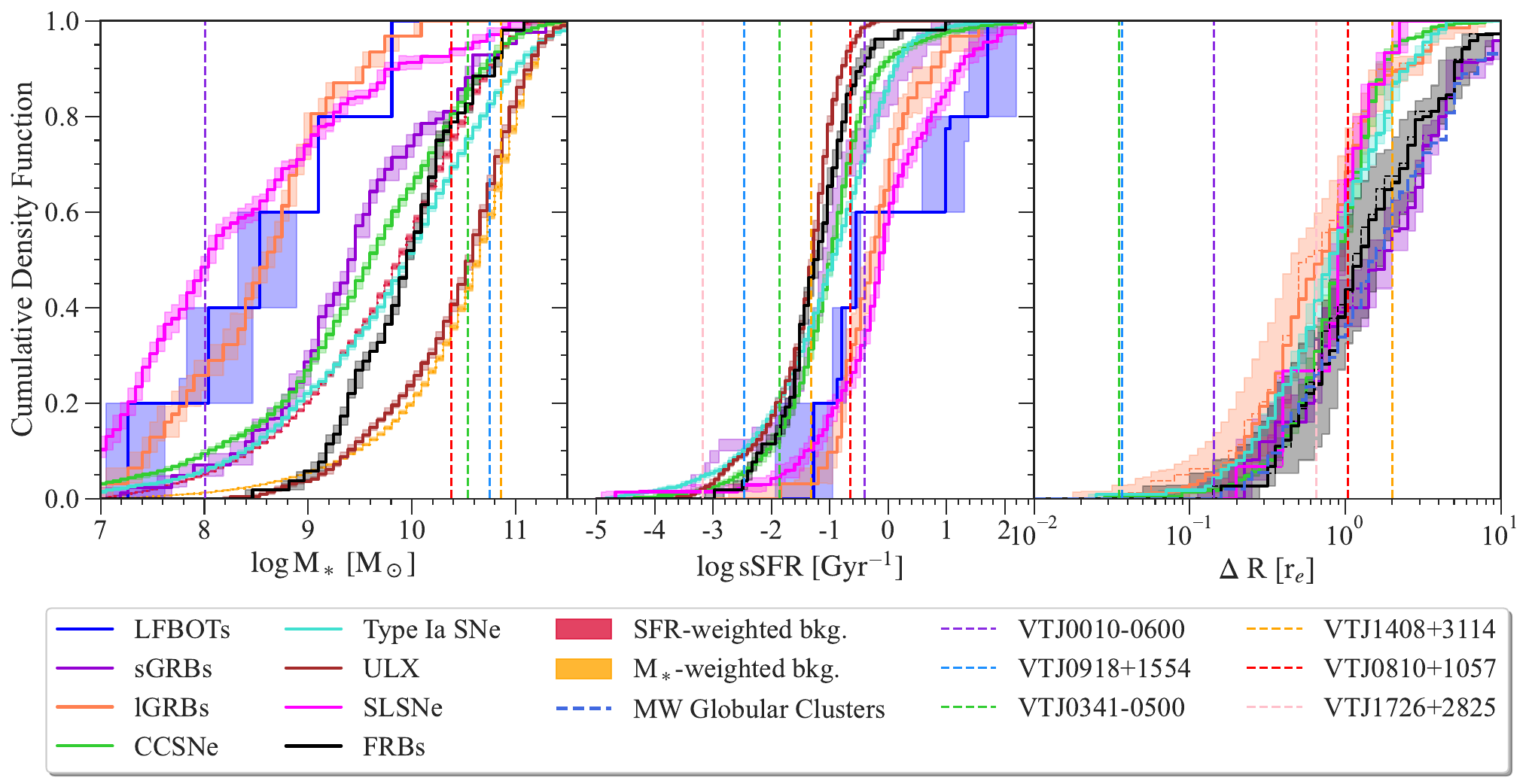}
\caption{Comparison of the VLASS transient's host galaxy properties and host normalized offsets with known transient classes. Since the redshift distribution of the comparison samples is very different from our sample, we correct the host galaxy properties of Fast Radio Bursts~\citep[FRB;][]{2024Natur.635...61S, 2023arXiv230205465G, 2023arXiv231010018B, 2021ApJ...917...75M, 2023arXiv231201578W}, Type Ia SNe~\citep{2010ApJ...722..566L, 2020ApJ...901..143U}, ULX sources~\citep[][]{2020MNRAS.498.4790K}, superluminous SNe~\citep[SLSNe;][]{2021ApJS..255...29S, 2015ApJ...804...90L}, core-collapse SNe~\citep[CCSNe;][]{2021ApJS..255...29S, 2012ApJ...759..107K}, short-duration GRBs~\citep[sGRBs;][]{2022ApJ...940...57N, 2022ApJ...940...56F}, long-duration GRBs~\citep[lGRBs;][]{2015A&A...581A.102V, 2021MNRAS.503.3931T, 2016ApJ...817..144B} and LFBOTs~\citep[][]{2024A&A...691A.329C, 2019MNRAS.484.1031P, 2020ApJ...895...49H, 2020ApJ...895L..23C, 2021MNRAS.508.5138P} for redshift evolution, following the techniques in \citet{2024Natur.635...61S}. The errors on cumulative distributions are computed using $1,000$ Monte Carlo samples of each measured property of the transients assuming a normal distribution with asymmetric errors as quoted in the literature.}
\label{fig:compare_with_other_transients}
\end{figure*}

Comparing the host galaxy stellar mass and SFR distribution of ULX sources~\citep{2020MNRAS.498.4790K} with these transients in Figure~\ref{fig:compare_with_other_transients}, we find that the probability of each of these transients being drawn from ULX sources population are 
$p_\mathrm{VTJ0918+1554} = 0.114$, $p_\mathrm{VTJ0341-0500} = 0.223$, $p_\mathrm{VTJ1408+3114} = 0.352$, $p_\mathrm{VTJ0810+1057} = 0.228$ and $p_\mathrm{VTJ1726+2825} = 0.002$, thus leaving ULX sources as a viable possibility. The host galaxies of these transients, with stellar mass exceeding $10^{10}~M_\odot$, are expected to contain supermassive black holes with masses $M_\mathrm{BH} \sim 10^{7}-10^8~M_\odot$ \citep{2020ARA&A..58..257G}. In this regime, highly sub-Eddington accretion is the most plausible scenario, as it aligns with the absence of AGN features in the host spectra and lack of variability at infrared wavelengths in WISE data.

To summarize, we exclude the prospect of stellar-mass black holes as sources because of the required extreme and rapidly fluctuating accretion rates or significant relativistic beaming. Intermediate-mass black holes could theoretically produce these transients through TDEs or RIAF accretion, but the lack of AGN features and variability at infrared wavelengths in the host galaxies remains a significant constraint. Conversely, it is plausible that the observed transients represent extreme flares from low-luminosity AGN that offers a potential interpretation for the fast extragalactic VLASS transients, at least in the absence of contextual information from the host galaxies.

\section{Discussion}\label{sec:discussion}

In this section, we discuss the plausible classifications of VLASS transients based on their host galaxy properties considered together with the constrained radio emission energetics (see \S\,\ref{sec:astrophysical_radio_transients_and_their_emission_mechanisms}). We then compute the implied detection rates for various transient classes based on our radio sample selection.

\subsection{Accretion-Induced Collapse of White Dwarfs}\label{subsubsec:aics}

Since there have been no confirmed detections of AIC events, we approximate their host galaxy environments and offset distributions to mirror those of Type Ia SNe \citep{2015MNRAS.454.3311M}. Therefore, for the five transients considered, we calculate the likelihood of their occurrence in their respective host galaxies at observed host-normalized offsets and with measured stellar population properties (see Figure~\ref{fig:compare_with_other_transients}), using the distributions of Type Ia SNe~\citep{2010ApJ...722..566L, 2020ApJ...901..143U}. We estimate the probabilities to be $p_\mathrm{VTJ0918+1554} < 0.001$, $p_\mathrm{VTJ0341-0500} < 0.001$, $p_\mathrm{VTJ1408+3114} = 0.016$, $p_\mathrm{VTJ0810+1057} = 0.055$ and $p_\mathrm{VTJ1726+2825} = 0.012$. The similarity of host galaxies of VTJ0918+1554 and VTJ0341-0500 with typical AIC environments can be excluded with more than $6\sigma$ confidence. Thus, the transients VTJ1408+3114, VTJ0810+1057, and VTJ1726+2825 are the only likely candidates for AIC events. 

The observed luminosities of these transients are also consistent with the \citet{2013ApJ...762L..17P} AIC/PWN model (see \S\,\ref{subsec:neutron_star_nebulae} and Figure~\ref{fig:ppdot_pulsar_wind_nebulae}). While it is impossible to definitively ascertain if the three transients, VTJ1408+3114, VTJ0810+1057, and VTJ1726+2825, are indeed such PWNe, we can estimate an upper limit on the AIC event rate for this model. By following the methodology outlined in \S\,\ref{subsec:detection_rates}, we calculate the upper limit on the rate of AIC events with luminosities $\geq 3\times 10^{29}$~erg~s$^{-1}$~Hz$^{-1}$ as $\mathcal{R}_\mathrm{AIC, PWNe} \lesssim 46.66_{-28.72}^{+174.3} \left({\tau}/{6~\mathrm{months}} \right)^{-1}$~Gpc$^{-3}$~yr$^{-1}$, where the constrained power-law index of the luminosity function is $\alpha = -1.57_{-1.10}^{+1.12}$. In our rate calculations, we assume a timescale of 6~months based on the median expected timescale of these transients above VLASS detection limits. This rate corresponds to a fraction $f \lesssim 0.002_{-0.001}^{+0.058}$ of the local volumetric rate of Type Ia SNe, $\mathcal{R}_\mathrm{Type~Ia~SNe} \approx 3 \times 10^4$~Gpc$^{-3}$~yr$^{-1}$~\citep{2011MNRAS.412.1473L}, aligning with theoretical models predicting $f \lesssim 1\%$~\citep{1998ApJ...497..168Y}.

Another compelling model for AIC events is presented by \citet{2016ApJ...830L..38M}, which predicts synchrotron emission resulting from the interaction between AIC ejecta and a dense CSM. The expected peak luminosities for velocity-normalized mass-loss rates $(\dot{M}_\mathrm{loss}/10^{-4}~M_\odot~\mathrm{yr}^{-1})/ (v_\mathrm{wind}/1000~\mathrm{km~s}^{-1}) \in (0.01, 100)$ and ejecta mass $M_\mathrm{ej} \in (0.001, 0.1)~M_\odot$ fall within the range $\sim 10^{26} - 10^{28}$~erg~s$^{-1}$~Hz$^{-1}$ (see \S\,\ref{subsec:expectations}). The only two transients identified within this luminosity range are VTJ0010-0600 and VTJ0918+1554. Of these, VTJ0010-0600 is a confirmed core-collapse SN, whereas VTJ0918+1554's host galaxy characteristics and offsets rule out similarity with Type Ia SNe with 6$\sigma$ confidence (computed above). Therefore, none of the VLASS transients are consistent with this dense CSM interaction scenario of AIC events. Thus, we can estimate an upper limit on the volumetric rate of such transients using their maximum possible luminosity. A source emitting at $10^{28}$~erg~s$^{-1}$~Hz$^{-1}$ would be detectable in the VLASS up to a maximum distance of $d_L = 100$~Mpc. Therefore, the inferred upper limit on the volumetric rate of these transients is $\mathcal{R}_\mathrm{AIC, CSM} \lesssim 339.211_{-280.6}^{+780.02} \left({\tau}/{2~\mathrm{yrs}} \right)^{-1}$~Gpc$^{-3}$~yr$^{-1}$, where the choice of timescale is based on the typical visibility period of these transients, given the VLASS sensitivity. This rate corresponds to a fraction $f \lesssim 0.011_{-0.009}^{+0.026}$ of the Type Ia SN local volumetric rate~\citep{2011MNRAS.412.1473L}, also consistent with the theoretical models~\citep{1998ApJ...497..168Y}.

\subsection{Long-duration Gamma Ray Bursts}\label{subsubsec:lGRBs}

We compare the host stellar population properties and host-normalized galactocentric offsets of our five transients with the known distributions of long-duration GRBs~\citep[see Figure~\ref{fig:compare_with_other_transients};][]{2015A&A...581A.102V, 2021MNRAS.503.3931T, 2016ApJ...817..144B}. We estimate the probability of each of these transients being drawn from the long-duration GRB population to be $p_\mathrm{VTJ0918+1554} < 0.001$, $p_\mathrm{VTJ0341-0500} < 0.001$, $p_\mathrm{VTJ1408+3114} < 0.001$, $p_\mathrm{VTJ0810+1057} = 0.010$ and $p_\mathrm{VTJ1726+2825} < 0.001$. Therefore, similarities with long-duration GRB population can be rejected with more than $6\sigma$ confidence for all but VTJ0810+1057.

Next, we examine whether the energetics of this transient comply with the observed radio afterglows of long-duration GRBs. The beaming corrected energy of long-duration GRB radio afterglows with typical jet opening angles of $\theta_j \approx 0.1$, are in the range $\sim 10^{50}-10^{52}$~ergs. Based on the energetics of the afterglow, derived using the \citet{1997ApJ...476..232M} model of the deceleration of an ultra-relativistic jet by interaction with the circumburst medium, we measure the shock energy of VTJ0810+1057 to be $U \sim 3 \times 10^{50}$~erg, thus implying that this transient is consistent with the population of observed long-duration GRB afterglows. For typical jet opening angles of long-duration GRBs, $\theta_\mathrm{jet} = 0.2$~radians, this measured shock energy can be explained by an on-axis jet, $\theta_\mathrm{obs} \lesssim 0.4$~radians, with high circumburst densities $n_e \gtrsim 16$~cm$^{-3}$. This circumburst density is consistent with our constraints using equation~\ref{eqn:num_density} ($n_e \lesssim 230$~cm$^{-3}$). Alternatively, the observed transient properties can also be explained by an off-axis jet with wide jet opening angles for moderate circumburst medium density of $n_e \gtrsim 3$~cm$^{-3}$. The non-detection in archival GRB catalogs may support the latter scenario.

Based on this single potential candidate long-duration GRB in our sample with observed luminosity of $9\times 10^{29}$~erg~s$^{-1}$~Hz$^{-1}$, the inferred upper limit on the rate of long-duration GRBs is $\mathcal{R}_\mathrm{lGRB} \lesssim 0.95_{-0.79}^{+2.19} \left({\tau}/{10~\mathrm{months}} \right)^{-1}$~Gpc$^{-3}$~yr$^{-1}$, where the choice of transient timescale is based on the typical visibility of long-duration GRBs, given the VLASS sensitivity. This estimated rate is consistent with the long-duration GRB rates reported in literature~\citep{2016A&A...587A..40P}. However, we caution that measuring the volumetric rate of any transient with jetted emission, at radio wavelengths need to be performed very carefully, by first identifying what observing angles the survey is sensitive to. The corrected event rate ($\mathcal{R}_\mathrm{true}$) is a factor of $f^{-1}$ larger than the observed rate $\mathcal{R}_\mathrm{obs}$, where $f = 1-\cos\theta_\mathrm{obs}$ is the correction factor. For VLASS, the median observing angle is $\langle \theta_\mathrm{obs} \rangle = 0.41$~radians (see \S\,\ref{subsec:expectations}), thus implying a correction factor of $f_b \approx 10$. The beaming corrected rate of long-duration GRBs measured using the VLASS sample for luminous extragalactic transients thus is $\mathcal{R}_\mathrm{lGRB, true} \lesssim 11.46_{-9.48}^{+26.28} \left({\tau}/{10~\mathrm{months}} \right)^{-1}$~Gpc$^{-3}$~yr$^{-1}$.

\subsection{Short-duration Gamma Ray Bursts}\label{subsubsec:sGRBs}

We first compare the host galaxy stellar properties and host-normalized offsets of the five transients with the known distributions for short-duration GRBs~\citep[see Figure~\ref{fig:compare_with_other_transients};][]{2022ApJ...940...57N, 2022ApJ...940...56F}. We estimate the probabilities to be $p_\mathrm{VTJ0918+1554} < 0.001$, $p_\mathrm{VTJ0341-0500} < 0.001$, $p_\mathrm{VTJ1408+3114} = 0.008$, $p_\mathrm{VTJ0810+1057} = 0.025$ and $p_\mathrm{VTJ1726+2825} < 0.001$, thus implying that the transients VTJ1408+3114 and VTJ0810+1057 are consistent with short-duration GRBs. Next, considering the energetics of the interaction of the relativistic jet with the circumburst medium, only VTJ1408+3114 is consistent with short-duration GRBs, which have been observed to have beaming corrected energies in the range $10^{49}-10^{50}$~erg. However, we note that given the shock energy of VTJ1408+3114, assuming typical jet opening angles and on-axis orientation, the circumburst density will have to be significantly higher than typically observed circumburst densities for short-duration GRBs. In the unlikely event of it being a short-duration GRB afterglow, based on this single candidate with observed luminosity of $3\times 10^{29}$~erg~s$^{-1}$~Hz$^{-1}$, we compute the upper limit on their rate as $\mathcal{R}_\mathrm{sGRB} \lesssim 4.63_{-3.83}^{+10.64} \left({\tau}/{10~\mathrm{months}} \right)^{-1}$~Gpc$^{-3}$~yr$^{-1}$. 

Similar to long-duration GRBs, the short-duration GRB afterglows discovered by VLASS are expected to have a median observing angle of $\theta_\mathrm{obs} = 0.34$~radians (see \S\,\ref{subsec:expectations}). Thus, the beaming corrected rate of short-duration GRBs is $\mathcal{R}_\mathrm{sGRB, true} \lesssim 80.88_{-66.90}^{+185.87} \left({\tau}/{10~\mathrm{months}} \right)^{-1}$~Gpc$^{-3}$~yr$^{-1}$, consistent with the current constraints from the LIGO Scientific and Virgo Collaboration~\citep{2023PhRvX..13a1048A}.

\subsection{Luminous Fast Blue Optical Transients}\label{subsubsec:fbots}

Since there is just one publicly available host-normalized offset measurement of LFBOTs~\citep{2024MNRAS.527L..47C}, we only use their host galaxy properties~\citep{2024A&A...691A.329C, 2019MNRAS.484.1031P, 2020ApJ...895...49H, 2020ApJ...895L..23C, 2021MNRAS.508.5138P} to compute the likelihood of our VLASS transients residing in an LFBOT-like host galaxy environment. We estimate these probabilities to be $p_\mathrm{VTJ0918+1554} < 0.001$, $p_\mathrm{VTJ0341-0500} < 0.001$, $p_\mathrm{VTJ1408+3114} = 0.024$, $p_\mathrm{VTJ0810+1057} = 0.041$ and $p_\mathrm{VTJ1726+2825} < 0.001$ for our five unknown transients. These transients are also consistent with LFBOTs based on the shock energy and velocity constrained using the constraints on peak luminosities and timescales. 

Following the methodology discussed in \S\,\ref{subsec:detection_rates}, we calculate the upper limit on the rate of AT2018cow-like events with luminosities $\gtrsim 2\times 10^{28}$~erg~s$^{-1}$~Hz$^{-1}$ as $\mathcal{R}_\mathrm{FBOTs} \lesssim 59.92_{-34.14}^{+972.18} \left({\tau}/{1.5~\mathrm{yr}} \right)^{-1}$~Gpc$^{-3}$~yr$^{-1}$, where the constrained power-law index of the luminosity function is $\alpha = -1.29_{-0.43}^{+0.38}$. This rate corresponds to a fraction $f \lesssim 0.02_{-0.01}^{+0.32}$\% of the local volumetric rate of core-collapse SNe, $\mathcal{R}_\mathrm{CCSNe} \approx 3 \times 10^5$~Gpc$^{-3}$~yr$^{-1}$~\citep{2012ApJ...757...70D}. Although the volumetric rate of LFBOTs, based on their optical emission, has been estimated to be $\sim 1$\% of the core-collapse SN rate~\citep{2018MNRAS.481..894P}, similar to \citet{2023ApJ...949..120H}, we find that the actual rate of AT2018cow-like events with luminous radio emission is even smaller. These findings underscore the rarity and unique nature of LFBOTs, particularly those with luminous radio emission.

\section{Conclusion}\label{sec:conclusion}

We present a search for luminous radio transients in the local universe ($z \leq 0.3$) on timescales of $\lesssim 3$ years, focusing on observations at GHz frequencies from the VLASS. Our search resulted in the discovery of six luminous radio transients, associated with host galaxies exhibiting a wide range of stellar population properties and host-normalized offsets. 

We developed a probabilistic framework to classify the six transients in our sample, among which VTJ0010-0600 has been previously identified as a core-collapse SN. Using this framework, we place upper limits on the volumetric rates for all classes of transients under consideration. Our analysis indicates that the rate of AIC events involving dense CSM interactions is $\lesssim 1.10_{-0.90}^{+2.60}$\% of the local volumetric rate of Type Ia SNe. Additionally, we find that the rate of AIC events that produce radio-luminous, fast-evolving PWNe is $\lesssim 0.20_{-0.10}^{+5.80}$\% of the local volumetric rate of Type Ia SNe. Both of these estimates align with theoretical models for AIC events, which predict that their occurrence should constitute less than 1\% of the local Type Ia SN rate. 

Additionally, based on the four potential LFBOT candidates in our sample, we estimate an upper limit on the rate of radio-luminous LFBOTs to be $\lesssim 0.02_{-0.01}^{+0.32}$\% of the local volumetric rate of core-collapse SNe, highlighting their rarity. Simulations of GRB light curves with isotropic observing angles indicate that the VLASS is sensitive to off-axis GRBs with median observing angle, $\langle \theta_\mathrm{obs} \rangle = 0.4$~radians. From our analysis, we identify at most one potential long-duration GRB and one potential short-duration GRB candidate. Based on these findings, we estimate the beaming-corrected local volumetric rates to be $\langle \mathcal{R}_\mathrm{lGRB, true} \rangle \lesssim 11.46_{-9.48}^{+26.28} \left({\tau}/{10~\mathrm{months}} \right)^{-1}$~Gpc$^{-3}$~yr$^{-1}$ and $\langle \mathcal{R}_\mathrm{sGRB, true} \rangle \lesssim 80.88_{-66.90}^{+185.87} \left({\tau}/{10~\mathrm{months}} \right)^{-1}$~Gpc$^{-3}$~yr$^{-1}$ for long- and short-duration GRBs, respectively.

With the forthcoming generation of highly sensitive surveys, such as the Square Kilometer Array~\citep[SKA;][]{2004NewAR..48..979C} and the Deep Synoptic Array~\citep[DSA-2000;][]{2019BAAS...51g.255H}, it is imperative to establish a framework capable of facilitating the study of extensive samples of radio transients. This work also underscores that mere detection in radio surveys is inadequate for a comprehensive understanding of these transients; multi-wavelength counterparts are essential for their full characterization. In the absence of such counterparts, the properties of the host galaxy, host-normalized offset measurements, and constraints on the energetics of the radio emission under various hypotheses offer the most reliable insights into the nature of these transients. In future, the overlap of SKA and DSA-2000 surveys with near-future surveys at other wavelengths, such as the Vera Rubin Observatory~\citep{2019ApJ...873..111I}, the Dark Energy Spectroscopic Instrument (DESI) Legacy Imaging surveys~\citep{2019AJ....157..168D}, the Nancy Grace Roman Space Telescope~\citep[Roman;][]{2020JATIS...6d6001M} and SPHEREx~\citep{2020SPIE11443E..0IC}, will provide contextual information.

\begin{acknowledgments}
The National Radio Astronomy Observatory is a facility of the National Science Foundation operated under cooperative agreement by Associated Universities, Inc. 

Some of the data presented herein were obtained at Keck Observatory, which is a private 501(c)3 non-profit organization operated as a scientific partnership among the California Institute of Technology, the University of California, and the National Aeronautics and Space Administration. The Observatory was made possible by the generous financial support of the W. M. Keck Foundation. 
\end{acknowledgments}

\appendix

\section{Dense Circumstellar Medium Interaction}\label{appendix:dense_csm_interaction}

This section summarizes the theoretical background for \S\ref{subsec:dense_csm_interaction}. The high accretion rates onto the receiver in interacting binaries inevitably lead to high mass-loss rates ($\dot{M}$) with a high wind velocity ($v_{\mathrm{w}}$). As a result, the evolution of the binary system towards the transient event may produce a dense CSM, with density\footnote{The assumption of a steady wind density profile directly influences the derived shock radius and post-shock magnetic field, and hence the inferred environment properties. While deviations from the $r^{-2}$ profile -- due to episodic mass loss or complex CSM structures -- can introduce systematic uncertainties, we adopt this form as it is standard in the literature and provides a consistent basis for comparison with previous work~\citep{2019ApJ...871...73H}.} given by

\begin{equation}
\rho_\mathrm{CSM}(r) = \frac{\dot{M}}{4\pi v_\mathrm{w}} \frac{1}{r^2}.
\label{eqn:rho_csm}
\end{equation}

The interaction between the ejecta from the transient event ($M_\mathrm{ej}$), and the dense CSM generates a strong shock, which can be observed at radio frequencies as synchrotron emission emerging from relativistic electrons accelerated at the shock. The nonthermal emission can be described by the source spectrum arising from a power-law distribution of electron Lorentz factors ($\gamma_e$) with power-law index $p$, defined as

\begin{equation}
    \frac{dN (\gamma_e)}{d\gamma_e} \propto \gamma_e^{-p}, \quad \gamma_e \geq \gamma_m,
\end{equation}

where the minimum Lorentz factor ($\gamma_m$) is determined by conserving the shock energy flux. In this scenario, an $\epsilon_e$ fraction of the total energy density is allocated to accelerating electrons

\begin{equation}
    \gamma_m - 1 \approx \epsilon_e \frac{m_p}{m_e} \frac{v^2}{c^2},
\end{equation}

with $m_p$ and $m_e$ being the proton and electron masses, respectively, and $v$ being the shock velocity. Typically, the observed power-law index for non-relativistic astrophysical shocks is $2.5 < p < 3$ and the constraint on the minimum Lorentz factor is $\gamma_m - 1 < 1$.

Similar to \citet{2019ApJ...871...73H}, following the methodology of \citet{1998ApJ...499..810C}, we estimate the shock properties, such as the outer shock radius ($R_p$)
\begin{equation}
    \begin{split}
    R_p & ~= \left[ \frac{ 6 c_6^{p+5} F_p^{p+6} D^{2p+12} }{ (\epsilon_e/\epsilon_B) f (p-2) \pi^{p+5} c_5^{p+6} E_l^{p-2} } \right]^{1/(2p+13)} \\
    & ~~~~ \left( \frac{\nu_p}{2c_1} \right)^{-1} \\
    & ~= 8.8 \times 10^{15} \left( \frac{\epsilon_e}{\epsilon_B} \right)^{-1/19} \left( \frac{f}{0.5} \right)^{-1/19} \left( \frac{F_p}{\mathrm{Jy}} \right)^{9/19} \\
    & ~~~~ \left( \frac{D}{\mathrm{Mpc}} \right)^{18/19} \left( \frac{\nu_p}{5~\mathrm{GHz}} \right)^{-1} ~\mathrm{cm}, \\
    \end{split}
\end{equation}
and the magnetic field ($B_p$)
\begin{equation}
    \begin{split}
    B_p & ~= \left[ \frac{ 36 \pi^2 c_5 }{ (\epsilon_e/\epsilon_B)^2 f^2 (p-2)^2 c_6^3 E_l^{2(p-2)} F_p D^2 } \right]^{2/(2p+13)} \\
    & ~~~~ \left( \frac{\nu_p}{2c_1} \right) \\
    & ~= 0.58 \left( \frac{\epsilon_e}{\epsilon_B} \right)^{-4/19} \left( \frac{f}{0.5} \right)^{-4/19} \left( \frac{F_p}{\mathrm{Jy}} \right)^{-2/19} \\
    & ~~~~ \left( \frac{D}{\mathrm{Mpc}} \right)^{-4/19} \left( \frac{\nu_p}{5~\mathrm{GHz}} \right) ~\mathrm{G}, \\
    \end{split}
\end{equation}
where we use the constants for $p = 3$, as tabulated in \citet{1970ranp.book.....P}. Here, $E_l = 0.51$~MeV is the electron rest mass energy, $\epsilon_B$ is the fraction of energy density in magnetic fields, $f$ is the filling factor, $F_p$ is the observed peak source flux at peak frequency $\nu_p$, and $D$ is the distance to the source. The total energy $U = U_B/\epsilon_B$ can be estimated as
\begin{equation}
    \begin{split}
        U & ~= \frac{1}{\epsilon_B} f \frac{4\pi}{3} R^3 \left( \frac{B^2}{8 \pi} \right) \\
        & ~= 1.9 \times 10^{46} \frac{1}{\epsilon_B} \left( \frac{\epsilon_e}{\epsilon_B} \right)^{-11/19} \left( \frac{f}{0.5} \right)^{8/19} \left( \frac{F_p}{\mathrm{Jy}} \right)^{23/19} \\
    & ~~~~ \left( \frac{D}{\mathrm{Mpc}} \right)^{46/19} \left( \frac{\nu_p}{5~\mathrm{GHz}} \right)^{-1} ~\mathrm{erg},
    \end{split}
\end{equation}
and the mean velocity $v = R_p/t_p$ can be written as
\begin{equation}
    \begin{split}
        v/c & ~\approx \left( \frac{\epsilon_e}{\epsilon_B} \right)^{-1/19} \left( \frac{L_p}{10^{26}~\mathrm{erg}~\mathrm{s}^{-1}~\mathrm{Hz}^{-1}} \right)^{9/19} \\
    & ~~~~ \left( \frac{f}{0.5} \right)^{-1/19} \left( \frac{\nu_p}{5~\mathrm{GHz}} \right)^{-1} \left( \frac{t_p}{1~\mathrm{days}} \right)^{-1}.
    \end{split}
\end{equation}

If the surrounding medium of the transient was formed from pre-explosion steady winds, then the density can be parameterized in terms of wind velocity-normalized mass-loss rate, as in eqn.~\ref{eqn:rho_csm}. The density of the ambient medium can also be independently computed under the strong shock limit as 
\(\frac{3 \rho v^2}{4} = P\), where \(P = \frac{1}{\epsilon_B} \frac{B^2}{8\pi}\) is the downstream medium (shocked ejecta) pressure and \(\rho = \mu_p m_p n_e\) is the upstream (ambient) medium density. Assuming \(\mu_p = 1\) for a fully ionized hydrogen medium and \(n_p = n_e\), we get:
\begin{equation}
    \label{eqn:num_density}
    \begin{split}
        n_e & ~\approx 20 \left( \frac{1}{\epsilon_B} \right) \left( \frac{\epsilon_e}{\epsilon_B} \right)^{-6/19} \left( \frac{L_p}{10^{26}~\mathrm{erg}~\mathrm{s}^{-1}~\mathrm{Hz}^{-1}} \right)^{-22/19} \\
        & ~~~~ \left( \frac{f}{0.5} \right)^{-6/19} \left( \frac{\nu_p}{5~\mathrm{GHz}} \right)^{4} \left( \frac{t_p}{1~\mathrm{days}} \right)^{2}~\mathrm{cm}^{-3}.
    \end{split}
\end{equation}

Combining the two parameterizations, the mass-loss rate can be estimated as
\begin{equation}
    \begin{split}
        & \left( \frac{\dot{M}}{10^{-4}~M_\odot~\mathrm{yr}^{-1}} \right) \left( \frac{v_w}{1000~\mathrm{km~s}^{-1}} \right)^{-1} \\
        & = 0.0005 \cdot \frac{1}{\epsilon_B} \left( \frac{\epsilon_e}{\epsilon_B} \right)^{-8/19} \left( \frac{f}{0.5} \right)^{-1/19} \left( \frac{\nu_p}{5~\mathrm{GHz}} \right)^{2} \\
        & ~~~ \left( \frac{L_p}{10^{26}\mathrm{erg~s}^{-1}~\mathrm{Hz}^{-1}} \right)^{-4/19} \left( \frac{t_p}{1~\mathrm{days}} \right)^{2} \\
    \end{split}
\end{equation}

\section{Neutron Star Nebulae}\label{appendix:neutron_star_nebulae}

This section summarizes theoretical background for \S\ref{subsec:neutron_star_nebulae}. The spin down of a newly born magnetar injects energy to power a PWN within the $M_{\mathrm{ej}} \sim 10^{-2}~M_\odot$ ejecta, freely expanding at a velocity of $v_{\mathrm{ej}} \sim 0.1~c$~\citep{2013ApJ...762L..17P}. The time evolution of the magnetar dipole spin-down luminosity is $L(t) = L_0/(1+t/\tau)^2$, where for a magnetar with initial magnetic field $B_0 = 10^{14}$~G, initial spin period $P_0 = 3$~ms, radius $R = 12$~km and mass $M_\ast = 1.4~M_\odot$, the initial spin-down luminosity $L_0$~\citep{2001ApJ...550..426L} is given by
\begin{equation}
    L_0 = 1.2 \times 10^{47} \left( \frac{B_0}{10^{14}~\mathrm{G}} \right)^2 \left( \frac{P_0}{3~\mathrm{ms}} \right)^{-4}~\mathrm{erg~s}^{-1},
\end{equation}
and the characteristic spin-down timescale ($\tau$) is
\begin{equation}
    \tau = 4.8 \times 10^4 \left( \frac{B_0}{10^{14}~\mathrm{G}} \right)^{-2} \left( \frac{P_0}{3~\mathrm{ms}} \right)^{2}~\mathrm{s}.
\end{equation}
The exact solution for the time evolution of the radius of the PWN ($R_{\mathrm{PWN}}$), magnetic field ($B$) and electron pressure ($P_e$) can be obtained by numerically solving the relativistic equation of state and momentum conservation~\citep{1984ApJ...278..630R}. However, it can be reasonably analytically approximated~\citep{1977ASSL...66...53C}:
\begin{equation}
\begin{split}
    R_{\mathrm{PWN}} \approx & ~2 \times 10^{17} \left( \frac{L}{10^{47}~\mathrm{erg~s}^{-1}} \right)^{1/5} \left( \frac{E}{10^{50}~\mathrm{erg}} \right)^{3/10} \\
    & \left( \frac{M_\mathrm{ej}}{10^{-2}~M_\odot} \right)^{-1/2} \left( \frac{t}{10^{7}~\mathrm{s}} \right)^{6/5} \mathrm{cm},
\end{split}
\end{equation}
\begin{equation}
\begin{split}
    B \approx & ~0.3 \left( \frac{\eta_B}{10^{-3}} \right)^{1/5} \left( \frac{L}{10^{47}~\mathrm{erg~s}^{-1}} \right)^{1/5} \left( \frac{E}{10^{50}~\mathrm{erg}} \right)^{-9/20} \\
    & \left( \frac{M_\mathrm{ej}}{10^{-2}~M_\odot} \right)^{-3/4} \left( \frac{t}{10^{7}~\mathrm{s}} \right)^{-13/10}~\mathrm{G},
\end{split}
\end{equation}
\begin{equation}
\begin{split}
    P \approx & ~4 \left( \frac{L}{10^{47}~\mathrm{erg~s}^{-1}} \right)^{2/5} \left( \frac{E}{10^{50}~\mathrm{erg}} \right)^{-9/10} \\
    & \left( \frac{M_\mathrm{ej}}{10^{-2}~M_\odot} \right)^{3/2} \left( \frac{t}{10^{7}~\mathrm{s}} \right)^{-13/5}~\mathrm{Ba},
\end{split}
\end{equation}
where $P = P_e + P_B$ is the PWN pressure, $E$ is the ejecta kinetic energy and $\eta_B \approx 10^{-3}$ is the fraction of magnetar luminosity that goes into magnetic fields. 

The power-law spectrum of relativistic electrons in the PWN are accelerated at the wind termination shock, where the PWN pressure equals the ram pressure of the pulsar wind, thus leading to synchrotron radio emission~\citep{1970ranp.book.....P}. For an average pitch angle $\sin \theta = (2/3)^{1/2}$ and power-law spectral index $s = 1.5$ for relativistic electrons (as motivated by Galactic PWNe~\citep{2009ApJ...703.2051G}), the synchrotron emission is self-absorbed below the frequency
\begin{equation}
\begin{split}
    \nu_{\mathrm{SA}} \approx & ~2.6 \left( \frac{R_\mathrm{PWN} (t)}{3 \times 10^{16}~\mathrm{cm}} \right)^{4/11} \left( \frac{B(t) \sin \theta}{0.04~\mathrm{G}} \right)^{8/11} \\
    & \left( \frac{P_e(t)}{0.09~\mathrm{Ba}} \right)^{4/11} \left( \frac{\nu_c(t)}{3.5 \times 10^5~\mathrm{GHz}} \right)^{-1/11} ~\mathrm{GHz} ,
\end{split}
\end{equation}
where $\nu_c$ is the characteristic frequency above which synchrotron cooling beats adiabatic expansion~\citep{1984ApJ...278..630R}
\begin{equation}
\begin{split}
    \nu_c \approx & ~3.5 \times 10^5 \left( \frac{B(t) \sin \theta}{0.04~\mathrm{G}} \right)^{-3} 
    \left( \frac{v_\mathrm{PWN} (t)}{1 \times 10^8~\mathrm{cm~s}^{-1}} \right)^{2} \\
    & \left( \frac{R_\mathrm{PWN} (t)}{3 \times 10^{16}~\mathrm{cm}} \right)^{-2} 
    ~\mathrm{GHz},
\end{split}
\end{equation}
and all the normalizations are computed for the parameters at $10^7$~seconds after the event. The synchrotron emission luminosity in the optically thick limit ($\nu \leq \nu_{\mathrm{SA}}$) and optically thin limit ($\nu > \nu_{\mathrm{SA}}$) is given by
\begin{equation}
L_\nu = 4 \times 10^{28}~\mathrm{erg~s}^{-1}\mathrm{Hz}^{-1} \begin{cases} \left( \frac{\nu_\mathrm{obs}}{3~\mathrm{GHz}} \right)^{5/2}
\left( \frac{R_\mathrm{PWN} (t)}{3 \times 10^{16}~\mathrm{cm}} \right)^{2} \\ \left( \frac{B(t) \sin \theta}{0.04~\mathrm{G}} \right)^{-1/2}, ~~~~~ \nu \leq \nu_{\mathrm{SA}} \\ 
\left( \frac{\nu_\mathrm{obs}}{3~\mathrm{GHz}} \right)^{-1/4} \left( \frac{R_\mathrm{PWN} (t)}{3 \times 10^{16}~\mathrm{cm}} \right)^{3} \\ \left( \frac{B(t) \sin \theta}{0.04~\mathrm{G}} \right)^{3/2} \left( \frac{P_e(t)}{0.09~\mathrm{Ba}} \right)  \\
\left( \frac{\nu_c(t)}{3.5 \times 10^5~\mathrm{GHz}} \right)^{-1/4} , \nu > \nu_{\mathrm{SA}} \end{cases}. 
\label{eqn:pwn_lum}
\end{equation}
Although the temperatures at the front end of the forward shock are too high ($T \gg 10^9$~K) for free-free absorption to have a significant impact, the temperatures at the front end of the ejecta at relevant timescales of $t\sim10^7$~s are lower ($T \sim 10^5$~K). Thus, the free-free absorption at the front edge of the ejecta can be accounted for by approximating the observed luminosity as $L_{\nu,\mathrm{obs}} \approx L_\nu e^{-\tau_{\mathrm{ff}}}$, where $\tau_{\mathrm{ff}} \approx \alpha_{\mathrm{ff}} \Delta R_{\mathrm{PWN}} \approx 0.1 \alpha_{\mathrm{ff}} R_{\mathrm{PWN}}$ and the free-free absorption coefficient is given by~\citep{1979rpa..book.....R}
\begin{equation}
\begin{split}
   \alpha_{\mathrm{ff}} &~\approx 1.9 \times 10^{-2} T^{-3/2} Z^2 n_e n_i \nu^{-2} g_\mathrm{ff} \\
   &~\approx 7 \times 10^{-19} \left( \frac{T}{10^5~\mathrm{K}} \right)^{-3/2} \left( \frac{M_\mathrm{ej}}{10^{-2}~M_\odot} \right)^2 \\
   &\quad \times \left( \frac{R_\mathrm{PWN}(t)}{3 \times 10^{16}~\mathrm{cm}} \right)^{-6} \left( \frac{\nu_\mathrm{obs}}{3~\mathrm{GHz}} \right)^{-2} g_\mathrm{ff}~\mathrm{cm}^{-1},
\end{split}
\end{equation}
where the pressure continuity across the contact discontinuity requires  $\rho_\mathrm{ej} k_B T / m_p \sim n_0 m_p v_s^2$ and we approximate $Z^2 n_e n_i \sim (\rho_\mathrm{ej}/m_p)^2$.

\bibliography{manuscript}{}

\begin{thebibliography}{}
\expandafter\ifx\csname natexlab\endcsname\relax\def\natexlab#1{#1}\fi
\providecommand{\url}[1]{\href{#1}{#1}}
\providecommand{\dodoi}[1]{doi:~\href{http://doi.org/#1}{\nolinkurl{#1}}}
\providecommand{\doeprint}[1]{\href{http://ascl.net/#1}{\nolinkurl{http://ascl.net/#1}}}
\providecommand{\doarXiv}[1]{\href{https://arxiv.org/abs/#1}{\nolinkurl{https://arxiv.org/abs/#1}}}

\bibitem[{{Abbott} {et~al.}(2023){Abbott}, {Abbott}, {Acernese}, {Ackley}, {Adams}, {Adhikari}, {Adhikari}, {Adya}, {Affeldt}, {Agarwal}, {Agathos}, {Agatsuma}, {Aggarwal}, {Aguiar}, {Aiello}, {Ain}, {Ajith}, {Akutsu}, {de Alarc{\'o}n}, {Akcay}, {Albanesi}, {Allocca}, {Altin}, {Amato}, {Anand}, {Anand}, {Ananyeva}, {Anderson}, {Anderson}, {Ando}, {Andrade}, {Andres}, {Andri{\'c}}, {Angelova}, {Ansoldi}, {Antelis}, {Antier}, {Antonini}, {Appert}, {Arai}, {Arai}, {Arai}, {Araki}, {Araya}, {Araya}, {Areeda}, {Ar{\`e}ne}, {Aritomi}, {Arnaud}, {Arogeti}, {Aronson}, {Arun}, {Asada}, {Asali}, {Ashton}, {Aso}, {Assiduo}, {Aston}, {Astone}, {Aubin}, {Austin}, {Babak}, {Badaracco}, {Bader}, {Badger}, {Bae}, {Bae}, {Baer}, {Bagnasco}, {Bai}, {Baiotti}, {Baird}, {Bajpai}, {Ball}, {Ballardin}, {Ballmer}, {Balsamo}, {Baltus}, {Banagiri}, {Bankar}, {Barayoga}, {Barbieri}, {Barish}, {Barker}, {Barneo}, {Barone}, {Barr}, {Barsotti}, {Barsuglia}, {Barta}, {Bartlett}, {Barton}, {Bartos}, {Bassiri}, {Basti}, {Bawaj}, {Bayley},
  {Baylor}, {Bazzan}, {B{\'e}csy}, {Bedakihale}, {Bejger}, {Belahcene}, {Benedetto}, {Beniwal}, {Bennett}, {Bentley}, {Benyaala}, {Bergamin}, {Berger}, {Bernuzzi}, {Berry}, {Bersanetti}, {Bertolini}, {Betzwieser}, {Beveridge}, {Bhandare}, {Bhardwaj}, {Bhattacharjee}, {Bhaumik}, {Bilenko}, {Billingsley}, {Bini}, {Birney}, {Birnholtz}, {Biscans}, {Bischi}, {Biscoveanu}, {Bisht}, {Biswas}, {Bitossi}, {Bizouard}, {Blackburn}, {Blair}, {Blair}, {Blair}, {Bobba}, {Bode}, {Boer}, {Bogaert}, {Boldrini}, {Bonavena}, {Bondu}, {Bonilla}, {Bonnand}, {Booker}, {Boom}, {Bork}, {Boschi}, {Bose}, {Bose}, {Bossilkov}, {Boudart}, {Bouffanais}, {Bozzi}, {Bradaschia}, {Brady}, {Bramley}, {Branch}, {Branchesi}, {Brandt}, {Brau}, {Breschi}, {Briant}, {Briggs}, {Brillet}, {Brinkmann}, {Brockill}, {Brooks}, {Brooks}, {Brown}, {Brunett}, {Bruno}, {Bruntz}, {Bryant}, {Bulik}, {Bulten}, {Buonanno}, {Buscicchio}, {Buskulic}, {Buy}, {Byer}, {Cadonati}, {Cagnoli}, {Cahillane}, {Bustillo}, {Callaghan}, {Callister}, {Calloni}, {Cameron},
  {Camp}, {Canepa}, {Canevarolo}, {Cannavacciuolo}, {Cannon}, {Cao}, {Cao}, {Capocasa}, {Capote}, \& {Carapella}}]{2023PhRvX..13a1048A}
{Abbott}, R., {Abbott}, T.~D., {Acernese}, F., {et~al.} 2023, Physical Review X, 13, 011048, \dodoi{10.1103/PhysRevX.13.011048}

\bibitem[{{Abbott} {et~al.}(2018){Abbott}, {Abdalla}, {Allam}, {Amara}, {Annis}, {Asorey}, {Avila}, {Ballester}, {Banerji}, {Barkhouse}, {Baruah}, {Baumer}, {Bechtol}, {Becker}, {Benoit-L{\'e}vy}, {Bernstein}, {Bertin}, {Blazek}, {Bocquet}, {Brooks}, {Brout}, {Buckley-Geer}, {Burke}, {Busti}, {Campisano}, {Cardiel-Sas}, {Carnero Rosell}, {Carrasco Kind}, {Carretero}, {Castander}, {Cawthon}, {Chang}, {Chen}, {Conselice}, {Costa}, {Crocce}, {Cunha}, {D'Andrea}, {da Costa}, {Das}, {Daues}, {Davis}, {Davis}, {De Vicente}, {DePoy}, {DeRose}, {Desai}, {Diehl}, {Dietrich}, {Dodelson}, {Doel}, {Drlica-Wagner}, {Eifler}, {Elliott}, {Evrard}, {Farahi}, {Fausti Neto}, {Fernandez}, {Finley}, {Flaugher}, {Foley}, {Fosalba}, {Friedel}, {Frieman}, {Garc{\'\i}a-Bellido}, {Gaztanaga}, {Gerdes}, {Giannantonio}, {Gill}, {Glazebrook}, {Goldstein}, {Gower}, {Gruen}, {Gruendl}, {Gschwend}, {Gupta}, {Gutierrez}, {Hamilton}, {Hartley}, {Hinton}, {Hislop}, {Hollowood}, {Honscheid}, {Hoyle}, {Huterer}, {Jain}, {James}, {Jeltema},
  {Johnson}, {Johnson}, {Kacprzak}, {Kent}, {Khullar}, {Klein}, {Kovacs}, {Koziol}, {Krause}, {Kremin}, {Kron}, {Kuehn}, {Kuhlmann}, {Kuropatkin}, {Lahav}, {Lasker}, {Li}, {Li}, {Liddle}, {Lima}, {Lin}, {L{\'o}pez-Reyes}, {MacCrann}, {Maia}, {Maloney}, {Manera}, {March}, {Marriner}, {Marshall}, {Martini}, {McClintock}, {McKay}, {McMahon}, {Melchior}, {Menanteau}, {Miller}, {Miquel}, {Mohr}, {Morganson}, {Mould}, {Neilsen}, {Nichol}, {Nogueira}, {Nord}, {Nugent}, {Nunes}, {Ogando}, {Old}, {Pace}, {Palmese}, {Paz-Chinch{\'o}n}, {Peiris}, {Percival}, {Petravick}, {Plazas}, {Poh}, {Pond}, {Porredon}, {Pujol}, {Refregier}, {Reil}, {Ricker}, {Rollins}, {Romer}, {Roodman}, {Rooney}, {Ross}, {Rykoff}, {Sako}, {Sanchez}, {Sanchez}, {Santiago}, {Saro}, {Scarpine}, {Scolnic}, {Serrano}, {Sevilla-Noarbe}, {Sheldon}, {Shipp}, {Silveira}, {Smith}, {Smith}, {Smith}, {Soares-Santos}, {Sobreira}, {Song}, {Stebbins}, {Suchyta}, {Sullivan}, {Swanson}, {Tarle}, {Thaler}, {Thomas}, {Thomas}, {Troxel}, {Tucker}, {Vikram}, {Vivas},
  {Walker}, {Wechsler}, {Weller}, {Wester}, {Wolf}, {Wu}, {Yanny}, {Zenteno}, {Zhang}, {Zuntz}, {DES Collaboration}, {Juneau}, {Fitzpatrick}, {Nikutta}, {Nidever}, {Olsen}, {Scott}, \& {NOAO Data Lab}}]{2018ApJS..239...18A}
{Abbott}, T.~M.~C., {Abdalla}, F.~B., {Allam}, S., {et~al.} 2018, \apjs, 239, 18, \dodoi{10.3847/1538-4365/aae9f0}

\bibitem[{{Ajello} {et~al.}(2021){Ajello}, {Atwood}, {Axelsson}, {Bagagli}, {Bagni}, {Baldini}, {Bastieri}, {Bellardi}, {Bellazzini}, {Bissaldi}, {Bloom}, {Bonino}, {Bregeon}, {Brez}, {Bruel}, {Buehler}, {Buson}, {Cameron}, {Caraveo}, {Cavazzuti}, {Ceccanti}, {Chen}, {Cheung}, {Ciprini}, {Cognard}, {Cohen-Tanugi}, {Cutini}, {D'Ammando}, {de la Torre Luque}, {de Palma}, {Digel}, {Dirirsa}, {Di Lalla}, {Di Venere}, {Dom{\'\i}nguez}, {Fabiani}, {Ferrara}, {Fiori}, {Foglia}, {Fukazawa}, {Fusco}, {Gargano}, {Gasparrini}, {Giroletti}, {Glanzman}, {Green}, {Griffin}, {Grondin}, {Grove}, {Guillemot}, {Guiriec}, {Gustafsson}, {Hays}, {Horan}, {J{\'o}hannesson}, {Johnson}, {Kamae}, {Kerr}, {Kuss}, {Larsson}, {Latronico}, {Lemoine-Goumard}, {Li}, {Liodakis}, {Longo}, {Loparco}, {Lovellette}, {Lubrano}, {Maldera}, {Manfreda}, {Mart{\'\i}-Devesa}, {Mazziotta}, {Menon}, {Mereu}, {Meyer}, {Michelson}, {Minuti}, {Mitthumsiri}, {Mizuno}, {Mongelli}, {Monzani}, {Moskalenko}, {Negro}, {Nuss}, {Ojha}, {Orienti}, {Orlando},
  {Paccagnella}, {Paliya}, {Paneque}, {Pei}, {Perkins}, {Pesce-Rollins}, {Pinchera}, {Piron}, {Poon}, {Porter}, {Primavera}, {Principe}, {Racusin}, {Rain{\`o}}, {Rando}, {Rani}, {Rapposelli}, {Razzano}, {Razzaque}, {Reimer}, {Reimer}, {Russell}, {Saggini}, {Saz Parkinson}, {Scolieri}, {Serini}, {Sgr{\`o}}, {Siskind}, {Smith}, {Spandre}, {Spinelli}, {Suson}, {Tajima}, {Thayer}, {Thompson}, {Tibaldo}, {Torres}, {Tosti}, {Valverde}, {Vigiani}, \& {Zaharijas}}]{2021ApJS..256...12A}
{Ajello}, M., {Atwood}, W.~B., {Axelsson}, M., {et~al.} 2021, \apjs, 256, 12, \dodoi{10.3847/1538-4365/ac0ceb}

\bibitem[{{Alam} {et~al.}(2015){Alam}, {Albareti}, {Allende Prieto}, {Anders}, {Anderson}, {Anderton}, {Andrews}, {Armengaud}, {Aubourg}, {Bailey}, \& et~al.}]{2015ApJS..219...12A}
{Alam}, S., {Albareti}, F.~D., {Allende Prieto}, C., {et~al.} 2015, \apjs, 219, 12, \dodoi{10.1088/0067-0049/219/1/12}

\bibitem[{{Balasubramanian} {et~al.}(2021){Balasubramanian}, {Corsi}, {Mooley}, {Brightman}, {Hallinan}, {Hotokezaka}, {Kaplan}, {Lazzati}, \& {Murphy}}]{2021ApJ...914L..20B}
{Balasubramanian}, A., {Corsi}, A., {Mooley}, K.~P., {et~al.} 2021, \apjl, 914, L20, \dodoi{10.3847/2041-8213/abfd38}

\bibitem[{{Baldwin} {et~al.}(1981){Baldwin}, {Phillips}, \& {Terlevich}}]{1981PASP...93....5B}
{Baldwin}, J.~A., {Phillips}, M.~M., \& {Terlevich}, R. 1981, \pasp, 93, 5, \dodoi{10.1086/130766}

\bibitem[{{Beck} {et~al.}(2021){Beck}, {Szapudi}, {Flewelling}, {Holmberg}, {Magnier}, \& {Chambers}}]{2021MNRAS.500.1633B}
{Beck}, R., {Szapudi}, I., {Flewelling}, H., {et~al.} 2021, \mnras, 500, 1633, \dodoi{10.1093/mnras/staa2587}

\bibitem[{{Bellm} {et~al.}(2019){Bellm}, {Kulkarni}, {Graham}, {Dekany}, {Smith}, {Riddle}, {Masci}, {Helou}, {Prince}, {Adams}, {Barbarino}, {Barlow}, {Bauer}, {Beck}, {Belicki}, {Biswas}, {Blagorodnova}, {Bodewits}, {Bolin}, {Brinnel}, {Brooke}, {Bue}, {Bulla}, {Burruss}, {Cenko}, {Chang}, {Connolly}, {Coughlin}, {Cromer}, {Cunningham}, {De}, {Delacroix}, {Desai}, {Duev}, {Eadie}, {Farnham}, {Feeney}, {Feindt}, {Flynn}, {Franckowiak}, {Frederick}, {Fremling}, {Gal-Yam}, {Gezari}, {Giomi}, {Goldstein}, {Golkhou}, {Goobar}, {Groom}, {Hacopians}, {Hale}, {Henning}, {Ho}, {Hover}, {Howell}, {Hung}, {Huppenkothen}, {Imel}, {Ip}, {Ivezi{\'c}}, {Jackson}, {Jones}, {Juric}, {Kasliwal}, {Kaspi}, {Kaye}, {Kelley}, {Kowalski}, {Kramer}, {Kupfer}, {Landry}, {Laher}, {Lee}, {Lin}, {Lin}, {Lunnan}, {Giomi}, {Mahabal}, {Mao}, {Miller}, {Monkewitz}, {Murphy}, {Ngeow}, {Nordin}, {Nugent}, {Ofek}, {Patterson}, {Penprase}, {Porter}, {Rauch}, {Rebbapragada}, {Reiley}, {Rigault}, {Rodriguez}, {van Roestel}, {Rusholme}, {van
  Santen}, {Schulze}, {Shupe}, {Singer}, {Soumagnac}, {Stein}, {Surace}, {Sollerman}, {Szkody}, {Taddia}, {Terek}, {Van Sistine}, {van Velzen}, {Vestrand}, {Walters}, {Ward}, {Ye}, {Yu}, {Yan}, \& {Zolkower}}]{2019PASP..131a8002B}
{Bellm}, E.~C., {Kulkarni}, S.~R., {Graham}, M.~J., {et~al.} 2019, \pasp, 131, 018002, \dodoi{10.1088/1538-3873/aaecbe}

\bibitem[{{Bhardwaj} {et~al.}(2023){Bhardwaj}, {Michilli}, {Kirichenko}, {Modilim}, {Shin}, {Kaspi}, {Andersen}, {Cassanelli}, {Brar}, {Chatterjee}, {Cook}, {Dong}, {Fonseca}, {Gaensler}, {Ibik}, {Kaczmarek}, {Lanman}, {Leung}, {Masui}, {Pandhi}, {Pearlman}, {Pleunis}, {Prochaska}, {Rafiei-Ravandi}, {Sand}, {Scholz}, \& {Smith}}]{2023arXiv231010018B}
{Bhardwaj}, M., {Michilli}, D., {Kirichenko}, A.~Y., {et~al.} 2023, arXiv e-prints, arXiv:2310.10018, \dodoi{10.48550/arXiv.2310.10018}

\bibitem[{{Blanchard} {et~al.}(2016){Blanchard}, {Berger}, \& {Fong}}]{2016ApJ...817..144B}
{Blanchard}, P.~K., {Berger}, E., \& {Fong}, W.-f. 2016, \apj, 817, 144, \dodoi{10.3847/0004-637X/817/2/144}

\bibitem[{{Bloom} {et~al.}(2002){Bloom}, {Kulkarni}, \& {Djorgovski}}]{2002AJ....123.1111B}
{Bloom}, J.~S., {Kulkarni}, S.~R., \& {Djorgovski}, S.~G. 2002, \aj, 123, 1111, \dodoi{10.1086/338893}

\bibitem[{{Calzetti} {et~al.}(2000){Calzetti}, {Armus}, {Bohlin}, {Kinney}, {Koornneef}, \& {Storchi-Bergmann}}]{2000ApJ...533..682C}
{Calzetti}, D., {Armus}, L., {Bohlin}, R.~C., {et~al.} 2000, \apj, 533, 682, \dodoi{10.1086/308692}

\bibitem[{{Cappellari}(2017)}]{2017MNRAS.466..798C}
{Cappellari}, M. 2017, \mnras, 466, 798, \dodoi{10.1093/mnras/stw3020}

\bibitem[{{Cappellari}(2022)}]{2022arXiv220814974C}
---. 2022, arXiv e-prints, arXiv:2208.14974.
\newblock \doarXiv{2208.14974}

\bibitem[{{Carilli} \& {Rawlings}(2004)}]{2004NewAR..48..979C}
{Carilli}, C.~L., \& {Rawlings}, S. 2004, \nar, 48, 979, \dodoi{10.1016/j.newar.2004.09.001}

\bibitem[{{Cavagnolo} {et~al.}(2010){Cavagnolo}, {McNamara}, {Nulsen}, {Carilli}, {Jones}, \& {B{\^\i}rzan}}]{2010ApJ...720.1066C}
{Cavagnolo}, K.~W., {McNamara}, B.~R., {Nulsen}, P.~E.~J., {et~al.} 2010, \apj, 720, 1066, \dodoi{10.1088/0004-637X/720/2/1066}

\bibitem[{{Chambers} {et~al.}(2016){Chambers}, {Magnier}, {Metcalfe}, {Flewelling}, {Huber}, {Waters}, {Denneau}, {Draper}, {Farrow}, {Finkbeiner}, {Holmberg}, {Koppenhoefer}, {Price}, {Rest}, {Saglia}, {Schlafly}, {Smartt}, {Sweeney}, {Wainscoat}, {Burgett}, {Chastel}, {Grav}, {Heasley}, {Hodapp}, {Jedicke}, {Kaiser}, {Kudritzki}, {Luppino}, {Lupton}, {Monet}, {Morgan}, {Onaka}, {Shiao}, {Stubbs}, {Tonry}, {White}, {Ba{\~n}ados}, {Bell}, {Bender}, {Bernard}, {Boegner}, {Boffi}, {Botticella}, {Calamida}, {Casertano}, {Chen}, {Chen}, {Cole}, {Deacon}, {Frenk}, {Fitzsimmons}, {Gezari}, {Gibbs}, {Goessl}, {Goggia}, {Gourgue}, {Goldman}, {Grant}, {Grebel}, {Hambly}, {Hasinger}, {Heavens}, {Heckman}, {Henderson}, {Henning}, {Holman}, {Hopp}, {Ip}, {Isani}, {Jackson}, {Keyes}, {Koekemoer}, {Kotak}, {Le}, {Liska}, {Long}, {Lucey}, {Liu}, {Martin}, {Masci}, {McLean}, {Mindel}, {Misra}, {Morganson}, {Murphy}, {Obaika}, {Narayan}, {Nieto-Santisteban}, {Norberg}, {Peacock}, {Pier}, {Postman}, {Primak}, {Rae}, {Rai},
  {Riess}, {Riffeser}, {Rix}, {R{\"o}ser}, {Russel}, {Rutz}, {Schilbach}, {Schultz}, {Scolnic}, {Strolger}, {Szalay}, {Seitz}, {Small}, {Smith}, {Soderblom}, {Taylor}, {Thomson}, {Taylor}, {Thakar}, {Thiel}, {Thilker}, {Unger}, {Urata}, {Valenti}, {Wagner}, {Walder}, {Walter}, {Watters}, {Werner}, {Wood-Vasey}, \& {Wyse}}]{2016arXiv161205560C}
{Chambers}, K.~C., {Magnier}, E.~A., {Metcalfe}, N., {et~al.} 2016, arXiv e-prints, arXiv:1612.05560.
\newblock \doarXiv{1612.05560}

\bibitem[{{Chandra} {et~al.}(2020){Chandra}, {Chevalier}, {Chugai}, {Milisavljevic}, \& {Fransson}}]{2020ApJ...902...55C}
{Chandra}, P., {Chevalier}, R.~A., {Chugai}, N., {Milisavljevic}, D., \& {Fransson}, C. 2020, \apj, 902, 55, \dodoi{10.3847/1538-4357/abb460}

\bibitem[{{Chandra} \& {Frail}(2012)}]{2012ApJ...746..156C}
{Chandra}, P., \& {Frail}, D.~A. 2012, \apj, 746, 156, \dodoi{10.1088/0004-637X/746/2/156}

\bibitem[{{Chevalier}(1977)}]{1977ASSL...66...53C}
{Chevalier}, R.~A. 1977, in Astrophysics and Space Science Library, Vol.~66, Supernovae, ed. D.~N. {Schramm}, 53, \dodoi{10.1007/978-94-010-1229-4_5}

\bibitem[{{Chevalier}(1998)}]{1998ApJ...499..810C}
{Chevalier}, R.~A. 1998, \apj, 499, 810, \dodoi{10.1086/305676}

\bibitem[{{Chrimes} {et~al.}(2024{\natexlab{a}}){Chrimes}, {Coppejans}, {Jonker}, {Levan}, {Groot}, {Mummery}, \& {Stanway}}]{2024A&A...691A.329C}
{Chrimes}, A.~A., {Coppejans}, D.~L., {Jonker}, P.~G., {et~al.} 2024{\natexlab{a}}, \aap, 691, A329, \dodoi{10.1051/0004-6361/202451172}

\bibitem[{{Chrimes} {et~al.}(2024{\natexlab{b}}){Chrimes}, {Jonker}, {Levan}, {Coppejans}, {Gaspari}, {Gompertz}, {Groot}, {Malesani}, {Mummery}, {Stanway}, \& {Wiersema}}]{2024MNRAS.527L..47C}
{Chrimes}, A.~A., {Jonker}, P.~G., {Levan}, A.~J., {et~al.} 2024{\natexlab{b}}, \mnras, 527, L47, \dodoi{10.1093/mnrasl/slad145}

\bibitem[{{Coppejans} {et~al.}(2020){Coppejans}, {Margutti}, {Terreran}, {Nayana}, {Coughlin}, {Laskar}, {Alexander}, {Bietenholz}, {Caprioli}, {Chandra}, {Drout}, {Frederiks}, {Frohmaier}, {Hurley}, {Kochanek}, {MacLeod}, {Meisner}, {Nugent}, {Ridnaia}, {Sand}, {Svinkin}, {Ward}, {Yang}, {Baldeschi}, {Chilingarian}, {Dong}, {Esquivia}, {Fong}, {Guidorzi}, {Lundqvist}, {Milisavljevic}, {Paterson}, {Reichart}, {Shappee}, {Stroh}, {Valenti}, {Zauderer}, \& {Zhang}}]{2020ApJ...895L..23C}
{Coppejans}, D.~L., {Margutti}, R., {Terreran}, G., {et~al.} 2020, \apjl, 895, L23, \dodoi{10.3847/2041-8213/ab8cc7}

\bibitem[{{Corbel} {et~al.}(2012){Corbel}, {Dubus}, {Tomsick}, {Szostek}, {Corbet}, {Miller-Jones}, {Richards}, {Pooley}, {Trushkin}, {Dubois}, {Hill}, {Kerr}, {Max-Moerbeck}, {Readhead}, {Bodaghee}, {Tudose}, {Parent}, {Wilms}, \& {Pottschmidt}}]{2012MNRAS.421.2947C}
{Corbel}, S., {Dubus}, G., {Tomsick}, J.~A., {et~al.} 2012, \mnras, 421, 2947, \dodoi{10.1111/j.1365-2966.2012.20517.x}

\bibitem[{{Corsi} {et~al.}(2014){Corsi}, {Ofek}, {Gal-Yam}, {Frail}, {Kulkarni}, {Fox}, {Kasliwal}, {Sullivan}, {Horesh}, {Carpenter}, {Maguire}, {Arcavi}, {Cenko}, {Cao}, {Mooley}, {Pan}, {Sesar}, {Sternberg}, {Xu}, {Bersier}, {James}, {Bloom}, \& {Nugent}}]{2014ApJ...782...42C}
{Corsi}, A., {Ofek}, E.~O., {Gal-Yam}, A., {et~al.} 2014, \apj, 782, 42, \dodoi{10.1088/0004-637X/782/1/42}

\bibitem[{{Crill} {et~al.}(2020){Crill}, {Werner}, {Akeson}, {Ashby}, {Bleem}, {Bock}, {Bryan}, {Burnham}, {Byunh}, {Chang}, {Chiang}, {Cook}, {Cooray}, {Davis}, {Dor{\'e}}, {Dowell}, {Dubois-Felsmann}, {Eifler}, {Faisst}, {Habib}, {Heinrich}, {Heitmann}, {Heaton}, {Hirata}, {Hristov}, {Hui}, {Jeong}, {Kang}, {Kecman}, {Kirkpatrick}, {Korngut}, {Krause}, {Lee}, {Lisse}, {Masters}, {Mauskopf}, {Melnick}, {Miyasaka}, {Nayyeri}, {Nguyen}, {{\"O}berg}, {Padin}, {Paladini}, {Pourrahmani}, {Pyo}, {Smith}, {Song}, {Symons}, {Teplitz}, {Tolls}, {Unwin}, {Windhorst}, {Yang}, \& {Zemcov}}]{2020SPIE11443E..0IC}
{Crill}, B.~P., {Werner}, M., {Akeson}, R., {et~al.} 2020, in Society of Photo-Optical Instrumentation Engineers (SPIE) Conference Series, Vol. 11443, Space Telescopes and Instrumentation 2020: Optical, Infrared, and Millimeter Wave, ed. M.~{Lystrup} \& M.~D. {Perrin}, 114430I, \dodoi{10.1117/12.2567224}

\bibitem[{{Cutri} {et~al.}(2021){Cutri}, {Wright}, {Conrow}, {Fowler}, {Eisenhardt}, {Grillmair}, {Kirkpatrick}, {Masci}, {McCallon}, {Wheelock}, {Fajardo-Acosta}, {Yan}, {Benford}, {Harbut}, {Jarrett}, {Lake}, {Leisawitz}, {Ressler}, {Stanford}, {Tsai}, {Liu}, {Helou}, {Mainzer}, {Gettngs}, {Gonzalez}, {Hoffman}, {Marsh}, {Padgett}, {Skrutskie}, {Beck}, {Papin}, \& {Wittman}}]{2014yCat.2328....0C}
{Cutri}, R.~M., {Wright}, E.~L., {Conrow}, T., {et~al.} 2021, VizieR Online Data Catalog, II/328

\bibitem[{{Dahlen} {et~al.}(2012){Dahlen}, {Strolger}, {Riess}, {Mattila}, {Kankare}, \& {Mobasher}}]{2012ApJ...757...70D}
{Dahlen}, T., {Strolger}, L.-G., {Riess}, A.~G., {et~al.} 2012, \apj, 757, 70, \dodoi{10.1088/0004-637X/757/1/70}

\bibitem[{{Dall'Osso} \& {Stella}(2022)}]{2022ASSL..465..245D}
{Dall'Osso}, S., \& {Stella}, L. 2022, in Astrophysics and Space Science Library, Vol. 465, Astrophysics and Space Science Library, ed. S.~{Bhattacharyya}, A.~{Papitto}, \& D.~{Bhattacharya}, 245--280, \dodoi{10.1007/978-3-030-85198-9_8}

\bibitem[{{Dey} {et~al.}(2019){Dey}, {Schlegel}, {Lang}, {Blum}, {Burleigh}, {Fan}, {Findlay}, {Finkbeiner}, {Herrera}, {Juneau}, {Landriau}, {Levi}, {McGreer}, {Meisner}, {Myers}, {Moustakas}, {Nugent}, {Patej}, {Schlafly}, {Walker}, {Valdes}, {Weaver}, {Y{\`e}che}, {Zou}, {Zhou}, {Abareshi}, {Abbott}, {Abolfathi}, {Aguilera}, {Alam}, {Allen}, {Alvarez}, {Annis}, {Ansarinejad}, {Aubert}, {Beechert}, {Bell}, {BenZvi}, {Beutler}, {Bielby}, {Bolton}, {Brice{\~n}o}, {Buckley-Geer}, {Butler}, {Calamida}, {Carlberg}, {Carter}, {Casas}, {Castander}, {Choi}, {Comparat}, {Cukanovaite}, {Delubac}, {DeVries}, {Dey}, {Dhungana}, {Dickinson}, {Ding}, {Donaldson}, {Duan}, {Duckworth}, {Eftekharzadeh}, {Eisenstein}, {Etourneau}, {Fagrelius}, {Farihi}, {Fitzpatrick}, {Font-Ribera}, {Fulmer}, {G{\"a}nsicke}, {Gaztanaga}, {George}, {Gerdes}, {Gontcho}, {Gorgoni}, {Green}, {Guy}, {Harmer}, {Hernandez}, {Honscheid}, {Huang}, {James}, {Jannuzi}, {Jiang}, {Joyce}, {Karcher}, {Karkar}, {Kehoe}, {Kneib}, {Kueter-Young}, {Lan},
  {Lauer}, {Le Guillou}, {Le Van Suu}, {Lee}, {Lesser}, {Perreault Levasseur}, {Li}, {Mann}, {Marshall}, {Mart{\'\i}nez-V{\'a}zquez}, {Martini}, {du Mas des Bourboux}, {McManus}, {Meier}, {M{\'e}nard}, {Metcalfe}, {Mu{\~n}oz-Guti{\'e}rrez}, {Najita}, {Napier}, {Narayan}, {Newman}, {Nie}, {Nord}, {Norman}, {Olsen}, {Paat}, {Palanque-Delabrouille}, {Peng}, {Poppett}, {Poremba}, {Prakash}, {Rabinowitz}, {Raichoor}, {Rezaie}, {Robertson}, {Roe}, {Ross}, {Ross}, {Rudnick}, {Safonova}, {Saha}, {S{\'a}nchez}, {Savary}, {Schweiker}, {Scott}, {Seo}, {Shan}, {Silva}, {Slepian}, {Soto}, {Sprayberry}, {Staten}, {Stillman}, {Stupak}, {Summers}, {Sien Tie}, {Tirado}, {Vargas-Maga{\~n}a}, {Vivas}, {Wechsler}, {Williams}, {Yang}, {Yang}, {Yapici}, {Zaritsky}, {Zenteno}, {Zhang}, {Zhang}, {Zhou}, \& {Zhou}}]{2019AJ....157..168D}
{Dey}, A., {Schlegel}, D.~J., {Lang}, D., {et~al.} 2019, \aj, 157, 168, \dodoi{10.3847/1538-3881/ab089d}

\bibitem[{Dong(2023)}]{Dong2023}
Dong, D.~Z. 2023, Dissertation (ph.d.), California Institute of Technology, \dodoi{10.7907/xwy9-tc17}

\bibitem[{{Dong} \& {Hallinan}(2023)}]{2023ApJ...948..119D}
{Dong}, D.~Z., \& {Hallinan}, G. 2023, \apj, 948, 119, \dodoi{10.3847/1538-4357/acc06c}

\bibitem[{{Dong} {et~al.}(2021){Dong}, {Hallinan}, {Nakar}, {Ho}, {Hughes}, {Hotokezaka}, {Myers}, {De}, {Mooley}, {Ravi}, {Horesh}, {Kasliwal}, \& {Kulkarni}}]{2021Sci...373.1125D}
{Dong}, D.~Z., {Hallinan}, G., {Nakar}, E., {et~al.} 2021, Science, 373, 1125, \dodoi{10.1126/science.abg6037}

\bibitem[{{Draine} \& {Li}(2007)}]{2007ApJ...657..810D}
{Draine}, B.~T., \& {Li}, A. 2007, \apj, 657, 810, \dodoi{10.1086/511055}

\bibitem[{{Driver} {et~al.}(2016){Driver}, {Andrews}, {Davies}, {Robotham}, {Wright}, {Windhorst}, {Cohen}, {Emig}, {Jansen}, \& {Dunne}}]{2016ApJ...827..108D}
{Driver}, S.~P., {Andrews}, S.~K., {Davies}, L.~J., {et~al.} 2016, \apj, 827, 108, \dodoi{10.3847/0004-637X/827/2/108}

\bibitem[{{Duncan}(2022)}]{2022MNRAS.512.3662D}
{Duncan}, K.~J. 2022, \mnras, 512, 3662, \dodoi{10.1093/mnras/stac608}

\bibitem[{{Duncan} \& {Thompson}(1992)}]{1992ApJ...392L...9D}
{Duncan}, R.~C., \& {Thompson}, C. 1992, \apjl, 392, L9, \dodoi{10.1086/186413}

\bibitem[{{Eftekhari} \& {Berger}(2017)}]{2017ApJ...849..162E}
{Eftekhari}, T., \& {Berger}, E. 2017, \apj, 849, 162, \dodoi{10.3847/1538-4357/aa90b9}

\bibitem[{{Eftekhari} {et~al.}(2018){Eftekhari}, {Berger}, {Zauderer}, {Margutti}, \& {Alexander}}]{2018ApJ...854...86E}
{Eftekhari}, T., {Berger}, E., {Zauderer}, B.~A., {Margutti}, R., \& {Alexander}, K.~D. 2018, \apj, 854, 86, \dodoi{10.3847/1538-4357/aaa8e0}

\bibitem[{{Ferrario} {et~al.}(2015){Ferrario}, {Melatos}, \& {Zrake}}]{2015SSRv..191...77F}
{Ferrario}, L., {Melatos}, A., \& {Zrake}, J. 2015, \ssr, 191, 77, \dodoi{10.1007/s11214-015-0138-y}

\bibitem[{{Fong} {et~al.}(2015){Fong}, {Berger}, {Margutti}, \& {Zauderer}}]{2015ApJ...815..102F}
{Fong}, W., {Berger}, E., {Margutti}, R., \& {Zauderer}, B.~A. 2015, \apj, 815, 102, \dodoi{10.1088/0004-637X/815/2/102}

\bibitem[{{Fong} {et~al.}(2022){Fong}, {Nugent}, {Dong}, {Berger}, {Paterson}, {Chornock}, {Levan}, {Blanchard}, {Alexander}, {Andrews}, {Cobb}, {Cucchiara}, {Fox}, {Fryer}, {Gordon}, {Kilpatrick}, {Lunnan}, {Margutti}, {Miller}, {Milne}, {Nicholl}, {Perley}, {Rastinejad}, {Escorial}, {Schroeder}, {Smith}, {Tanvir}, \& {Terreran}}]{2022ApJ...940...56F}
{Fong}, W.-f., {Nugent}, A.~E., {Dong}, Y., {et~al.} 2022, \apj, 940, 56, \dodoi{10.3847/1538-4357/ac91d010.48550/arXiv.2206.01763}

\bibitem[{{Forster} {et~al.}(2019){Forster}, {Bauer}, {Pignata}, {Arredondo}, {Cabrera-Vives}, {Carrasco-Davis}, {Estevez}, {Huijse}, {Reyes}, {Reyes}, {Sanchez-Saez}, {Valenzuela}, {Castillo}, {Ruz-Mieres}, {Rodriguez-Mancini}, {Bauer}, {Catelan}, {Eyheramendy}, \& {Graham}}]{2019TNSTR2663....1F}
{Forster}, F., {Bauer}, F.~E., {Pignata}, G., {et~al.} 2019, Transient Name Server Discovery Report, 2019-2663, 1

\bibitem[{{Gaia Collaboration} {et~al.}(2016){Gaia Collaboration}, {Prusti}, {de Bruijne}, {Brown}, {Vallenari}, {Babusiaux}, {Bailer-Jones}, {Bastian}, {Biermann}, {Evans}, {Eyer}, {Jansen}, {Jordi}, {Klioner}, {Lammers}, {Lindegren}, {Luri}, {Mignard}, {Milligan}, {Panem}, {Poinsignon}, {Pourbaix}, {Randich}, {Sarri}, {Sartoretti}, {Siddiqui}, {Soubiran}, {Valette}, {van Leeuwen}, {Walton}, {Aerts}, {Arenou}, {Cropper}, {Drimmel}, {H{\o}g}, {Katz}, {Lattanzi}, {O'Mullane}, {Grebel}, {Holland}, {Huc}, {Passot}, {Bramante}, {Cacciari}, {Casta{\~n}eda}, {Chaoul}, {Cheek}, {De Angeli}, {Fabricius}, {Guerra}, {Hern{\'a}ndez}, {Jean-Antoine-Piccolo}, {Masana}, {Messineo}, {Mowlavi}, {Nienartowicz}, {Ord{\'o}{\~n}ez-Blanco}, {Panuzzo}, {Portell}, {Richards}, {Riello}, {Seabroke}, {Tanga}, {Th{\'e}venin}, {Torra}, {Els}, {Gracia-Abril}, {Comoretto}, {Garcia-Reinaldos}, {Lock}, {Mercier}, {Altmann}, {Andrae}, {Astraatmadja}, {Bellas-Velidis}, {Benson}, {Berthier}, {Blomme}, {Busso}, {Carry}, {Cellino}, {Clementini},
  {Cowell}, {Creevey}, {Cuypers}, {Davidson}, {De Ridder}, {de Torres}, {Delchambre}, {Dell'Oro}, {Ducourant}, {Fr{\'e}mat}, {Garc{\'\i}a-Torres}, {Gosset}, {Halbwachs}, {Hambly}, {Harrison}, {Hauser}, {Hestroffer}, {Hodgkin}, {Huckle}, {Hutton}, {Jasniewicz}, {Jordan}, {Kontizas}, {Korn}, {Lanzafame}, {Manteiga}, {Moitinho}, {Muinonen}, {Osinde}, {Pancino}, {Pauwels}, {Petit}, {Recio-Blanco}, {Robin}, {Sarro}, {Siopis}, {Smith}, {Smith}, {Sozzetti}, {Thuillot}, {van Reeven}, {Viala}, {Abbas}, {Abreu Aramburu}, {Accart}, {Aguado}, {Allan}, {Allasia}, {Altavilla}, {{\'A}lvarez}, {Alves}, {Anderson}, {Andrei}, {Anglada Varela}, {Antiche}, {Antoja}, {Ant{\'o}n}, {Arcay}, {Atzei}, {Ayache}, {Bach}, {Baker}, {Balaguer-N{\'u}{\~n}ez}, {Barache}, {Barata}, {Barbier}, {Barblan}, {Baroni}, {Barrado y Navascu{\'e}s}, {Barros}, {Barstow}, {Becciani}, {Bellazzini}, {Bellei}, {Bello Garc{\'\i}a}, {Belokurov}, {Bendjoya}, {Berihuete}, {Bianchi}, {Bienaym{\'e}}, {Billebaud}, {Blagorodnova}, {Blanco-Cuaresma}, {Boch},
  {Bombrun}, {Borrachero}, {Bouquillon}, {Bourda}, {Bouy}, {Bragaglia}, {Breddels}, {Brouillet}, {Br{\"u}semeister}, {Bucciarelli}, {Budnik}, {Burgess}, {Burgon}, {Burlacu}, {Busonero}, {Buzzi}, {Caffau}, {Cambras}, {Campbell}, {Cancelliere}, {Cantat-Gaudin}, {Carlucci}, {Carrasco}, {Castellani}, {Charlot}, {Charnas}, {Charvet}, {Chassat}, {Chiavassa}, {Clotet}, {Cocozza}, {Collins}, {Collins}, {Costigan}, {Crifo}, {Cross}, {Crosta}, {Crowley}, {Dafonte}, {Damerdji}, {Dapergolas}, {David}, {David}, {De Cat}, {de Felice}, {de Laverny}, {De Luise}, {De March}, {de Martino}, {de Souza}, {Debosscher}, {del Pozo}, {Delbo}, {Delgado}, {Delgado}, {di Marco}, {Di Matteo}, {Diakite}, {Distefano}, {Dolding}, {Dos Anjos}, {Drazinos}, {Dur{\'a}n}, {Dzigan}, {Ecale}, {Edvardsson}, {Enke}, {Erdmann}, {Escolar}, {Espina}, {Evans}, {Eynard Bontemps}, {Fabre}, {Fabrizio}, {Faigler}, {Falc{\~a}o}, {Farr{\`a}s Casas}, {Faye}, {Federici}, {Fedorets}, {Fern{\'a}ndez-Hern{\'a}ndez}, {Fernique}, {Fienga}, {Figueras}, {Filippi},
  {Findeisen}, {Fonti}, {Fouesneau}, {Fraile}, {Fraser}, {Fuchs}, {Furnell}, {Gai}, {Galleti}, {Galluccio}, {Garabato}, {Garc{\'\i}a-Sedano}, {Gar{\'e}}, {Garofalo}, {Garralda}, {Gavras}, {Gerssen}, {Geyer}, {Gilmore}, {Girona}, {Giuffrida}, {Gomes}, {Gonz{\'a}lez-Marcos}, {Gonz{\'a}lez-N{\'u}{\~n}ez}, {Gonz{\'a}lez-Vidal}, {Granvik}, {Guerrier}, {Guillout}, {Guiraud}, {G{\'u}rpide}, {Guti{\'e}rrez-S{\'a}nchez}, {Guy}, {Haigron}, {Hatzidimitriou}, {Haywood}, {Heiter}, {Helmi}, {Hobbs}, {Hofmann}, {Holl}, {Holland}, {Hunt}, {Hypki}, {Icardi}, {Irwin}, {Jevardat de Fombelle}, {Jofr{\'e}}, {Jonker}, {Jorissen}, {Julbe}, {Karampelas}, {Kochoska}, {Kohley}, {Kolenberg}, {Kontizas}, {Koposov}, {Kordopatis}, {Koubsky}, {Kowalczyk}, {Krone-Martins}, {Kudryashova}, {Kull}, {Bachchan}, {Lacoste-Seris}, {Lanza}, {Lavigne}, {Le Poncin-Lafitte}, {Lebreton}, {Lebzelter}, {Leccia}, {Leclerc}, {Lecoeur-Taibi}, {Lemaitre}, {Lenhardt}, {Leroux}, {Liao}, {Licata}, {Lindstr{\o}m}, {Lister}, {Livanou}, {Lobel}, {L{\"o}ffler},
  {L{\'o}pez}, {Lopez-Lozano}, {Lorenz}, {Loureiro}, {MacDonald}, {Magalh{\~a}es Fernandes}, {Managau}, {Mann}, {Mantelet}, {Marchal}, {Marchant}, {Marconi}, {Marie}, {Marinoni}, {Marrese}, {Marschalk{\'o}}, {Marshall}, {Mart{\'\i}n-Fleitas}, {Martino}, {Mary}, {Matijevi{\v{c}}}, {Mazeh}, {McMillan}, {Messina}, {Mestre}, {Michalik}, {Millar}, {Miranda}, {Molina}, {Molinaro}, {Molinaro}, {Moln{\'a}r}, {Moniez}, {Montegriffo}, {Monteiro}, {Mor}, {Mora}, {Morbidelli}, {Morel}, {Morgenthaler}, {Morley}, {Morris}, {Mulone}, {Muraveva}, {Musella}, {Narbonne}, {Nelemans}, {Nicastro}, {Noval}, {Ord{\'e}novic}, {Ordieres-Mer{\'e}}, {Osborne}, {Pagani}, {Pagano}, {Pailler}, {Palacin}, {Palaversa}, {Parsons}, {Paulsen}, {Pecoraro}, {Pedrosa}, {Pentik{\"a}inen}, {Pereira}, {Pichon}, {Piersimoni}, {Pineau}, {Plachy}, {Plum}, {Poujoulet}, {Pr{\v{s}}a}, {Pulone}, {Ragaini}, {Rago}, {Rambaux}, {Ramos-Lerate}, {Ranalli}, {Rauw}, {Read}, {Regibo}, {Renk}, {Reyl{\'e}}, {Ribeiro}, {Rimoldini}, {Ripepi}, {Riva}, {Rixon},
  {Roelens}, {Romero-G{\'o}mez}, {Rowell}, {Royer}, {Rudolph}, {Ruiz-Dern}, {Sadowski}, {Sagrist{\`a} Sell{\'e}s}, {Sahlmann}, {Salgado}, {Salguero}, {Sarasso}, {Savietto}, {Schnorhk}, {Schultheis}, {Sciacca}, {Segol}, {Segovia}, {Segransan}, {Serpell}, {Shih}, {Smareglia}, {Smart}, {Smith}, {Solano}, {Solitro}, {Sordo}, {Soria Nieto}, {Souchay}, {Spagna}, {Spoto}, {Stampa}, {Steele}, {Steidelm{\"u}ller}, {Stephenson}, {Stoev}, {Suess}, {S{\"u}veges}, {Surdej}, {Szabados}, {Szegedi-Elek}, {Tapiador}, {Taris}, {Tauran}, {Taylor}, {Teixeira}, {Terrett}, {Tingley}, {Trager}, {Turon}, {Ulla}, {Utrilla}, {Valentini}, {van Elteren}, {Van Hemelryck}, {van Leeuwen}, {Varadi}, {Vecchiato}, {Veljanoski}, {Via}, {Vicente}, {Vogt}, {Voss}, {Votruba}, {Voutsinas}, {Walmsley}, {Weiler}, {Weingrill}, {Werner}, {Wevers}, {Whitehead}, {Wyrzykowski}, {Yoldas}, {{\v{Z}}erjal}, {Zucker}, {Zurbach}, {Zwitter}, {Alecu}, {Allen}, {Allende Prieto}, {Amorim}, {Anglada-Escud{\'e}}, {Arsenijevic}, {Azaz}, {Balm}, {Beck}, {Bernstein},
  {Bigot}, {Bijaoui}, {Blasco}, {Bonfigli}, {Bono}, {Boudreault}, {Bressan}, {Brown}, {Brunet}, {Bunclark}, {Buonanno}, {Butkevich}, {Carret}, {Carrion}, {Chemin}, {Ch{\'e}reau}, {Corcione}, {Darmigny}, {de Boer}, {de Teodoro}, {de Zeeuw}, {Delle Luche}, {Domingues}, {Dubath}, {Fodor}, {Fr{\'e}zouls}, {Fries}, {Fustes}, {Fyfe}, {Gallardo}, {Gallegos}, {Gardiol}, {Gebran}, {Gomboc}, {G{\'o}mez}, {Grux}, {Gueguen}, {Heyrovsky}, {Hoar}, {Iannicola}, {Isasi Parache}, {Janotto}, {Joliet}, {Jonckheere}, {Keil}, {Kim}, {Klagyivik}, {Klar}, {Knude}, {Kochukhov}, {Kolka}, {Kos}, {Kutka}, {Lainey}, {LeBouquin}, {Liu}, {Loreggia}, {Makarov}, {Marseille}, {Martayan}, {Martinez-Rubi}, {Massart}, {Meynadier}, {Mignot}, {Munari}, {Nguyen}, {Nordlander}, {Ocvirk}, {O'Flaherty}, {Olias Sanz}, {Ortiz}, {Osorio}, {Oszkiewicz}, {Ouzounis}, {Palmer}, {Park}, {Pasquato}, {Peltzer}, {Peralta}, {P{\'e}turaud}, {Pieniluoma}, {Pigozzi}, {Poels}, {Prat}, {Prod'homme}, {Raison}, {Rebordao}, {Risquez}, {Rocca-Volmerange}, {Rosen},
  {Ruiz-Fuertes}, {Russo}, {Sembay}, {Serraller Vizcaino}, {Short}, {Siebert}, {Silva}, {Sinachopoulos}, {Slezak}, {Soffel}, {Sosnowska}, {Strai{\v{z}}ys}, {ter Linden}, {Terrell}, {Theil}, {Tiede}, {Troisi}, {Tsalmantza}, {Tur}, {Vaccari}, {Vachier}, {Valles}, {Van Hamme}, {Veltz}, {Virtanen}, {Wallut}, {Wichmann}, {Wilkinson}, {Ziaeepour}, \& {Zschocke}}]{2016A&A...595A...1G}
{Gaia Collaboration}, {Prusti}, T., {de Bruijne}, J.~H.~J., {et~al.} 2016, \aap, 595, A1, \dodoi{10.1051/0004-6361/201629272}

\bibitem[{{Gaia Collaboration} {et~al.}(2023){Gaia Collaboration}, {Vallenari}, {Brown}, {Prusti}, {de Bruijne}, {Arenou}, {Babusiaux}, {Biermann}, {Creevey}, {Ducourant}, {Evans}, {Eyer}, {Guerra}, {Hutton}, {Jordi}, {Klioner}, {Lammers}, {Lindegren}, {Luri}, {Mignard}, {Panem}, {Pourbaix}, {Randich}, {Sartoretti}, {Soubiran}, {Tanga}, {Walton}, {Bailer-Jones}, {Bastian}, {Drimmel}, {Jansen}, {Katz}, {Lattanzi}, {van Leeuwen}, {Bakker}, {Cacciari}, {Casta{\~n}eda}, {De Angeli}, {Fabricius}, {Fouesneau}, {Fr{\'e}mat}, {Galluccio}, {Guerrier}, {Heiter}, {Masana}, {Messineo}, {Mowlavi}, {Nicolas}, {Nienartowicz}, {Pailler}, {Panuzzo}, {Riclet}, {Roux}, {Seabroke}, {Sordo}, {Th{\'e}venin}, {Gracia-Abril}, {Portell}, {Teyssier}, {Altmann}, {Andrae}, {Audard}, {Bellas-Velidis}, {Benson}, {Berthier}, {Blomme}, {Burgess}, {Busonero}, {Busso}, {C{\'a}novas}, {Carry}, {Cellino}, {Cheek}, {Clementini}, {Damerdji}, {Davidson}, {de Teodoro}, {Nu{\~n}ez Campos}, {Delchambre}, {Dell'Oro}, {Esquej},
  {Fern{\'a}ndez-Hern{\'a}ndez}, {Fraile}, {Garabato}, {Garc{\'\i}a-Lario}, {Gosset}, {Haigron}, {Halbwachs}, {Hambly}, {Harrison}, {Hern{\'a}ndez}, {Hestroffer}, {Hodgkin}, {Holl}, {Jan{\ss}en}, {Jevardat de Fombelle}, {Jordan}, {Krone-Martins}, {Lanzafame}, {L{\"o}ffler}, {Marchal}, {Marrese}, {Moitinho}, {Muinonen}, {Osborne}, {Pancino}, {Pauwels}, {Recio-Blanco}, {Reyl{\'e}}, {Riello}, {Rimoldini}, {Roegiers}, {Rybizki}, {Sarro}, {Siopis}, {Smith}, {Sozzetti}, {Utrilla}, {van Leeuwen}, {Abbas}, {{\'A}brah{\'a}m}, {Abreu Aramburu}, {Aerts}, {Aguado}, {Ajaj}, {Aldea-Montero}, {Altavilla}, {{\'A}lvarez}, {Alves}, {Anders}, {Anderson}, {Anglada Varela}, {Antoja}, {Baines}, {Baker}, {Balaguer-N{\'u}{\~n}ez}, {Balbinot}, {Balog}, {Barache}, {Barbato}, {Barros}, {Barstow}, {Bartolom{\'e}}, {Bassilana}, {Bauchet}, {Becciani}, {Bellazzini}, {Berihuete}, {Bernet}, {Bertone}, {Bianchi}, {Binnenfeld}, {Blanco-Cuaresma}, {Blazere}, {Boch}, {Bombrun}, {Bossini}, {Bouquillon}, {Bragaglia}, {Bramante}, {Breedt},
  {Bressan}, {Brouillet}, {Brugaletta}, {Bucciarelli}, {Burlacu}, {Butkevich}, {Buzzi}, {Caffau}, {Cancelliere}, {Cantat-Gaudin}, {Carballo}, {Carlucci}, {Carnerero}, {Carrasco}, {Casamiquela}, {Castellani}, {Castro-Ginard}, {Chaoul}, {Charlot}, {Chemin}, {Chiaramida}, {Chiavassa}, {Chornay}, {Comoretto}, {Contursi}, {Cooper}, {Cornez}, {Cowell}, {Crifo}, {Cropper}, {Crosta}, {Crowley}, {Dafonte}, {Dapergolas}, {David}, {David}, {de Laverny}, {De Luise}, {De March}, {De Ridder}, {de Souza}, {de Torres}, {del Peloso}, {del Pozo}, {Delbo}, {Delgado}, {Delisle}, {Demouchy}, {Dharmawardena}, {Di Matteo}, {Diakite}, {Diener}, {Distefano}, {Dolding}, {Edvardsson}, {Enke}, {Fabre}, {Fabrizio}, {Faigler}, {Fedorets}, {Fernique}, {Fienga}, {Figueras}, {Fournier}, {Fouron}, {Fragkoudi}, {Gai}, {Garcia-Gutierrez}, {Garcia-Reinaldos}, {Garc{\'\i}a-Torres}, {Garofalo}, {Gavel}, {Gavras}, {Gerlach}, {Geyer}, {Giacobbe}, {Gilmore}, {Girona}, {Giuffrida}, {Gomel}, {Gomez}, {Gonz{\'a}lez-N{\'u}{\~n}ez},
  {Gonz{\'a}lez-Santamar{\'\i}a}, {Gonz{\'a}lez-Vidal}, {Granvik}, {Guillout}, {Guiraud}, {Guti{\'e}rrez-S{\'a}nchez}, {Guy}, {Hatzidimitriou}, {Hauser}, {Haywood}, {Helmer}, {Helmi}, {Sarmiento}, {Hidalgo}, {Hilger}, {H{\l}adczuk}, {Hobbs}, {Holland}, {Huckle}, {Jardine}, {Jasniewicz}, {Jean-Antoine Piccolo}, {Jim{\'e}nez-Arranz}, {Jorissen}, {Juaristi Campillo}, {Julbe}, {Karbevska}, {Kervella}, {Khanna}, {Kontizas}, {Kordopatis}, {Korn}, {K{\'o}sp{\'a}l}, {Kostrzewa-Rutkowska}, {Kruszy{\'n}ska}, {Kun}, {Laizeau}, {Lambert}, {Lanza}, {Lasne}, {Le Campion}, {Lebreton}, {Lebzelter}, {Leccia}, {Leclerc}, {Lecoeur-Taibi}, {Liao}, {Licata}, {Lindstr{\o}m}, {Lister}, {Livanou}, {Lobel}, {Lorca}, {Loup}, {Madrero Pardo}, {Magdaleno Romeo}, {Managau}, {Mann}, {Manteiga}, {Marchant}, {Marconi}, {Marcos}, {Marcos Santos}, {Mar{\'\i}n Pina}, {Marinoni}, {Marocco}, {Marshall}, {Martin Polo}, {Mart{\'\i}n-Fleitas}, {Marton}, {Mary}, {Masip}, {Massari}, {Mastrobuono-Battisti}, {Mazeh}, {McMillan}, {Messina}, {Michalik},
  {Millar}, {Mints}, {Molina}, {Molinaro}, {Moln{\'a}r}, {Monari}, {Mongui{\'o}}, {Montegriffo}, {Montero}, {Mor}, {Mora}, {Morbidelli}, {Morel}, {Morris}, {Muraveva}, {Murphy}, {Musella}, {Nagy}, {Noval}, {Oca{\~n}a}, {Ogden}, {Ordenovic}, {Osinde}, {Pagani}, {Pagano}, {Palaversa}, {Palicio}, {Pallas-Quintela}, {Panahi}, {Payne-Wardenaar}, {Pe{\~n}alosa Esteller}, {Penttil{\"a}}, {Pichon}, {Piersimoni}, {Pineau}, {Plachy}, {Plum}, {Poggio}, {Pr{\v{s}}a}, {Pulone}, {Racero}, {Ragaini}, {Rainer}, {Raiteri}, {Rambaux}, {Ramos}, {Ramos-Lerate}, {Re Fiorentin}, {Regibo}, {Richards}, {Rios Diaz}, {Ripepi}, {Riva}, {Rix}, {Rixon}, {Robichon}, {Robin}, {Robin}, {Roelens}, {Rogues}, {Rohrbasser}, {Romero-G{\'o}mez}, {Rowell}, {Royer}, {Ruz Mieres}, {Rybicki}, {Sadowski}, {S{\'a}ez N{\'u}{\~n}ez}, {Sagrist{\`a} Sell{\'e}s}, {Sahlmann}, {Salguero}, {Samaras}, {Sanchez Gimenez}, {Sanna}, {Santove{\~n}a}, {Sarasso}, {Schultheis}, {Sciacca}, {Segol}, {Segovia}, {S{\'e}gransan}, {Semeux}, {Shahaf}, {Siddiqui}, {Siebert},
  {Siltala}, {Silvelo}, {Slezak}, {Slezak}, {Smart}, {Snaith}, {Solano}, {Solitro}, {Souami}, {Souchay}, {Spagna}, {Spina}, {Spoto}, {Steele}, {Steidelm{\"u}ller}, {Stephenson}, {S{\"u}veges}, {Surdej}, {Szabados}, {Szegedi-Elek}, {Taris}, {Taylor}, {Teixeira}, {Tolomei}, {Tonello}, {Torra}, {Torra}, {Torralba Elipe}, {Trabucchi}, {Tsounis}, {Turon}, {Ulla}, {Unger}, {Vaillant}, {van Dillen}, {van Reeven}, {Vanel}, {Vecchiato}, {Viala}, {Vicente}, {Voutsinas}, {Weiler}, {Wevers}, {Wyrzykowski}, {Yoldas}, {Yvard}, {Zhao}, {Zorec}, {Zucker}, \& {Zwitter}}]{2023A&A...674A...1G}
{Gaia Collaboration}, {Vallenari}, A., {Brown}, A.~G.~A., {et~al.} 2023, \aap, 674, A1, \dodoi{10.1051/0004-6361/202243940}

\bibitem[{{Gallazzi} {et~al.}(2005){Gallazzi}, {Charlot}, {Brinchmann}, {White}, \& {Tremonti}}]{2005MNRAS.362...41G}
{Gallazzi}, A., {Charlot}, S., {Brinchmann}, J., {White}, S. D.~M., \& {Tremonti}, C.~A. 2005, \mnras, 362, 41, \dodoi{10.1111/j.1365-2966.2005.09321.x}

\bibitem[{{Gelfand} {et~al.}(2009){Gelfand}, {Slane}, \& {Zhang}}]{2009ApJ...703.2051G}
{Gelfand}, J.~D., {Slane}, P.~O., \& {Zhang}, W. 2009, \apj, 703, 2051, \dodoi{10.1088/0004-637X/703/2/2051}

\bibitem[{{Gompertz} {et~al.}(2015){Gompertz}, {van der Horst}, {O'Brien}, {Wynn}, \& {Wiersema}}]{2015MNRAS.448..629G}
{Gompertz}, B.~P., {van der Horst}, A.~J., {O'Brien}, P.~T., {Wynn}, G.~A., \& {Wiersema}, K. 2015, \mnras, 448, 629, \dodoi{10.1093/mnras/stu2752}

\bibitem[{{Gordon} {et~al.}(2023){Gordon}, {Fong}, {Kilpatrick}, {Eftekhari}, {Leja}, {Prochaska}, {Nugent}, {Bhandari}, {Blanchard}, {Caleb}, {Day}, {Deller}, {Dong}, {Glowacki}, {Gourdji}, {Mannings}, {Mahoney}, {Marnoch}, {Miller}, {Paterson}, {Rastinejad}, {Ryder}, {Sadler}, {Scott}, {Sears}, {Shannon}, {Simha}, {Stappers}, \& {Tejos}}]{2023arXiv230205465G}
{Gordon}, A.~C., {Fong}, W.-f., {Kilpatrick}, C.~D., {et~al.} 2023, \apj, 954, 80, \dodoi{10.3847/1538-4357/ace5aa}

\bibitem[{{Green}(2019)}]{2019JApA...40...36G}
{Green}, D.~A. 2019, Journal of Astrophysics and Astronomy, 40, 36, \dodoi{10.1007/s12036-019-9601-6}

\bibitem[{{Greene} {et~al.}(2020){Greene}, {Strader}, \& {Ho}}]{2020ARA&A..58..257G}
{Greene}, J.~E., {Strader}, J., \& {Ho}, L.~C. 2020, \araa, 58, 257, \dodoi{10.1146/annurev-astro-032620-021835}

\bibitem[{{Hallinan} {et~al.}(2019){Hallinan}, {Ravi}, {Weinreb}, {Kocz}, {Huang}, {Woody}, {Lamb}, {D'Addario}, {Catha}, {Law}, {Kulkarni}, {Phinney}, {Eastwood}, {Bouman}, {McLaughlin}, {Ransom}, {Siemens}, {Cordes}, {Lynch}, {Kaplan}, {Brazier}, {Bhatnagar}, {Myers}, {Walter}, \& {Gaensler}}]{2019BAAS...51g.255H}
{Hallinan}, G., {Ravi}, V., {Weinreb}, S., {et~al.} 2019, in Bulletin of the American Astronomical Society, Vol.~51, 255, \dodoi{10.48550/arXiv.1907.07648}

\bibitem[{{Hallinan} {et~al.}(2020){Hallinan}, {Mooley}, {Dong}, {Ravi}, {Law}, {Myers}, {Sivakoff}, {Scheers}, {Schiebelbein}, {Chatterjee}, {Clarke}, {Eisenhardt}, {Stern}, {Jun}, \& {Diaz-Santos}}]{2020ATel14020....1H}
{Hallinan}, G., {Mooley}, K., {Dong}, D., {et~al.} 2020, The Astronomer's Telegram, 14020, 1

\bibitem[{{Ho} {et~al.}(2019){Ho}, {Phinney}, {Ravi}, {Kulkarni}, {Petitpas}, {Emonts}, {Bhalerao}, {Blundell}, {Cenko}, {Dobie}, {Howie}, {Kamraj}, {Kasliwal}, {Murphy}, {Perley}, {Sridharan}, \& {Yoon}}]{2019ApJ...871...73H}
{Ho}, A. Y.~Q., {Phinney}, E.~S., {Ravi}, V., {et~al.} 2019, \apj, 871, 73, \dodoi{10.3847/1538-4357/aaf473}

\bibitem[{{Ho} {et~al.}(2020){Ho}, {Perley}, {Kulkarni}, {Dong}, {De}, {Chandra}, {Andreoni}, {Bellm}, {Burdge}, {Coughlin}, {Dekany}, {Feeney}, {Frederiks}, {Fremling}, {Golkhou}, {Graham}, {Hale}, {Helou}, {Horesh}, {Kasliwal}, {Laher}, {Masci}, {Miller}, {Porter}, {Ridnaia}, {Rusholme}, {Shupe}, {Soumagnac}, \& {Svinkin}}]{2020ApJ...895...49H}
{Ho}, A. Y.~Q., {Perley}, D.~A., {Kulkarni}, S.~R., {et~al.} 2020, \apj, 895, 49, \dodoi{10.3847/1538-4357/ab8bcf}

\bibitem[{{Ho} {et~al.}(2022){Ho}, {Perley}, {Yao}, {Svinkin}, {de Ugarte Postigo}, {Perley}, {Kann}, {Burns}, {Andreoni}, {Bellm}, {Bissaldi}, {Bloom}, {Brink}, {Dekany}, {Drake}, {Ag{\"u}{\'\i} Fern{\'a}ndez}, {Filippenko}, {Frederiks}, {Graham}, {Hristov}, {Kasliwal}, {Kulkarni}, {Kumar}, {Laher}, {Lysenko}, {Mailyan}, {Malacaria}, {Miller}, {Poolakkil}, {Riddle}, {Ridnaia}, {Rusholme}, {Savchenko}, {Sollerman}, {Th{\"o}ne}, {Tsvetkova}, {Ulanov}, \& {von Kienlin}}]{2022ApJ...938...85H}
{Ho}, A. Y.~Q., {Perley}, D.~A., {Yao}, Y., {et~al.} 2022, \apj, 938, 85, \dodoi{10.3847/1538-4357/ac8bd0}

\bibitem[{{Ho} {et~al.}(2023){Ho}, {Perley}, {Gal-Yam}, {Lunnan}, {Sollerman}, {Schulze}, {Das}, {Dobie}, {Yao}, {Fremling}, {Adams}, {Anand}, {Andreoni}, {Bellm}, {Bruch}, {Burdge}, {Castro-Tirado}, {Dahiwale}, {De}, {Dekany}, {Drake}, {Duev}, {Graham}, {Helou}, {Kaplan}, {Karambelkar}, {Kasliwal}, {Kool}, {Kulkarni}, {Mahabal}, {Medford}, {Miller}, {Nordin}, {Ofek}, {Petitpas}, {Riddle}, {Sharma}, {Smith}, {Stewart}, {Taggart}, {Tartaglia}, {Tzanidakis}, \& {Winters}}]{2023ApJ...949..120H}
{Ho}, A. Y.~Q., {Perley}, D.~A., {Gal-Yam}, A., {et~al.} 2023, \apj, 949, 120, \dodoi{10.3847/1538-4357/acc533}

\bibitem[{{Ivezi{\'c}} {et~al.}(2019){Ivezi{\'c}}, {Kahn}, {Tyson}, {Abel}, {Acosta}, {Allsman}, {Alonso}, {AlSayyad}, {Anderson}, {Andrew}, {Angel}, {Angeli}, {Ansari}, {Antilogus}, {Araujo}, {Armstrong}, {Arndt}, {Astier}, {Aubourg}, {Auza}, {Axelrod}, {Bard}, {Barr}, {Barrau}, {Bartlett}, {Bauer}, {Bauman}, {Baumont}, {Bechtol}, {Bechtol}, {Becker}, {Becla}, {Beldica}, {Bellavia}, {Bianco}, {Biswas}, {Blanc}, {Blazek}, {Blandford}, {Bloom}, {Bogart}, {Bond}, {Booth}, {Borgland}, {Borne}, {Bosch}, {Boutigny}, {Brackett}, {Bradshaw}, {Brandt}, {Brown}, {Bullock}, {Burchat}, {Burke}, {Cagnoli}, {Calabrese}, {Callahan}, {Callen}, {Carlin}, {Carlson}, {Chandrasekharan}, {Charles-Emerson}, {Chesley}, {Cheu}, {Chiang}, {Chiang}, {Chirino}, {Chow}, {Ciardi}, {Claver}, {Cohen-Tanugi}, {Cockrum}, {Coles}, {Connolly}, {Cook}, {Cooray}, {Covey}, {Cribbs}, {Cui}, {Cutri}, {Daly}, {Daniel}, {Daruich}, {Daubard}, {Daues}, {Dawson}, {Delgado}, {Dellapenna}, {de Peyster}, {de Val-Borro}, {Digel}, {Doherty}, {Dubois},
  {Dubois-Felsmann}, {Durech}, {Economou}, {Eifler}, {Eracleous}, {Emmons}, {Fausti Neto}, {Ferguson}, {Figueroa}, {Fisher-Levine}, {Focke}, {Foss}, {Frank}, {Freemon}, {Gangler}, {Gawiser}, {Geary}, {Gee}, {Geha}, {Gessner}, {Gibson}, {Gilmore}, {Glanzman}, {Glick}, {Goldina}, {Goldstein}, {Goodenow}, {Graham}, {Gressler}, {Gris}, {Guy}, {Guyonnet}, {Haller}, {Harris}, {Hascall}, {Haupt}, {Hernandez}, {Herrmann}, {Hileman}, {Hoblitt}, {Hodgson}, {Hogan}, {Howard}, {Huang}, {Huffer}, {Ingraham}, {Innes}, {Jacoby}, {Jain}, {Jammes}, {Jee}, {Jenness}, {Jernigan}, {Jevremovi{\'c}}, {Johns}, {Johnson}, {Johnson}, {Jones}, {Juramy-Gilles}, {Juri{\'c}}, {Kalirai}, {Kallivayalil}, {Kalmbach}, {Kantor}, {Karst}, {Kasliwal}, {Kelly}, {Kessler}, {Kinnison}, {Kirkby}, {Knox}, {Kotov}, {Krabbendam}, {Krughoff}, {Kub{\'a}nek}, {Kuczewski}, {Kulkarni}, {Ku}, {Kurita}, {Lage}, {Lambert}, {Lange}, {Langton}, {Le Guillou}, {Levine}, {Liang}, {Lim}, {Lintott}, {Long}, {Lopez}, {Lotz}, {Lupton}, {Lust}, {MacArthur}, {Mahabal},
  {Mandelbaum}, {Markiewicz}, {Marsh}, {Marshall}, {Marshall}, {May}, {McKercher}, {McQueen}, {Meyers}, {Migliore}, {Miller}, {Mills}, {Miraval}, {Moeyens}, {Moolekamp}, {Monet}, {Moniez}, {Monkewitz}, {Montgomery}, {Morrison}, {Mueller}, {Muller}, {Mu{\~n}oz Arancibia}, {Neill}, {Newbry}, {Nief}, {Nomerotski}, {Nordby}, {O'Connor}, {Oliver}, {Olivier}, {Olsen}, {O'Mullane}, {Ortiz}, {Osier}, {Owen}, {Pain}, {Palecek}, {Parejko}, {Parsons}, {Pease}, {Peterson}, {Peterson}, {Petravick}, {Libby Petrick}, {Petry}, {Pierfederici}, {Pietrowicz}, {Pike}, {Pinto}, {Plante}, {Plate}, {Plutchak}, {Price}, {Prouza}, {Radeka}, {Rajagopal}, {Rasmussen}, {Regnault}, {Reil}, {Reiss}, {Reuter}, {Ridgway}, {Riot}, {Ritz}, {Robinson}, {Roby}, {Roodman}, {Rosing}, {Roucelle}, {Rumore}, {Russo}, {Saha}, {Sassolas}, {Schalk}, {Schellart}, {Schindler}, {Schmidt}, {Schneider}, {Schneider}, {Schoening}, {Schumacher}, {Schwamb}, {Sebag}, {Selvy}, {Sembroski}, {Seppala}, {Serio}, {Serrano}, {Shaw}, {Shipsey}, {Sick}, {Silvestri},
  {Slater}, {Smith}, {Smith}, {Sobhani}, {Soldahl}, {Storrie-Lombardi}, {Stover}, {Strauss}, {Street}, {Stubbs}, {Sullivan}, {Sweeney}, {Swinbank}, {Szalay}, {Takacs}, {Tether}, {Thaler}, {Thayer}, {Thomas}, {Thornton}, {Thukral}, {Tice}, {Trilling}, {Turri}, {Van Berg}, {Vanden Berk}, {Vetter}, {Virieux}, {Vucina}, {Wahl}, {Walkowicz}, {Walsh}, {Walter}, {Wang}, {Wang}, {Warner}, {Wiecha}, {Willman}, {Winters}, {Wittman}, {Wolff}, {Wood-Vasey}, {Wu}, {Xin}, {Yoachim}, \& {Zhan}}]{2019ApJ...873..111I}
{Ivezi{\'c}}, {\v{Z}}., {Kahn}, S.~M., {Tyson}, J.~A., {et~al.} 2019, \apj, 873, 111, \dodoi{10.3847/1538-4357/ab042c}

\bibitem[{{Johnson} {et~al.}(2021){Johnson}, {Leja}, {Conroy}, \& {Speagle}}]{2021ApJS..254...22J}
{Johnson}, B.~D., {Leja}, J., {Conroy}, C., \& {Speagle}, J.~S. 2021, \apjs, 254, 22, \dodoi{10.3847/1538-4365/abef67}

\bibitem[{{Kaspi} \& {Beloborodov}(2017)}]{2017ARA&A..55..261K}
{Kaspi}, V.~M., \& {Beloborodov}, A.~M. 2017, \araa, 55, 261, \dodoi{10.1146/annurev-astro-081915-023329}

\bibitem[{{Kauffmann} {et~al.}(2003){Kauffmann}, {Heckman}, {Tremonti}, {Brinchmann}, {Charlot}, {White}, {Ridgway}, {Brinkmann}, {Fukugita}, {Hall}, {Ivezi{\'c}}, {Richards}, \& {Schneider}}]{2003MNRAS.346.1055K}
{Kauffmann}, G., {Heckman}, T.~M., {Tremonti}, C., {et~al.} 2003, \mnras, 346, 1055, \dodoi{10.1111/j.1365-2966.2003.07154.x}

\bibitem[{{Kelly} \& {Kirshner}(2012)}]{2012ApJ...759..107K}
{Kelly}, P.~L., \& {Kirshner}, R.~P. 2012, \apj, 759, 107, \dodoi{10.1088/0004-637X/759/2/107}

\bibitem[{{Kennicutt} {et~al.}(1994){Kennicutt}, {Tamblyn}, \& {Congdon}}]{1994ApJ...435...22K}
{Kennicutt}, Robert~C., J., {Tamblyn}, P., \& {Congdon}, C.~E. 1994, \apj, 435, 22, \dodoi{10.1086/174790}

\bibitem[{{Kewley} {et~al.}(2001){Kewley}, {Dopita}, {Sutherland}, {Heisler}, \& {Trevena}}]{2001ApJ...556..121K}
{Kewley}, L.~J., {Dopita}, M.~A., {Sutherland}, R.~S., {Heisler}, C.~A., \& {Trevena}, J. 2001, \apj, 556, 121, \dodoi{10.1086/321545}

\bibitem[{{Kewley} {et~al.}(2006){Kewley}, {Groves}, {Kauffmann}, \& {Heckman}}]{2006MNRAS.372..961K}
{Kewley}, L.~J., {Groves}, B., {Kauffmann}, G., \& {Heckman}, T. 2006, \mnras, 372, 961, \dodoi{10.1111/j.1365-2966.2006.10859.x}

\bibitem[{{Kovlakas} {et~al.}(2020){Kovlakas}, {Zezas}, {Andrews}, {Basu-Zych}, {Fragos}, {Hornschemeier}, {Lehmer}, \& {Ptak}}]{2020MNRAS.498.4790K}
{Kovlakas}, K., {Zezas}, A., {Andrews}, J.~J., {et~al.} 2020, \mnras, 498, 4790, \dodoi{10.1093/mnras/staa2481}

\bibitem[{{Kroupa}(2001)}]{2001MNRAS.322..231K}
{Kroupa}, P. 2001, \mnras, 322, 231, \dodoi{10.1046/j.1365-8711.2001.04022.x}

\bibitem[{{Lacy} {et~al.}(2020){Lacy}, {Baum}, {Chandler}, {Chatterjee}, {Clarke}, {Deustua}, {English}, {Farnes}, {Gaensler}, {Gugliucci}, {Hallinan}, {Kent}, {Kimball}, {Law}, {Lazio}, {Marvil}, {Mao}, {Medlin}, {Mooley}, {Murphy}, {Myers}, {Osten}, {Richards}, {Rosolowsky}, {Rudnick}, {Schinzel}, {Sivakoff}, {Sjouwerman}, {Taylor}, {White}, {Wrobel}, {Andernach}, {Beasley}, {Berger}, {Bhatnager}, {Birkinshaw}, {Bower}, {Brandt}, {Brown}, {Burke-Spolaor}, {Butler}, {Comerford}, {Demorest}, {Fu}, {Giacintucci}, {Golap}, {G{\"u}th}, {Hales}, {Hiriart}, {Hodge}, {Horesh}, {Ivezi{\'c}}, {Jarvis}, {Kamble}, {Kassim}, {Liu}, {Loinard}, {Lyons}, {Masters}, {Mezcua}, {Moellenbrock}, {Mroczkowski}, {Nyland}, {O'Dea}, {O'Sullivan}, {Peters}, {Radford}, {Rao}, {Robnett}, {Salcido}, {Shen}, {Sobotka}, {Witz}, {Vaccari}, {van Weeren}, {Vargas}, {Williams}, \& {Yoon}}]{2020PASP..132c5001L}
{Lacy}, M., {Baum}, S.~A., {Chandler}, C.~J., {et~al.} 2020, \pasp, 132, 035001, \dodoi{10.1088/1538-3873/ab63eb}

\bibitem[{{Laigle} {et~al.}(2016){Laigle}, {McCracken}, {Ilbert}, {Hsieh}, {Davidzon}, {Capak}, {Hasinger}, {Silverman}, {Pichon}, {Coupon}, {Aussel}, {Le Borgne}, {Caputi}, {Cassata}, {Chang}, {Civano}, {Dunlop}, {Fynbo}, {Kartaltepe}, {Koekemoer}, {Le F{\`e}vre}, {Le Floc'h}, {Leauthaud}, {Lilly}, {Lin}, {Marchesi}, {Milvang-Jensen}, {Salvato}, {Sanders}, {Scoville}, {Smolcic}, {Stockmann}, {Taniguchi}, {Tasca}, {Toft}, {Vaccari}, \& {Zabl}}]{2016ApJS..224...24L}
{Laigle}, C., {McCracken}, H.~J., {Ilbert}, O., {et~al.} 2016, \apjs, 224, 24, \dodoi{10.3847/0067-0049/224/2/24}

\bibitem[{{Lampeitl} {et~al.}(2010){Lampeitl}, {Smith}, {Nichol}, {Bassett}, {Cinabro}, {Dilday}, {Foley}, {Frieman}, {Garnavich}, {Goobar}, {Im}, {Jha}, {Marriner}, {Miquel}, {Nordin}, {{\"O}stman}, {Riess}, {Sako}, {Schneider}, {Sollerman}, \& {Stritzinger}}]{2010ApJ...722..566L}
{Lampeitl}, H., {Smith}, M., {Nichol}, R.~C., {et~al.} 2010, \apj, 722, 566, \dodoi{10.1088/0004-637X/722/1/56610.48550/arXiv.1005.4687}

\bibitem[{{Lattimer} \& {Prakash}(2001)}]{2001ApJ...550..426L}
{Lattimer}, J.~M., \& {Prakash}, M. 2001, \apj, 550, 426, \dodoi{10.1086/319702}

\bibitem[{{Law} {et~al.}(2018){Law}, {Gaensler}, {Metzger}, {Ofek}, \& {Sironi}}]{2018ApJ...866L..22L}
{Law}, C.~J., {Gaensler}, B.~M., {Metzger}, B.~D., {Ofek}, E.~O., \& {Sironi}, L. 2018, \apjl, 866, L22, \dodoi{10.3847/2041-8213/aae5f3}

\bibitem[{{Leja} {et~al.}(2020){Leja}, {Speagle}, {Johnson}, {Conroy}, {van Dokkum}, \& {Franx}}]{2020ApJ...893..111L}
{Leja}, J., {Speagle}, J.~S., {Johnson}, B.~D., {et~al.} 2020, \apj, 893, 111, \dodoi{10.3847/1538-4357/ab7e27}

\bibitem[{{Leja} {et~al.}(2022){Leja}, {Speagle}, {Ting}, {Johnson}, {Conroy}, {Whitaker}, {Nelson}, {van Dokkum}, \& {Franx}}]{2022ApJ...936..165L}
{Leja}, J., {Speagle}, J.~S., {Ting}, Y.-S., {et~al.} 2022, \apj, 936, 165, \dodoi{10.3847/1538-4357/ac887d}

\bibitem[{{Li} {et~al.}(2011){Li}, {Chornock}, {Leaman}, {Filippenko}, {Poznanski}, {Wang}, {Ganeshalingam}, \& {Mannucci}}]{2011MNRAS.412.1473L}
{Li}, W., {Chornock}, R., {Leaman}, J., {et~al.} 2011, \mnras, 412, 1473, \dodoi{10.1111/j.1365-2966.2011.18162.x}

\bibitem[{{Lien} {et~al.}(2016){Lien}, {Sakamoto}, {Barthelmy}, {Baumgartner}, {Cannizzo}, {Chen}, {Collins}, {Cummings}, {Gehrels}, {Krimm}, {Markwardt}, {Palmer}, {Stamatikos}, {Troja}, \& {Ukwatta}}]{2016ApJ...829....7L}
{Lien}, A., {Sakamoto}, T., {Barthelmy}, S.~D., {et~al.} 2016, \apj, 829, 7, \dodoi{10.3847/0004-637X/829/1/7}

\bibitem[{{Lunnan} {et~al.}(2015){Lunnan}, {Chornock}, {Berger}, {Rest}, {Fong}, {Scolnic}, {Jones}, {Soderberg}, {Challis}, {Drout}, {Foley}, {Huber}, {Kirshner}, {Leibler}, {Marion}, {McCrum}, {Milisavljevic}, {Narayan}, {Sanders}, {Smartt}, {Smith}, {Tonry}, {Burgett}, {Chambers}, {Flewelling}, {Kudritzki}, {Wainscoat}, \& {Waters}}]{2015ApJ...804...90L}
{Lunnan}, R., {Chornock}, R., {Berger}, E., {et~al.} 2015, \apj, 804, 90, \dodoi{10.1088/0004-637X/804/2/90}

\bibitem[{{Mainzer} {et~al.}(2011){Mainzer}, {Bauer}, {Grav}, {Masiero}, {Cutri}, {Dailey}, {Eisenhardt}, {McMillan}, {Wright}, {Walker}, {Jedicke}, {Spahr}, {Tholen}, {Alles}, {Beck}, {Brandenburg}, {Conrow}, {Evans}, {Fowler}, {Jarrett}, {Marsh}, {Masci}, {McCallon}, {Wheelock}, {Wittman}, {Wyatt}, {DeBaun}, {Elliott}, {Elsbury}, {Gautier}, {Gomillion}, {Leisawitz}, {Maleszewski}, {Micheli}, \& {Wilkins}}]{2011ApJ...731...53M}
{Mainzer}, A., {Bauer}, J., {Grav}, T., {et~al.} 2011, \apj, 731, 53, \dodoi{10.1088/0004-637X/731/1/53}

\bibitem[{{Manchester} {et~al.}(2005){Manchester}, {Hobbs}, {Teoh}, \& {Hobbs}}]{2005AJ....129.1993M}
{Manchester}, R.~N., {Hobbs}, G.~B., {Teoh}, A., \& {Hobbs}, M. 2005, \aj, 129, 1993, \dodoi{10.1086/428488}

\bibitem[{{Mannings} {et~al.}(2021){Mannings}, {Fong}, {Simha}, {Prochaska}, {Rafelski}, {Kilpatrick}, {Tejos}, {Heintz}, {Bannister}, {Bhandari}, {Day}, {Deller}, {Ryder}, {Shannon}, \& {Tendulkar}}]{2021ApJ...917...75M}
{Mannings}, A.~G., {Fong}, W.-f., {Simha}, S., {et~al.} 2021, \apj, 917, 75, \dodoi{10.3847/1538-4357/abff56}

\bibitem[{{Masci} {et~al.}(2023){Masci}, {Laher}, {Rusholme}, {Shupe}, {Paladini}, {Groom}, {Wold}, {Miller}, \& {Drake}}]{2023arXiv230516279M}
{Masci}, F.~J., {Laher}, R.~R., {Rusholme}, B., {et~al.} 2023, arXiv e-prints, arXiv:2305.16279, \dodoi{10.48550/arXiv.2305.16279}

\bibitem[{{M{\'e}sz{\'a}ros} \& {Rees}(1997)}]{1997ApJ...476..232M}
{M{\'e}sz{\'a}ros}, P., \& {Rees}, M.~J. 1997, \apj, 476, 232, \dodoi{10.1086/303625}

\bibitem[{{Metzger} {et~al.}(2015{\natexlab{a}}){Metzger}, {Margalit}, {Kasen}, \& {Quataert}}]{2015MNRAS.454.3311M}
{Metzger}, B.~D., {Margalit}, B., {Kasen}, D., \& {Quataert}, E. 2015{\natexlab{a}}, \mnras, 454, 3311, \dodoi{10.1093/mnras/stv2224}

\bibitem[{{Metzger} {et~al.}(2009){Metzger}, {Piro}, \& {Quataert}}]{2009MNRAS.396.1659M}
{Metzger}, B.~D., {Piro}, A.~L., \& {Quataert}, E. 2009, \mnras, 396, 1659, \dodoi{10.1111/j.1365-2966.2009.14909.x}

\bibitem[{{Metzger} {et~al.}(2015{\natexlab{b}}){Metzger}, {Williams}, \& {Berger}}]{2015ApJ...806..224M}
{Metzger}, B.~D., {Williams}, P.~K.~G., \& {Berger}, E. 2015{\natexlab{b}}, \apj, 806, 224, \dodoi{10.1088/0004-637X/806/2/224}

\bibitem[{{Moriya}(2016)}]{2016ApJ...830L..38M}
{Moriya}, T.~J. 2016, \apjl, 830, L38, \dodoi{10.3847/2041-8205/830/2/L38}

\bibitem[{{Mosby} {et~al.}(2020){Mosby}, {Rauscher}, {Bennett}, {Cheng}, {Cheung}, {Cillis}, {Content}, {Cottingham}, {Foltz}, {Gygax}, {Hill}, {Kruk}, {Mah}, {Meier}, {Merchant}, {Miko}, {Piquette}, {Waczynski}, \& {Wen}}]{2020JATIS...6d6001M}
{Mosby}, G., {Rauscher}, B.~J., {Bennett}, C., {et~al.} 2020, Journal of Astronomical Telescopes, Instruments, and Systems, 6, 046001, \dodoi{10.1117/1.JATIS.6.4.046001}

\bibitem[{{Narayan} {et~al.}(2022){Narayan}, {Chael}, {Chatterjee}, {Ricarte}, \& {Curd}}]{2022MNRAS.511.3795N}
{Narayan}, R., {Chael}, A., {Chatterjee}, K., {Ricarte}, A., \& {Curd}, B. 2022, \mnras, 511, 3795, \dodoi{10.1093/mnras/stac285}

\bibitem[{{Nenkova} {et~al.}(2008){Nenkova}, {Sirocky}, {Nikutta}, {Ivezi{\'c}}, \& {Elitzur}}]{2008ApJ...685..160N}
{Nenkova}, M., {Sirocky}, M.~M., {Nikutta}, R., {Ivezi{\'c}}, {\v{Z}}., \& {Elitzur}, M. 2008, \apj, 685, 160, \dodoi{10.1086/590483}

\bibitem[{{Nicholl} {et~al.}(2017){Nicholl}, {Guillochon}, \& {Berger}}]{2017ApJ...850...55N}
{Nicholl}, M., {Guillochon}, J., \& {Berger}, E. 2017, \apj, 850, 55, \dodoi{10.3847/1538-4357/aa9334}

\bibitem[{{Nugent} {et~al.}(2022){Nugent}, {Fong}, {Dong}, {Leja}, {Berger}, {Zevin}, {Chornock}, {Cobb}, {Kelley}, {Kilpatrick}, {Levan}, {Margutti}, {Paterson}, {Perley}, {Escorial}, {Smith}, \& {Tanvir}}]{2022ApJ...940...57N}
{Nugent}, A.~E., {Fong}, W.-F., {Dong}, Y., {et~al.} 2022, \apj, 940, 57, \dodoi{10.3847/1538-4357/ac91d1}

\bibitem[{{Oke} {et~al.}(1995){Oke}, {Cohen}, {Carr}, {Cromer}, {Dingizian}, {Harris}, {Labrecque}, {Lucinio}, {Schaal}, {Epps}, \& {Miller}}]{1995PASP..107..375O}
{Oke}, J.~B., {Cohen}, J.~G., {Carr}, M., {et~al.} 1995, \pasp, 107, 375, \dodoi{10.1086/133562}

\bibitem[{{Olausen} \& {Kaspi}(2014)}]{2014ApJS..212....6O}
{Olausen}, S.~A., \& {Kaspi}, V.~M. 2014, \apjs, 212, 6, \dodoi{10.1088/0067-0049/212/1/6}

\bibitem[{{OMullane} {et~al.}(2005){OMullane}, {Li}, {Nieto-Santisteban}, {Szalay}, {Thakar}, \& {Gray}}]{2005cs........2072O}
{OMullane}, W., {Li}, N., {Nieto-Santisteban}, M., {et~al.} 2005, arXiv e-prints, cs/0502072.
\newblock \doarXiv{cs/0502072}

\bibitem[{{Pacholczyk}(1970)}]{1970ranp.book.....P}
{Pacholczyk}, A.~G. 1970, {Radio astrophysics. Nonthermal processes in galactic and extragalactic sources}

\bibitem[{{Perley}(2019)}]{2019PASP..131h4503P}
{Perley}, D.~A. 2019, \pasp, 131, 084503, \dodoi{10.1088/1538-3873/ab215d}

\bibitem[{{Perley} {et~al.}(2019){Perley}, {Mazzali}, {Yan}, {Cenko}, {Gezari}, {Taggart}, {Blagorodnova}, {Fremling}, {Mockler}, {Singh}, {Tominaga}, {Tanaka}, {Watson}, {Ahumada}, {Anupama}, {Ashall}, {Becerra}, {Bersier}, {Bhalerao}, {Bloom}, {Butler}, {Copperwheat}, {Coughlin}, {De}, {Drake}, {Duev}, {Frederick}, {Gonz{\'a}lez}, {Goobar}, {Heida}, {Ho}, {Horst}, {Hung}, {Itoh}, {Jencson}, {Kasliwal}, {Kawai}, {Khanam}, {Kulkarni}, {Kumar}, {Kumar}, {Kutyrev}, {Lee}, {Maeda}, {Mahabal}, {Murata}, {Neill}, {Ngeow}, {Penprase}, {Pian}, {Quimby}, {Ramirez-Ruiz}, {Richer}, {Rom{\'a}n-Z{\'u}{\~n}iga}, {Sahu}, {Srivastav}, {Socia}, {Sollerman}, {Tachibana}, {Taddia}, {Tinyanont}, {Troja}, {Ward}, {Wee}, \& {Yu}}]{2019MNRAS.484.1031P}
{Perley}, D.~A., {Mazzali}, P.~A., {Yan}, L., {et~al.} 2019, \mnras, 484, 1031, \dodoi{10.1093/mnras/sty3420}

\bibitem[{{Perley} {et~al.}(2021){Perley}, {Ho}, {Yao}, {Fremling}, {Anderson}, {Schulze}, {Kumar}, {Anupama}, {Barway}, {Bellm}, {Bhalerao}, {Chen}, {Duev}, {Galbany}, {Graham}, {Gromadzki}, {Guti{\'e}rrez}, {Ihanec}, {Inserra}, {Kasliwal}, {Kool}, {Kulkarni}, {Laher}, {Masci}, {Neill}, {Nicholl}, {Pursiainen}, {van Roestel}, {Sharma}, {Sollerman}, {Walters}, \& {Wiseman}}]{2021MNRAS.508.5138P}
{Perley}, D.~A., {Ho}, A. Y.~Q., {Yao}, Y., {et~al.} 2021, \mnras, 508, 5138, \dodoi{10.1093/mnras/stab2785}

\bibitem[{{Pescalli} {et~al.}(2016){Pescalli}, {Ghirlanda}, {Salvaterra}, {Ghisellini}, {Vergani}, {Nappo}, {Salafia}, {Melandri}, {Covino}, \& {G{\"o}tz}}]{2016A&A...587A..40P}
{Pescalli}, A., {Ghirlanda}, G., {Salvaterra}, R., {et~al.} 2016, \aap, 587, A40, \dodoi{10.1051/0004-6361/201526760}

\bibitem[{{Piro} \& {Kulkarni}(2013)}]{2013ApJ...762L..17P}
{Piro}, A.~L., \& {Kulkarni}, S.~R. 2013, \apjl, 762, L17, \dodoi{10.1088/2041-8205/762/2/L17}

\bibitem[{{Planck Collaboration} {et~al.}(2020){Planck Collaboration}, {Aghanim}, {Akrami}, {Ashdown}, {Aumont}, {Baccigalupi}, {Ballardini}, {Banday}, {Barreiro}, {Bartolo}, {Basak}, {Battye}, {Benabed}, {Bernard}, {Bersanelli}, {Bielewicz}, {Bock}, {Bond}, {Borrill}, {Bouchet}, {Boulanger}, {Bucher}, {Burigana}, {Butler}, {Calabrese}, {Cardoso}, {Carron}, {Challinor}, {Chiang}, {Chluba}, {Colombo}, {Combet}, {Contreras}, {Crill}, {Cuttaia}, {de Bernardis}, {de Zotti}, {Delabrouille}, {Delouis}, {Di Valentino}, {Diego}, {Dor{\'e}}, {Douspis}, {Ducout}, {Dupac}, {Dusini}, {Efstathiou}, {Elsner}, {En{\ss}lin}, {Eriksen}, {Fantaye}, {Farhang}, {Fergusson}, {Fernandez-Cobos}, {Finelli}, {Forastieri}, {Frailis}, {Fraisse}, {Franceschi}, {Frolov}, {Galeotta}, {Galli}, {Ganga}, {G{\'e}nova-Santos}, {Gerbino}, {Ghosh}, {Gonz{\'a}lez-Nuevo}, {G{\'o}rski}, {Gratton}, {Gruppuso}, {Gudmundsson}, {Hamann}, {Handley}, {Hansen}, {Herranz}, {Hildebrandt}, {Hivon}, {Huang}, {Jaffe}, {Jones}, {Karakci}, {Keih{\"a}nen},
  {Keskitalo}, {Kiiveri}, {Kim}, {Kisner}, {Knox}, {Krachmalnicoff}, {Kunz}, {Kurki-Suonio}, {Lagache}, {Lamarre}, {Lasenby}, {Lattanzi}, {Lawrence}, {Le Jeune}, {Lemos}, {Lesgourgues}, {Levrier}, {Lewis}, {Liguori}, {Lilje}, {Lilley}, {Lindholm}, {L{\'o}pez-Caniego}, {Lubin}, {Ma}, {Mac{\'\i}as-P{\'e}rez}, {Maggio}, {Maino}, {Mandolesi}, {Mangilli}, {Marcos-Caballero}, {Maris}, {Martin}, {Martinelli}, {Mart{\'\i}nez-Gonz{\'a}lez}, {Matarrese}, {Mauri}, {McEwen}, {Meinhold}, {Melchiorri}, {Mennella}, {Migliaccio}, {Millea}, {Mitra}, {Miville-Desch{\^e}nes}, {Molinari}, {Montier}, {Morgante}, {Moss}, {Natoli}, {N{\o}rgaard-Nielsen}, {Pagano}, {Paoletti}, {Partridge}, {Patanchon}, {Peiris}, {Perrotta}, {Pettorino}, {Piacentini}, {Polastri}, {Polenta}, {Puget}, {Rachen}, {Reinecke}, {Remazeilles}, {Renzi}, {Rocha}, {Rosset}, {Roudier}, {Rubi{\~n}o-Mart{\'\i}n}, {Ruiz-Granados}, {Salvati}, {Sandri}, {Savelainen}, {Scott}, {Shellard}, {Sirignano}, {Sirri}, {Spencer}, {Sunyaev}, {Suur-Uski}, {Tauber}, {Tavagnacco},
  {Tenti}, {Toffolatti}, {Tomasi}, {Trombetti}, {Valenziano}, {Valiviita}, {Van Tent}, {Vibert}, {Vielva}, {Villa}, {Vittorio}, {Wandelt}, {Wehus}, {White}, {White}, {Zacchei}, \& {Zonca}}]{2020A&A...641A...6P}
{Planck Collaboration}, {Aghanim}, N., {Akrami}, Y., {et~al.} 2020, \aap, 641, A6, \dodoi{10.1051/0004-6361/201833910}

\bibitem[{{Pursiainen} {et~al.}(2018){Pursiainen}, {Childress}, {Smith}, {Prajs}, {Sullivan}, {Davis}, {Foley}, {Asorey}, {Calcino}, {Carollo}, {Curtin}, {D'Andrea}, {Glazebrook}, {Gutierrez}, {Hinton}, {Hoormann}, {Inserra}, {Kessler}, {King}, {Kuehn}, {Lewis}, {Lidman}, {Macaulay}, {M{\"o}ller}, {Nichol}, {Sako}, {Sommer}, {Swann}, {Tucker}, {Uddin}, {Wiseman}, {Zhang}, {Abbott}, {Abdalla}, {Allam}, {Annis}, {Avila}, {Brooks}, {Buckley-Geer}, {Burke}, {Carnero Rosell}, {Carrasco Kind}, {Carretero}, {Castander}, {Cunha}, {Davis}, {De Vicente}, {Diehl}, {Doel}, {Eifler}, {Flaugher}, {Fosalba}, {Frieman}, {Garc{\'\i}a-Bellido}, {Gruen}, {Gruendl}, {Gutierrez}, {Hartley}, {Hollowood}, {Honscheid}, {James}, {Jeltema}, {Kuropatkin}, {Li}, {Lima}, {Maia}, {Martini}, {Menanteau}, {Ogando}, {Plazas}, {Roodman}, {Sanchez}, {Scarpine}, {Schindler}, {Smith}, {Soares-Santos}, {Sobreira}, {Suchyta}, {Swanson}, {Tarle}, {Tucker}, {Walker}, \& {DES Collaboration}}]{2018MNRAS.481..894P}
{Pursiainen}, M., {Childress}, M., {Smith}, M., {et~al.} 2018, \mnras, 481, 894, \dodoi{10.1093/mnras/sty2309}

\bibitem[{{Rau} {et~al.}(2005){Rau}, {Kienlin}, {Hurley}, \& {Lichti}}]{2005A&A...438.1175R}
{Rau}, A., {Kienlin}, A.~V., {Hurley}, K., \& {Lichti}, G.~G. 2005, \aap, 438, 1175, \dodoi{10.1051/0004-6361:20053159}

\bibitem[{{Rea} {et~al.}(2010){Rea}, {Esposito}, {Turolla}, {Israel}, {Zane}, {Stella}, {Mereghetti}, {Tiengo}, {G{\"o}tz}, {G{\"o}{\u{g}}{\"u}{\c{s}}}, \& {Kouveliotou}}]{2010Sci...330..944R}
{Rea}, N., {Esposito}, P., {Turolla}, R., {et~al.} 2010, Science, 330, 944, \dodoi{10.1126/science.1196088}

\bibitem[{{Reynolds} \& {Chevalier}(1984)}]{1984ApJ...278..630R}
{Reynolds}, S.~P., \& {Chevalier}, R.~A. 1984, \apj, 278, 630, \dodoi{10.1086/161831}

\bibitem[{{Rowlinson} {et~al.}(2013){Rowlinson}, {O'Brien}, {Metzger}, {Tanvir}, \& {Levan}}]{2013MNRAS.430.1061R}
{Rowlinson}, A., {O'Brien}, P.~T., {Metzger}, B.~D., {Tanvir}, N.~R., \& {Levan}, A.~J. 2013, \mnras, 430, 1061, \dodoi{10.1093/mnras/sts683}

\bibitem[{{Rybicki} \& {Lightman}(1979)}]{1979rpa..book.....R}
{Rybicki}, G.~B., \& {Lightman}, A.~P. 1979, {Radiative processes in astrophysics}

\bibitem[{{S{\'a}nchez-Bl{\'a}zquez} {et~al.}(2006){S{\'a}nchez-Bl{\'a}zquez}, {Peletier}, {Jim{\'e}nez-Vicente}, {Cardiel}, {Cenarro}, {Falc{\'o}n-Barroso}, {Gorgas}, {Selam}, \& {Vazdekis}}]{2006MNRAS.371..703S}
{S{\'a}nchez-Bl{\'a}zquez}, P., {Peletier}, R.~F., {Jim{\'e}nez-Vicente}, J., {et~al.} 2006, \mnras, 371, 703, \dodoi{10.1111/j.1365-2966.2006.10699.x}

\bibitem[{{Schady} {et~al.}(2007){Schady}, {Mason}, {Page}, {de Pasquale}, {Morris}, {Romano}, {Roming}, {Immler}, \& {vanden Berk}}]{2007MNRAS.377..273S}
{Schady}, P., {Mason}, K.~O., {Page}, M.~J., {et~al.} 2007, \mnras, 377, 273, \dodoi{10.1111/j.1365-2966.2007.11592.x}

\bibitem[{{Schulze} {et~al.}(2021){Schulze}, {Yaron}, {Sollerman}, {Leloudas}, {Gal}, {Wright}, {Lunnan}, {Gal-Yam}, {Ofek}, {Perley}, {Filippenko}, {Kasliwal}, {Kulkarni}, {Neill}, {Nugent}, {Quimby}, {Sullivan}, {Strotjohann}, {Arcavi}, {Ben-Ami}, {Bianco}, {Bloom}, {De}, {Fraser}, {Fremling}, {Horesh}, {Johansson}, {Kelly}, {Kne{\v{z}}evi{\'c}}, {Kne{\v{z}}evi{\'c}}, {Maguire}, {Nyholm}, {Papadogiannakis}, {Petrushevska}, {Rubin}, {Yan}, {Yang}, {Adams}, {Bufano}, {Clubb}, {Foley}, {Green}, {Harmanen}, {Ho}, {Hook}, {Hosseinzadeh}, {Howell}, {Kong}, {Kotak}, {Matheson}, {McCully}, {Milisavljevic}, {Pan}, {Poznanski}, {Shivvers}, {van Velzen}, \& {Verbeek}}]{2021ApJS..255...29S}
{Schulze}, S., {Yaron}, O., {Sollerman}, J., {et~al.} 2021, \apjs, 255, 29, \dodoi{10.3847/1538-4365/abff5e}

\bibitem[{{Serino} {et~al.}(2014){Serino}, {Sakamoto}, {Kawai}, {Yoshida}, {Ohno}, {Ogawa}, {Nishimura}, {Fukushima}, {Higa}, {Ishikawa}, {Ishikawa}, {Kawamuro}, {Kimura}, {Matsuoka}, {Mihara}, {Morii}, {Nakagawa}, {Nakahira}, {Nakajima}, {Nakano}, {Negoro}, {Onodera}, {Sasaki}, {Shidatsu}, {Sugimoto}, {Sugizaki}, {Suwa}, {Suzuki}, {Tachibana}, {Takagi}, {Toizumi}, {Tomida}, {Tsuboi}, {Tsunemi}, {Ueda}, {Ueno}, {Usui}, {Yamada}, {Yamamoto}, {Yamaoka}, {Yamauchi}, {Yoshidome}, \& {Yoshii}}]{2014PASJ...66...87S}
{Serino}, M., {Sakamoto}, T., {Kawai}, N., {et~al.} 2014, \pasj, 66, 87, \dodoi{10.1093/pasj/psu063}

\bibitem[{{Sharma} {et~al.}(2024){Sharma}, {Ravi}, {Connor}, {Law}, {Ocker}, {Sherman}, {Kosogorov}, {Faber}, {Hallinan}, {Harnach}, {Hellbourg}, {Hobbs}, {Hodge}, {Hodges}, {Lamb}, {Rasmussen}, {Somalwar}, {Weinreb}, {Woody}, {Leja}, {Anand}, {Das}, {Qin}, {Rose}, {Dong}, {Miller}, \& {Yao}}]{2024Natur.635...61S}
{Sharma}, K., {Ravi}, V., {Connor}, L., {et~al.} 2024, \nat, 635, 61, \dodoi{10.1038/s41586-024-08074-9}

\bibitem[{{Shingles} {et~al.}(2021){Shingles}, {Smith}, {Young}, {Smartt}, {Tonry}, {Denneau}, {Heinze}, {Weiland}, {Flewelling}, {Stalder}, {Clocchiatti}, {F{\"o}rster}, {Pignata}, {Rest}, {Anderson}, {Stubbs}, \& {Erasmus}}]{2021TNSAN...7....1S}
{Shingles}, L., {Smith}, K.~W., {Young}, D.~R., {et~al.} 2021, Transient Name Server AstroNote, 7, 1

\bibitem[{{Skelton} {et~al.}(2014){Skelton}, {Whitaker}, {Momcheva}, {Brammer}, {van Dokkum}, {Labb{\'e}}, {Franx}, {van der Wel}, {Bezanson}, {Da Cunha}, {Fumagalli}, {F{\"o}rster Schreiber}, {Kriek}, {Leja}, {Lundgren}, {Magee}, {Marchesini}, {Maseda}, {Nelson}, {Oesch}, {Pacifici}, {Patel}, {Price}, {Rix}, {Tal}, {Wake}, \& {Wuyts}}]{2014ApJS..214...24S}
{Skelton}, R.~E., {Whitaker}, K.~E., {Momcheva}, I.~G., {et~al.} 2014, \apjs, 214, 24, \dodoi{10.1088/0067-0049/214/2/24}

\bibitem[{{Skrutskie} {et~al.}(2006){Skrutskie}, {Cutri}, {Stiening}, {Weinberg}, {Schneider}, {Carpenter}, {Beichman}, {Capps}, {Chester}, {Elias}, {Huchra}, {Liebert}, {Lonsdale}, {Monet}, {Price}, {Seitzer}, {Jarrett}, {Kirkpatrick}, {Gizis}, {Howard}, {Evans}, {Fowler}, {Fullmer}, {Hurt}, {Light}, {Kopan}, {Marsh}, {McCallon}, {Tam}, {Van Dyk}, \& {Wheelock}}]{2006AJ....131.1163S}
{Skrutskie}, M.~F., {Cutri}, R.~M., {Stiening}, R., {et~al.} 2006, \aj, 131, 1163, \dodoi{10.1086/498708}

\bibitem[{{Soderberg} {et~al.}(2006{\natexlab{a}}){Soderberg}, {Chevalier}, {Kulkarni}, \& {Frail}}]{2006ApJ...651.1005S}
{Soderberg}, A.~M., {Chevalier}, R.~A., {Kulkarni}, S.~R., \& {Frail}, D.~A. 2006{\natexlab{a}}, \apj, 651, 1005, \dodoi{10.1086/507571}

\bibitem[{{Soderberg} {et~al.}(2005){Soderberg}, {Kulkarni}, {Berger}, {Chevalier}, {Frail}, {Fox}, \& {Walker}}]{2005ApJ...621..908S}
{Soderberg}, A.~M., {Kulkarni}, S.~R., {Berger}, E., {et~al.} 2005, \apj, 621, 908, \dodoi{10.1086/427649}

\bibitem[{{Soderberg} {et~al.}(2006{\natexlab{b}}){Soderberg}, {Kulkarni}, {Nakar}, {Berger}, {Cameron}, {Fox}, {Frail}, {Gal-Yam}, {Sari}, {Cenko}, {Kasliwal}, {Chevalier}, {Piran}, {Price}, {Schmidt}, {Pooley}, {Moon}, {Penprase}, {Ofek}, {Rau}, {Gehrels}, {Nousek}, {Burrows}, {Persson}, \& {McCarthy}}]{2006Natur.442.1014S}
{Soderberg}, A.~M., {Kulkarni}, S.~R., {Nakar}, E., {et~al.} 2006{\natexlab{b}}, \nat, 442, 1014, \dodoi{10.1038/nature05087}

\bibitem[{{Soderberg} {et~al.}(2010){Soderberg}, {Chakraborti}, {Pignata}, {Chevalier}, {Chandra}, {Ray}, {Wieringa}, {Copete}, {Chaplin}, {Connaughton}, {Barthelmy}, {Bietenholz}, {Chugai}, {Stritzinger}, {Hamuy}, {Fransson}, {Fox}, {Levesque}, {Grindlay}, {Challis}, {Foley}, {Kirshner}, {Milne}, \& {Torres}}]{2010Natur.463..513S}
{Soderberg}, A.~M., {Chakraborti}, S., {Pignata}, G., {et~al.} 2010, \nat, 463, 513, \dodoi{10.1038/nature08714}

\bibitem[{{Somalwar} {et~al.}(2023){Somalwar}, {Ravi}, {Dong}, {Hammerstein}, {Hallinan}, {Law}, {Miller}, {Myers}, {Yao}, {Dekany}, {Graham}, {Groom}, {Purdum}, \& {Wold}}]{2023arXiv231003791S}
{Somalwar}, J.~J., {Ravi}, V., {Dong}, D.~Z., {et~al.} 2023, arXiv e-prints, arXiv:2310.03791, \dodoi{10.48550/arXiv.2310.03791}

\bibitem[{Speagle(2020)}]{Speagle_2020}
Speagle, J.~S. 2020, Monthly Notices of the Royal Astronomical Society, 493, 3132–3158, \dodoi{10.1093/mnras/staa278}

\bibitem[{{Tacchella} {et~al.}(2022){Tacchella}, {Conroy}, {Faber}, {Johnson}, {Leja}, {Barro}, {Cunningham}, {Deason}, {Guhathakurta}, {Guo}, {Hernquist}, {Koo}, {McKinnon}, {Rockosi}, {Speagle}, {van Dokkum}, \& {Yesuf}}]{2022ApJ...926..134T}
{Tacchella}, S., {Conroy}, C., {Faber}, S.~M., {et~al.} 2022, \apj, 926, 134, \dodoi{10.3847/1538-4357/ac449b}

\bibitem[{{Tachibana} \& {Miller}(2018)}]{2018PASP..130l8001T}
{Tachibana}, Y., \& {Miller}, A.~A. 2018, \pasp, 130, 128001, \dodoi{10.1088/1538-3873/aae3d9}

\bibitem[{{Taggart} \& {Perley}(2021)}]{2021MNRAS.503.3931T}
{Taggart}, K., \& {Perley}, D.~A. 2021, \mnras, 503, 3931, \dodoi{10.1093/mnras/stab174}

\bibitem[{{Tchekhovskoy} {et~al.}(2014){Tchekhovskoy}, {Metzger}, {Giannios}, \& {Kelley}}]{2014MNRAS.437.2744T}
{Tchekhovskoy}, A., {Metzger}, B.~D., {Giannios}, D., \& {Kelley}, L.~Z. 2014, \mnras, 437, 2744, \dodoi{10.1093/mnras/stt2085}

\bibitem[{{Teboul} \& {Metzger}(2023)}]{2023ApJ...957L...9T}
{Teboul}, O., \& {Metzger}, B.~D. 2023, \apjl, 957, L9, \dodoi{10.3847/2041-8213/ad0037}

\bibitem[{{Thompson} \& {Duncan}(1993)}]{1993ApJ...408..194T}
{Thompson}, C., \& {Duncan}, R.~C. 1993, \apj, 408, 194, \dodoi{10.1086/172580}

\bibitem[{{Tonry} {et~al.}(2018){Tonry}, {Denneau}, {Heinze}, {Stalder}, {Smith}, {Smartt}, {Stubbs}, {Weiland}, \& {Rest}}]{2018PASP..130f4505T}
{Tonry}, J.~L., {Denneau}, L., {Heinze}, A.~N., {et~al.} 2018, \pasp, 130, 064505, \dodoi{10.1088/1538-3873/aabadf}

\bibitem[{{Uddin} {et~al.}(2020){Uddin}, {Burns}, {Phillips}, {Suntzeff}, {Contreras}, {Hsiao}, {Morrell}, {Galbany}, {Stritzinger}, {Hoeflich}, {Ashall}, {Piro}, {Freedman}, {Persson}, {Krisciunas}, \& {Brown}}]{2020ApJ...901..143U}
{Uddin}, S.~A., {Burns}, C.~R., {Phillips}, M.~M., {et~al.} 2020, \apj, 901, 143, \dodoi{10.3847/1538-4357/abafb7}

\bibitem[{{van Eerten} {et~al.}(2012){van Eerten}, {van der Horst}, \& {MacFadyen}}]{2012ApJ...749...44V}
{van Eerten}, H., {van der Horst}, A., \& {MacFadyen}, A. 2012, \apj, 749, 44, \dodoi{10.1088/0004-637X/749/1/44}

\bibitem[{{Vergani} {et~al.}(2015){Vergani}, {Salvaterra}, {Japelj}, {Le Floc'h}, {D'Avanzo}, {Fernandez-Soto}, {Kr{\"u}hler}, {Melandri}, {Boissier}, {Covino}, {Puech}, {Greiner}, {Hunt}, {Perley}, {Petitjean}, {Vinci}, {Hammer}, {Levan}, {Mannucci}, {Campana}, {Flores}, {Gomboc}, \& {Tagliaferri}}]{2015A&A...581A.102V}
{Vergani}, S.~D., {Salvaterra}, R., {Japelj}, J., {et~al.} 2015, \aap, 581, A102, \dodoi{10.1051/0004-6361/20142501310.48550/arXiv.1409.7064}

\bibitem[{{Woodland} {et~al.}(2023){Woodland}, {Mannings}, {Prochaska}, {Ryder}, {Marnoch}, {Jorgenson}, {Simha}, {Tejos}, {Gordon}, {Fong}, {Kilpatrick}, {Deller}, \& {Glowacki}}]{2023arXiv231201578W}
{Woodland}, M.~N., {Mannings}, A.~G., {Prochaska}, J.~X., {et~al.} 2023, arXiv e-prints, arXiv:2312.01578, \dodoi{10.48550/arXiv.2312.01578}

\bibitem[{{Wright} {et~al.}(2010){Wright}, {Eisenhardt}, {Mainzer}, {Ressler}, {Cutri}, {Jarrett}, {Kirkpatrick}, {Padgett}, {McMillan}, {Skrutskie}, {Stanford}, {Cohen}, {Walker}, {Mather}, {Leisawitz}, {Gautier}, {McLean}, {Benford}, {Lonsdale}, {Blain}, {Mendez}, {Irace}, {Duval}, {Liu}, {Royer}, {Heinrichsen}, {Howard}, {Shannon}, {Kendall}, {Walsh}, {Larsen}, {Cardon}, {Schick}, {Schwalm}, {Abid}, {Fabinsky}, {Naes}, \& {Tsai}}]{2010AJ....140.1868W}
{Wright}, E.~L., {Eisenhardt}, P. R.~M., {Mainzer}, A.~K., {et~al.} 2010, \aj, 140, 1868, \dodoi{10.1088/0004-6256/140/6/1868}

\bibitem[{{Yuan} {et~al.}(2025){Yuan}, {Winter}, {Zhang}, {Murase}, \& {Zhang}}]{2025ApJ...982..196Y}
{Yuan}, C., {Winter}, W., {Zhang}, B.~T., {Murase}, K., \& {Zhang}, B. 2025, \apj, 982, 196, \dodoi{10.3847/1538-4357/adbbde}

\bibitem[{{Yungelson} \& {Livio}(1998)}]{1998ApJ...497..168Y}
{Yungelson}, L., \& {Livio}, M. 1998, \apj, 497, 168, \dodoi{10.1086/305455}

\bibitem[{{Zou} {et~al.}(2017){Zou}, {Zhou}, {Fan}, {Zhang}, {Zhou}, {Nie}, {Peng}, {McGreer}, {Jiang}, {Dey}, {Fan}, {He}, {Jiang}, {Lang}, {Lesser}, {Ma}, {Mao}, {Schlegel}, \& {Wang}}]{2017PASP..129f4101Z}
{Zou}, H., {Zhou}, X., {Fan}, X., {et~al.} 2017, \pasp, 129, 064101, \dodoi{10.1088/1538-3873/aa65ba}

\end{thebibliography}
\bibliographystyle{aasjournal}

\end{document}